\pdfoutput=1
\documentclass[twoside,letterpaper,titlepage,nofootinbib,prd,12pt,superscriptaddress,floatfix]{revtex4-1}

\usepackage{ifpdf}
\usepackage[english]{babel}
\usepackage{subfigure}
\usepackage{graphicx}
\usepackage{epsfig}
\usepackage{amsmath}
\usepackage{amssymb}
\usepackage{amsfonts}
\usepackage{slashed}
\usepackage{bbm}
\usepackage{ctable,longtable}
\usepackage{cancel}
\usepackage{fancyhdr,graphicx}
\usepackage{color}
\usepackage{lineno}
\def\mum{\,\mu\text{m}}

\def\SU2U1{{\rm SU}(2)\times{\rm U}(1)}

\mathchardef\qsm=63
\mathchardef\pls=43
\mathchardef\mns=512
\mathchardef\plm=518
\mathchardef\eql=61
\mathchardef\smallleft=300
\mathchardef\smallright=301
\mathchardef\perslsh=47
\mathchardef\les=316
\mathchardef\gre=318
\mathchardef\leq=532
\mathchardef\grq=533
\chardef\usc=95
\chardef\til=126

\def\sqr#1#2#3{{\vcenter{\hrule height.#3ex\hbox{\vrule width.#2ex height#1ex
    \kern#1ex\vrule width.#3ex}\hrule height.#2ex}}}

\def\angleto{\vrule width.035em height2.1ex depth-.56ex\unskip\kern-.6ex\to}
\def\perchc#1{{\raise.4ex\hbox{$\mkern4mu#1{\it\perslsh}_
             {\mkern-5mu\scriptscriptstyle{{\rm o}\!{\rm o}}}^
             {\mkern-12.8mu\scriptscriptstyle{\rm o}}$}}}

\catcode`\@=11 
\def\parenbar{\mathpalette\p@renb@r}
\def\p@renb@r#1#2{\vbox{%
  \ifx#1\scriptscriptstyle \dimen@.7em\dimen@ii.2em\else
  \ifx#1\scriptstyle \dimen@.8em\dimen@ii.25em\else
  \dimen@1em\dimen@ii.4em\fi\fi \offinterlineskip
  \ialign{\hfill##\hfill\cr
    \vbox{\hrule width\dimen@ii}\cr
    \noalign{\vskip-.3ex}%
    \hbox to\dimen@{$\mathchar300\hfil\mathchar301$}\cr
    \noalign{\vskip-.3ex}%
    $#1#2$\cr}}}
\catcode`\@=12 

\newbox\struttbox
\setbox\struttbox=\hbox{\vrule height1.65ex depth.485ex width0pt}
\def\strutt{\relax\ifmmode\copy\struttbox\else\unhcopy\struttbox\fi}
\def\stru#1#2{\relax\ifmmode\hbox{\vrule height#1 depth#2 width0pt}
\else\vrule height#1 depth#2 width0pt\fi}

\def\ronum#1{\uppercase\expandafter{\romannumeral#1}}
\def\ronuml#1{\expandafter{\romannumeral#1}}

\DeclareMathAlphabet{\mathbf}{OT1}{cmr}{bx}{sl}

 {\end{list}}
 {\end{list}}
 {\end{list}}

\catcode`\@=11 
\newlength{\@fninsert}
\newlength{\@fnwidth}
\renewcommand{\@makefntext}[1]%
  {\noindent\makebox[\@fninsert][r]{\@makefnmark}\hfil%
  \parbox[t]{\@fnwidth}{#1}}
\catcode`\@=12 
\catcode`\@=11 
\renewcommand\section{\@startsection{section}{1}{\z@}%
                                   {-3.5ex \@plus -1ex \@minus -.2ex}%
                                   {2.3ex \@plus.2ex}%
                                   {\normalfont\Large\bfseries}}
\renewcommand\subsection{\@startsection{subsection}{2}{\z@}%
                                   {-3.25ex\@plus -1ex \@minus -.2ex}%
                                   {1.5ex \@plus .2ex}%
                                   {\normalfont\large\bfseries}}
\renewcommand\subsubsection{\@startsection{subsubsection}{3}{\z@}%
                                   {-3.25ex\@plus -1ex \@minus -.2ex}%
                                   {1.5ex \@plus .2ex}%
                                   {\normalfont\large\bfseries}}
\renewcommand\paragraph{\@startsection{paragraph}{4}{\z@}%
                                   {3.25ex \@plus1ex \@minus.2ex}%
                                   {1.2ex \@plus .2ex}%
                                   {\normalfont\normalsize\bfseries}}
\catcode`\@=12 

\newcommand{\bi}{\begin{itemize}}
\newcommand{\ei}{\end{itemize}}
\newcommand{\be}{\begin{equation}}
\newcommand{\ee}{\end{equation}}
\newcommand{\bea}{\begin{eqnarray}}
\newcommand{\eea}{\end{eqnarray}}

\newcommand{\Ref}{Ref.}

\begin{document}
%
%
\selectlanguage{english}
\makeatletter
%
%
%
\pagestyle{empty}
\thispagestyle{empty}
%
%
\title{Neutrinos from STORed Muons\\ Letter of Intent\\ to the Fermilab Physics Advisory Committee}
\date{\today}
\clearpage
%
%
%
\vspace{-8mm}
\thispagestyle{plain}
\author{P.~Kyberd}
\author{D.R.~Smith}
\affiliation{Brunel University}
\author{L.~Coney}
\affiliation{University of California, Riverside}
\author{S.~Pascoli}
\affiliation{Institute for Particle Physics Phenomenology,  Durham University}
\author{C.~Ankenbrandt}
\author{S.J.~Brice}
\author{A.D.~Bross\footnote{Corresponding author: bross@fnal.gov}}
\author{H.~Cease}
\author{J.~Kopp}
\author{N.~Mokhov}
\author{J.~Morfin}
\author{D.~Neuffer}
\author{M.~Popovic}
\author{P.~Rubinov}
\author{S.~Striganov}
\affiliation{Fermi National Accelerator Laboratory}
\author{A.~Blondel}
\author{A.~Bravar}
\author{E.~Noah}
\affiliation{University of Geneva}
\author{R.~Bayes}
\author{F.J.P.~Soler}
\affiliation{University of Glasgow}
\author{A.~Dobbs}
\author{K.~Long}
\author{J.~Pasternak}
\author{E.~Santos}
\author{M.O.~Wascko}
\affiliation{Imperial College London}
\author{S.K.~Agarwalla}
\affiliation{Instituto de Fisica Corpuscular, CSIC and Universidad de Valencia}
\author{S.A.~Bogacz}
\affiliation{Thomas Jefferson National Accelerator Facility}
\author{Y.~Mori}
\author{J.B.~Lagrange}
\affiliation{Kyoto University}
\author{A.~de~Gouv\^ea}
\affiliation{Northwestern University}
\author{Y.~Kuno}
\author{A.~Sato}
\affiliation{Osaka University}
\author{V.~Blackmore}
\author{J.~Cobb}
\author{C.~D.~Tunnell}
\affiliation{Oxford University, Subdepartment of Particle Physics}
\author{J.M.~Link}
\author{P.~Huber}
\affiliation{Center for Neutrino Physics, Virginia Polytechnic Institute and State University}
\author{W.~Winter}
\affiliation{Institut f{\"u}r theoretische Physik und Astrophysik, Universit{\"a}t W{\"u}rzburg}
\thispagestyle{empty}
\maketitle
\clearpage
%
%
\parindent 10pt
\pagenumbering{roman}                   
\setcounter{page}{1}
\thispagestyle{plain}
\pagestyle{plain}
%
\tableofcontents
\pagenumbering{arabic}                   
\setcounter{page}{1}
\cleardoublepage
\pagestyle{plain}
\section{Overview}
\label{sec:Overview}
The idea of using a muon storage ring to produce a high-energy ($\simeq$ 50 GeV) neutrino beam for experiments
was first discussed by Koshkarev \cite{Koshkarev:1974my} in 1974.  A detailed description of a muon storage ring for neutrino oscillation experiments was first produced by Neuffer \cite{NeufferTelmark} in 1980.  In his paper, Neuffer studied muon decay rings 
with E$_\mu$ of 8, 4.5 and 1.5 GeV.  With his 4.5 GeV ring design, he achieved a figure of merit of $\simeq 6\times10^9$
useful neutrinos per $3\times10^{13}$ protons on target.  The facility we describe here ($\nu$STORM) is essentially the same 
facility proposed in 1980 and would utilize a 3-4 GeV/c muon storage ring to study eV-scale oscillation physics and, 
in addition, could add 
significantly to our understanding of $\nu_e$ and $\nu_\mu$ cross sections.  In particular the facility can:

\begin{enumerate}
\setlength{\leftmargin}{0.5in}
\item address the large $\Delta$m$^2$ oscillation regime and make a major contribution to the study of sterile 
neutrinos,\vspace{-3mm}
\item make precision $\nu_e$ and $\bar{\nu}_e$ cross-section measurements,\vspace{-3mm}
\item provide a technology ($\mu$ decay ring) test demonstration and $\mu$ beam diagnostics test bed,\vspace{-3mm}
\item provide a precisely understood $\nu$ beam for detector studies.\vspace{-3mm}
\end{enumerate}

\begin{figure}[htpb]
  \centering{
    \includegraphics[width=0.6\textwidth]{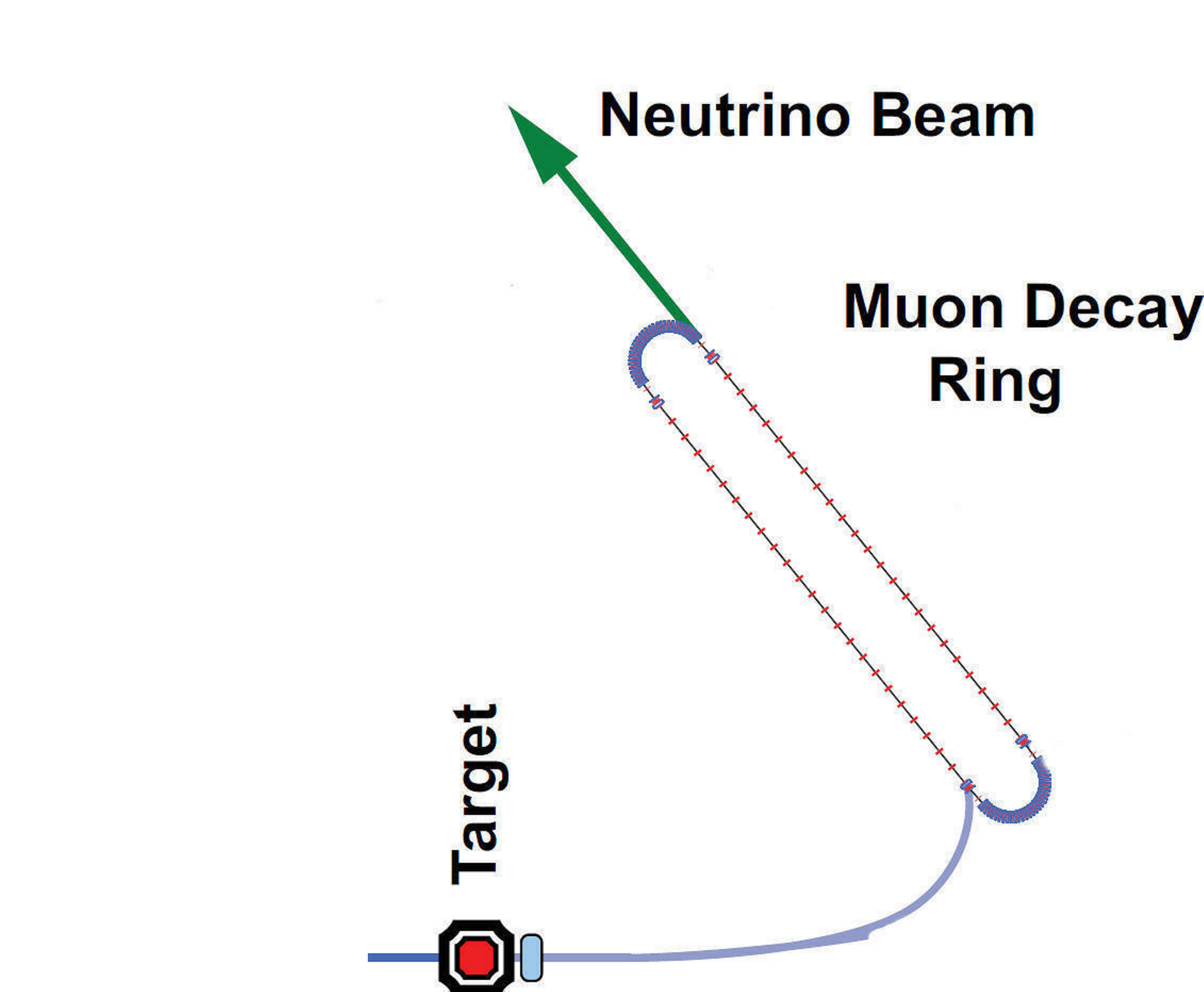}}
  \caption{Schematic of the facility}
  \label{fig:STORM}
\end{figure}

The facility is the simplest implementation of the Neutrino Factory concept \cite{Geer:1997iz}. In our case,
60 GeV/c protons are used to produce pions off a conventional solid target.  The pions are collected with a focusing device 
(horn or lithium lens) and are then transported to, and injected into, a storage ring.  The pions that decay in the first straight of the ring can yield a muon that is captured in the ring.  The circulating muons then 
subsequently decay into electrons and neutrinos.  We are starting with a storage ring design that
is optimized for 3.8 GeV/c muon momentum. This momentum was selected to maximize the physics reach for both oscillation and the cross section physics.  See Fig.~\ref{fig:STORM} for a schematic of the facility.

It would also be possible to create a $\pi \rightarrow \mu$ decay channel and inject the muons into the decay ring with a kicker magnet.  This scheme would have the advantage that the transport channel could be longer than the straight in the decay ring and thus allow for more $\pi$ decays to result in a useful $\mu$.  This does complicate the facility design, however, due to the need for the kicker magnet and the desire to use single-turn extraction from the Main Injector.

Muon decay yields a neutrino beam of precisely known flavor content and energy.  For example 
for positive muons:  $\mu^+ \rightarrow e^+$ + $\bar{\nu}_\mu$ + $\nu_e$.
In addition, if the circulating muon flux in the ring is measured
accurately (with beam-current transformers, for example), then the neutrino beam flux is also accurately known.
Near and far detectors are placed along the line of one of the straight sections of the racetrack decay ring.
The near detector can be placed at 20-50 meters from
the end of the straight.  A near detector for disappearance measurements will be identical to the 
far detector, but only about one tenth the fiducial mass.  It will require a $\mu$ catcher, however.
Additional purpose-specific near detectors can also be located in the near hall and will measure neutrino-nucleon cross sections. $\nu$STORM can provide the first precision measurements of $\nu_e$ and $\bar{\nu}_e$ cross sections which are important for future long-baseline experiments. 
A far detector at $\simeq$ 2000 m would study neutrino oscillation physics and would be capable
of performing searches in both appearance and disappearance channels.
The experiment will take advantage of the ``golden channel" of oscillation appearance $\nu_e \rightarrow \nu_\mu$, where the resulting 
final state has a muon of the wrong-sign from interactions of the $\bar{\nu}_\mu$ in the beam.
In the case of $\mu^+$s stored in the ring, this would mean the observation of an event with a $\mu^-$.
This detector would need to be magnetized for the
wrong-sign muon appearance channel, as is the case for the current baseline Neutrino Factory detector \cite{NF:2011aa}.
A number of possibilities for the far detector exist.  However, a magnetized iron detector similar to that used
in MINOS is likely to be the most straight forward approach for the far detector design.
We believe that it will meet the performance requirements needed to reach our physics goals.
For the purposes of the $\nu$STORM oscillation physics, a detector inspired by MINOS, but with 
thinner plates and much larger excitation current (larger B field) is assumed.

\section{Theoretical and Experimental Motivations}
\label{sec:theory}
\subsection{Sterile neutrinos in extensions of the Standard Model}
\label{sec:sterile-bsm}

Sterile neutrinos, fermions that are uncharged under the $SU(3) \times SU(2)
\times U(1)$ gauge group, arise naturally in many extensions to the Standard
Model. Even where they are not an integral part of a model, they can
usually be easily accommodated. A detailed overview of the phenomenology of
sterile neutrinos and of related model building considerations is given
in \cite{Abazajian:2012ys}.

For instance, in Grand Unified Theories (GUTs), fermions are grouped into
multiplets of a large gauge group, of which $SU(3) \times SU(2) \times U(1)$ is
a subgroup.  If these multiplets contain not only the known quarks and leptons,
but also additional fermions, these new fermions will, after the breaking of
the GUT symmetry, often behave like gauge singlets (see for
instance~\cite{Bando:1998ww, Ma:1995xk, Shafi:1999rm, Babu:2004mj} for GUT
models with sterile neutrinos).

Models attempting to explain the smallness of neutrino masses through a seesaw
mechanism generically contain sterile neutrinos. While in the most generic
seesaw scenarios, these sterile neutrinos are extremely heavy ($\sim
10^{14}$~GeV) and have very small mixing angles ($\sim 10^{-12}$) with the
active neutrinos, slightly non-minimal seesaw models can easily feature sterile
neutrinos with eV-scale masses and with percent level mixing with the active
neutrinos. Examples for non-minimal seesaw models with relatively light sterile
neutrinos include the split seesaw scenario~\cite{Kusenko:2010ik}, seesaw
models with additional flavor symmetries (see e.g.~\cite{Mohapatra:2001ns}),
models with a Froggatt-Nielsen mechanism~\cite{Froggatt:1978nt, Barry:2011fp},
and extended seesaw models that augment the mechanism by introducing more than
three singlet fermions, as well as additional
symmetries~\cite{Mohapatra:2005wk, Fong:2011xh, Zhang:2011vh}.

Furthermore, sterile neutrinos arise naturally in ``mirror models'', in which
the existence of an extended ``dark sector'', with nontrivial
dynamics of its own, is postulated. If the dark sector is similar to
the visible sector, as is the case, for instance, in string-inspired
$E_8 \times E_8$ models, it is natural to assume that it also contains
neutrinos~\cite{Berezhiani:1995yi, Foot:1995pa, Berezinsky:2002fa}.

Finally, sterile neutrinos also have an impact in cosmology on the evolution of the Early Universe and on astrophysical objects such as supernovae (for a review see ~\cite{Abazajian:2012ys} and 
references therein). By mixing with active neutrinos, they can be produced 
in the Early Universe by oscillations before neutrino decoupling. They could constitute 
the dark matter (DM) of the Universe, if they have masses in the keV range, or part of 
it in the case of lighter masses in the eV range, in which case they contribute to hot 
DM. A thermal population of a light sterile neutrino acts as an additional relativistic 
degree of freedom at sufficiently high temperatures. If present, they affect Big Bang 
Nucleosynthesis, the Cosmic Microwave Background (CMB) and the formation of large scale 
structures such as galaxies and clusters of galaxies. Their effect on the CMB anisotropies 
is due mainly to the change of the matter radiation equality redshift and the sound horizon 
at the time of CMB decoupling and to their anisotropic stress which suppresses the amplitude 
of higher harmonics in the temperature anisotropy spectrum. Interestingly, recent observations 
of the CMB by WMAP and of the CMB damping tail by ACT and SPT indicate a value of the effective 
number of relativistic degrees of freedom higher than 3 at a significant confidence level, 
suggesting the existence of sterile neutrinos or of a thermal population of other light 
particles, in addition to 3 active neutrinos. If future observations, and in particular 
Planck, confirm this result, testing the mixing angles required for a thermal distribution 
of sterile neutrinos to be produced in the Early Universe will be of paramount importance 
in order to establish the identity of the additional relativistic degrees of freedom in 
the Universe. $\nu$STORM could test a large part of the required parameter space, having 
sensitivity to the relevant masses and mixing angles with different flavors.

\subsection{Experimental hints for light sterile neutrinos}
\label{sec:sterile-hints}

While the theoretical motivation for the existence of sterile neutrinos
is certainly strong, what has mostly prompted the interest of the scientific
community in this topic are several experimental
results that show significant deviations from the Standard Model predictions.
These results can be interpreted as hints for oscillations involving sterile neutrinos.

The first of these hints was obtained by the LSND collaboration, who
carried out a search for $\bar\nu_\mu \to \bar\nu_e$ oscillations over a
baseline of $\sim 30$~m~\cite{Aguilar:2001ty}. Neutrinos were produced in a
stopped pion source in the decay $\pi^+ \to \mu^+ + \nu_\mu$
and the subsequent decay $\mu^+ \to e^+ \bar\nu_\mu \nu_e$.  Electron
antineutrinos are detected through the inverse beta decay reaction $\bar\nu_e p
\to e^+ n$ in a liquid scintillator detector.  Backgrounds to this search arise
from the decay chain $\pi^- \to \bar\nu_\mu + (\mu^- \to \nu_\mu \bar\nu_e
e^-)$ if negative pions produced in the target decay before they are captured
by a nucleus, and from the reaction $\bar\nu_\mu p \to \mu^+ n$, which is only
allowed for the small fraction of muon antineutrinos produced by pion decay
\emph{in flight} rather than stopped pion decay. The LSND collaboration finds
an excess of $\bar\nu_e$ candidate events above this background with a
significance of more than $3\sigma$. When interpreted as $\bar\nu_\mu \to
\bar\nu_e$ oscillations through an intermediate sterile state $\bar\nu_s$, this
result is best explained by sterile neutrinos with an effective mass squared
splitting $\Delta m^2 \gtrsim 0.2$~eV$^2$ relative to the active neutrinos, and
with an effective sterile-induced $\bar\nu_\mu$--$\bar\nu_e$ mixing angle
$\sin^2 2\theta_{e\mu, \rm eff} \gtrsim 2 \times 10^{-3}$, depending on $\Delta m^2$.

The MiniBooNE
experiment~\cite{AguilarArevalo:2007it,AguilarArevalo:2010wv} was designed to
test the neutrino oscillation interpretation of the LSND result using a
different technique, namely neutrinos from a horn-focused pion beam. While a
MiniBooNE search for $\nu_\mu \to \nu_e$ oscillations indeed disfavors most
(but not all) of the parameter region preferred by LSND in the simplest model
with only one sterile neutrino~\cite{AguilarArevalo:2007it}, the experiment
obtains results \emph{consistent} with LSND when running in antineutrino mode
and searching for $\bar\nu_\mu \to \bar\nu_e$.  Due to low statistics, however,
the antineutrino data favors LSND-like oscillations over the null hypothesis
only at the 90\% confidence level.  Moreover, MiniBooNE observes a yet
unexplained $3.0\sigma$ excess of $\nu_e$-like events (and, with smaller
significance also of $\bar\nu_e$ events) at low energies, 200~MeV~$\lesssim
E_\nu \lesssim$~475~MeV, outside the energy range where LSND-like oscillations
would be expected.

A third hint for the possible existence of sterile neutrinos is provided by the
so-called reactor antineutrino anomaly. In 2011, Mueller {\it et al.}\
published a new {\it ab initio} computation of the expected neutrino fluxes from
nuclear reactors~\cite{Mueller:2011nm}. Their results improve upon a 1985
calculation by Schreckenbach~\cite{Schreckenbach:1985ep} by using up-to-date
nuclear databases, a careful treatment of systematic uncertainties and various
other corrections and improvements that were neglected in the earlier
calculation. Mueller {\it et al.}\ find that the predicted antineutrino flux from a
nuclear reactor is about 3\% higher than previously thought. This result, which
was later confirmed by Huber~\cite{Huber:2011wv}, implies that short baseline
reactor experiments have observed a $3\sigma$ \emph{deficit} of antineutrinos
compared to the prediction~\cite{Mention:2011rk, Abazajian:2012ys}. 
It needs to be emphasized that the significance of the
deficit depends crucially on the systematic uncertainties associated with the
theoretical prediction, some of which are difficult to estimate reliably.  If
the reactor antineutrino deficit is interpreted as $\bar\nu_e \to \bar\nu_s$
disappearance via oscillation, the required 2-flavor oscillation parameters
are $\Delta m^2 \gtrsim 1$~eV$^2$ and $\sin^2 2\theta_{ee,\rm eff} \sim 0.1$.

Such short-baseline oscillations could also explain another experimental
result: the Gallium anomaly. The GALLEX and SAGE solar neutrino
experiments used electron neutrinos from intense artificial radioactive sources
to test their radiochemical detection principle~\cite{Anselmann:1994ar,
Hampel:1997fc, Abdurashitov:1996dp, Abdurashitov:1998ne, Abdurashitov:2005tb}.
Both experiments observed fewer $\nu_e$ from the source than expected. The
statistical significance of the deficit is above 99\% and can be interpreted in
terms of short-baseline $\bar\nu_e \to \bar\nu_s$ disappearance with $\Delta
m^2 \gtrsim 1$~eV$^2$ and $\sin^2 2\theta_{ee,\rm eff} \sim
0.1$--$0.8$.~\cite{Acero:2007su, Giunti:2010wz, Giunti:2010zu}.

\subsection{Constraints and global fit}
\label{sec:sterile-global-fit}

While the previous section shows that there is an intriguing accumulation of
hints for the existence of new oscillation effects---possibly related to
sterile neutrinos---in short-baseline experiments, these hints are not
undisputed. Several short-baseline oscillation experiments did \emph{not}
confirm the observations from LSND, MiniBooNE, reactor experiments, and Gallium
experiments, and place very strong limits on the relevant regions of parameter
space in sterile neutrino models. To assess the viability of these models it is
necessary to carry out a global fit to all relevant experimental data sets, and
several groups have endeavored to do so~\cite{Kopp:2011qd, Giunti:2011cp,
Karagiorgi:2011ut, Giunti:2011hn, Giunti:2011gz, Abazajian:2012ys}.
In Fig. 2 ~\cite{Kopp:2011qd, Abazajian:2012ys}, we
show the current constraints on the parameter space of a $3+1$ model (a model
with three active neutrinos and one sterile neutrino). We have projected the
parameter space onto a plane spanned by the mass squared difference $\Delta
m^2$ between the heavy, mostly sterile mass eigenstate and the light, most
active ones and by the effective amplitude $\sin^2 2\theta_{e\mu, \rm eff}$ for
$\nu_\mu \to \nu_e$ 2-flavor oscillations to which LSND and MiniBooNE are
sensitive.

\begin{figure}[htpb]
  \includegraphics[width=0.7\textwidth]{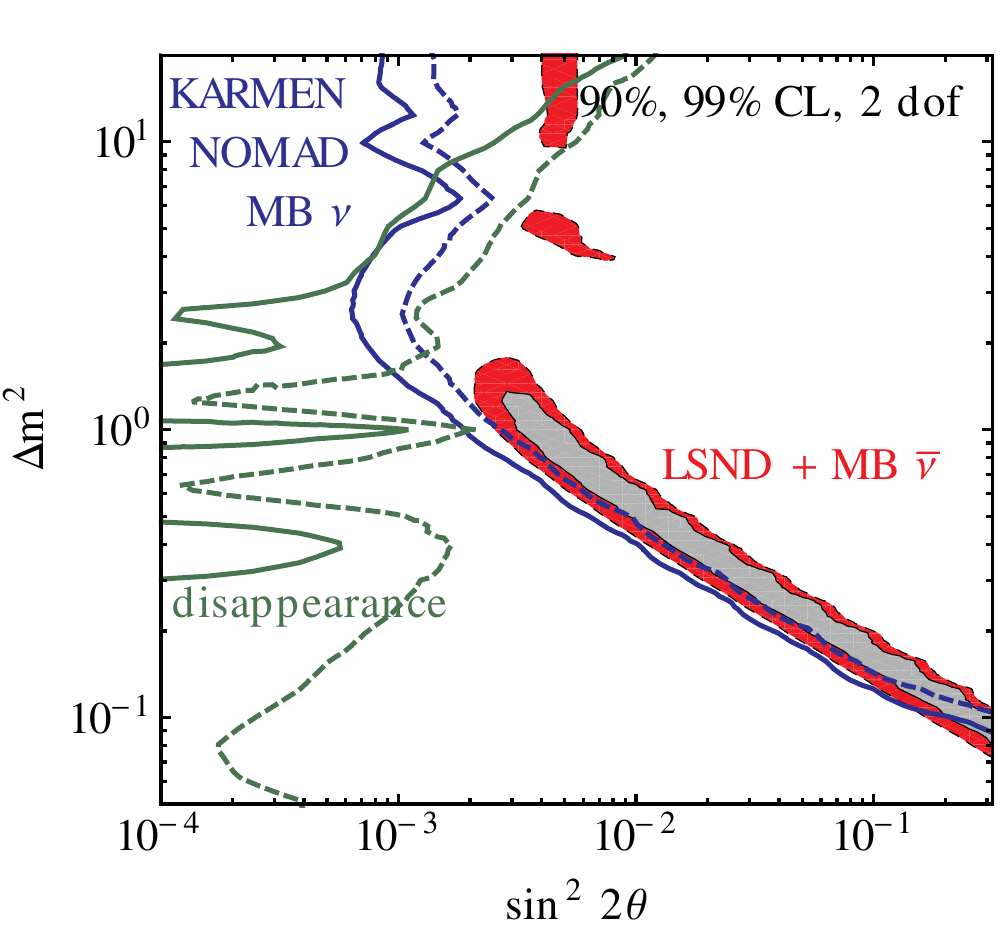}
  \begin{flushleft}
  \caption{Global constraints on sterile neutrinos in a 3+1 model.  We show the
    allowed regions at 90\% and 99\%~CL from a combined analysis of the
    LSND~\cite{Aguilar:2001ty} and MiniBooNE
    antineutrino~\cite{AguilarArevalo:2010wv} signals (filled regions), as well
    as the constraints from the null results of KARMEN~\cite{Armbruster:2002mp},
    NOMAD~\cite{Astier:2001yj} and MiniBooNE
    neutrino~\cite{AguilarArevalo:2007it} appearance searches (blue contour).
    The limit from disappearance experiments (green contours) includes data from
    CDHS~\cite{Dydak:1983zq}, atmospheric neutrinos~\cite{Ashie:2005ik},
    MINOS~\cite{Adamson:2010wi,Adamson:2011ku}, and from SBL reactor
    experiments~\cite{Declais:1994su, Declais:1994ma, Kuvshinnikov:1990ry,
    Vidyakin:1987ue, Kwon:1981ua, Zacek:1986cu, Apollonio:2002gd, Boehm:2001ik}.
    For the latter, we have used the new reactor flux predictions
    from~\cite{Mueller:2011nm}, but we have checked that the results, especially
    regarding consistency with LSND and MiniBooNE $\bar\nu$ data, are
    qualitatively unchanged when the old reactor fluxes are used.  Fits have been
    carried out in the GLoBES framework~\cite{Huber:2004ka, Huber:2007ji} using
    external modules discussed in~\cite{GonzalezGarcia:2007ib, Maltoni:2007zf,
    Akhmedov:2010vy}}
   \end{flushleft}
\label{fig:regions-3p1}
\end{figure}

We see that there is severe tension in the global data set: the parameter
region favored by LSND and MiniBooNE antineutrino data is disfavored at more
than 99\% confidence level by searches for $\nu_e$ ($\bar \nu_e$) and $\bar \nu_\mu$
disappearance. Using a parameter goodness-of-fit
test~\cite{Maltoni:2003cu} to quantify this tension, p-values on the order
of $\text{few} \times 10^{-6}$ are found for the compatibility of LSND
and MiniBooNe $\bar\nu$ data with the rest of the global data set, and
p-values smaller than $10^{-3}$ are found for the compatibility of appearance
data and disappearance data~\cite{Abazajian:2012ys}. The global fit improves somewhat in models
with more than one sterile neutrino, but significant tension remains~\cite{Kopp:2011qd, Abazajian:2012ys}.

One can imagine several possible resolutions to this puzzle:
\vspace{-2mm}
\begin{enumerate}
  \item One or several of the apparent deviations from the standard
    three neutrino oscillation framework discussed in
    section~\ref{sec:sterile-hints} have explanations not related to
    sterile neutrinos.\vspace{-3mm}
  \item One or several of the null results that favor the no-oscillation
    hypothesis are in error.\vspace{-3mm}
  \item There are more than two sterile neutrino flavors. Note that 
    scenarios with one sterile neutrino with an eV scale mass are already in some
    tension with cosmology, even though the existence of one sterile
    neutrino with a mass well below 1~eV is actually preferred by cosmological
    fits~\cite{GonzalezGarcia:2010un, Hamann:2010bk, Giusarma:2011ex,
    Mangano:2011ar}. Cosmological bounds on sterile neutrinos can be avoided in
    non-standard cosmologies~\cite{Hamann:2011ge} or by invoking mechanisms
    that suppress sterile neutrino production in the early
    universe~\cite{Bento:2001xi, Dolgov:2004jw}.\vspace{-3mm}
  \item There are sterile neutrinos plus some other kind of new physics
    at the eV scale. (See for instance~\cite{Akhmedov:2010vy,
    Karagiorgi:2012kw} for an attempt in this direction.)\vspace{-3mm}
\end{enumerate}

We conclude that our understanding of short baseline neutrino oscillations is
currently incomplete. On the one hand, several
experiments indicate deviations from the established three-neutrino framework.
However, none of these hints can be considered conclusive, and moreover, when
interpreted in the simplest sterile neutrino models, they are in severe tension
with existing constraints on the parameter space of these models. An experiment
searching for short-baseline neutrino oscillations with good sensitivity and
well-controlled systematic uncertainties has great potential to clarify the
situation by either finding a new type of neutrino oscillation or by deriving a
strong and robust constraint on any such oscillation. The requirements for this
proposed experiment are as follows:
\begin{itemize}
 \item 
  Direct test of the LSND and MiniBooNE anomalies.\vspace{-2mm}
\item 
 Provide stringent constraints for both $\nu_e$ and $\nu_\mu$ disappearance to
overconstrain $3+N$ oscillation models and to test the Gallium and reactor anomalies
directly.\vspace{-2mm}
\item
 Test the CP- and T-conjugated channels as well, in order to obtain the relevant clues 
for the underlying physics model, such as CP violation in $3+2$ models.\vspace{-2mm}
\end{itemize}
Neutrino production with a muon storage ring is the only option  which can fulfill 
these requirements simultaneously, since both $\nu_e$ ($\bar \nu_e$) and $\bar \nu_\mu$
($\nu_\mu$) are in the beam in equal quantities.
\subsection{Measurement of neutrino-nucleon scattering cross sections}
\label{sec:cross-section}

\begin{figure} 
  \begin{center}
    \includegraphics[width=0.9\linewidth]%
      {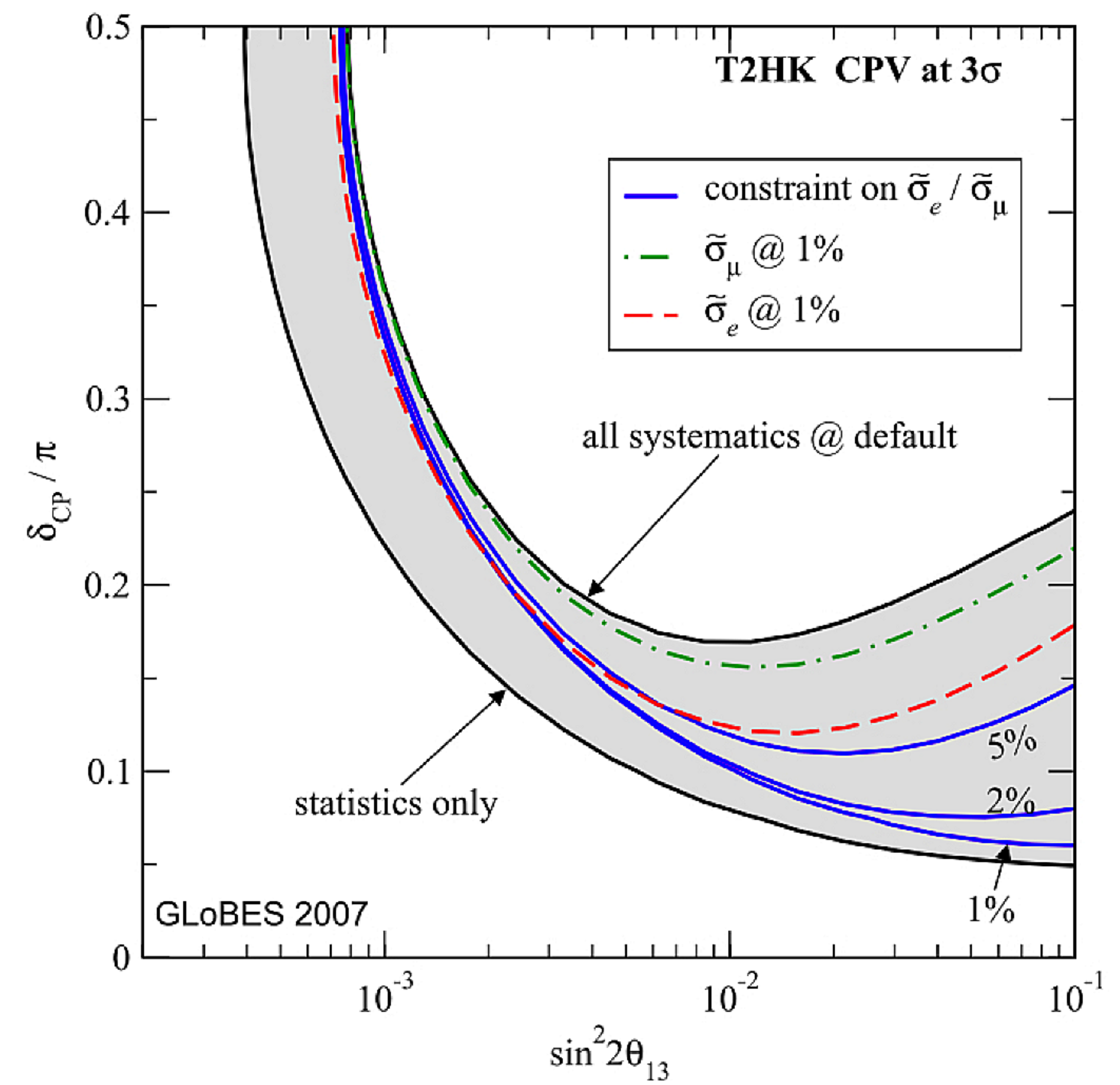}
  \end{center}
  \caption{
    T2HK sensitivity to CP-invariance violation at $3 \sigma$.
    The sensitivity that would be obtained in the absence of
    systematic uncertainties is shown by the lower solid black line.
    Taking systematic errors into account, as described in
    \cite{Huber:2007em} yields the sensitivity shown by the upper
    solid black line.
    The sensitivity that would pertain if the product of the
    efficiency and the (anti)neutrino scattering cross sections
    (denoted $\bar{\sigma}_{\mu,e}$ are known with a precision of 1\%
    are shown by the dashed red, and dot dashed green lines.
    The solid blue lines show the effect of an uncertainty of 1\%, 2\%
    and 5\% on the ratio of the electron- to muon-neutrino times the
    relevant efficiency.
    Figure taken from \cite{Huber:2007em}.
  }
  \label{Fig:T2HKSys}
\end{figure}

A number of recent articles have presented detailed reviews of the
status of neutrino-nucleon scattering cross section measurements in
the context of the oscillation-physics program (see for example
\cite{Hewett:2012ns} and references therein).
The effect of uncertainties in the neutrino scattering cross sections
is to reduce the sensitivity of the present and future short- and
long-baseline experiments.
The impact of the uncertainties on the cross sections is particularly
pernicious at large $\theta_{13}$.
This is illustrated in Fig.~\ref{Fig:T2HKSys} where the sensitivity
of the T2HK experiment to CP-invariance violation is plotted as a
function of $\sin^2 2 \theta_{13}$ \cite{Huber:2007em}.
The experiment considered in this analysis assumes a proton beam power
of 4\,MW is used to generate a conventional super-beam illuminating a
500\,kT water Cherenkov detector at a distance of 295\,km from the
source. 
The analysis assumes a 0.1kT water Cherenkov near detector at a
distance of 2\,km.
Fig.~\ref{Fig:T2HKSys} shows that, for $\theta_{13} \sim 0.1$, the
statistical power of the experiment can only be exploited if the
neutrino scattering cross sections times efficiencies are known with a
precision of $\sim 1\%$ and the ratio of the electron-neutrino cross
section times efficiency to the muon-neutrino cross section times
efficiency is known to $\sim 1\%$.
Experiments that exploit a wide-band neutrino beam with a near/far
detector combination that is capable of resolving the first and second
oscillation maxima are less severely affected by the cross section
errors. 
However, the sensitivity of such experiments to CP-invariance
violation is significantly enhanced if it is assumed that the cross
sections have been determined with a precision of 1\% or better
\cite{Coloma:2012wq}.

The search for the existence of sterile neutrinos through the
measurement of oscillations requires that an anomalous rate of
neutrino appearance, or neutrino disappearance, be
demonstrated. 
This requires that accurate predictions can be made of the neutrino
event rates that would be expected in the absence of active/sterile
neutrino mixing.
The experiment described in this LOI is conceived to rule out, at the 
level of at least $5 \sigma$, the hypothesis that the anomalies
observed in the LSND, MiniBOONE, MINOS and reactor experiments may be
attributed to statistical fluctuations or unexpected background
processes.
To do this requires that the neutrino-nucleon scattering cross
sections are measured accurately.

\begin{figure} 
  \begin{center}
    \includegraphics[width=0.9\linewidth]%
      {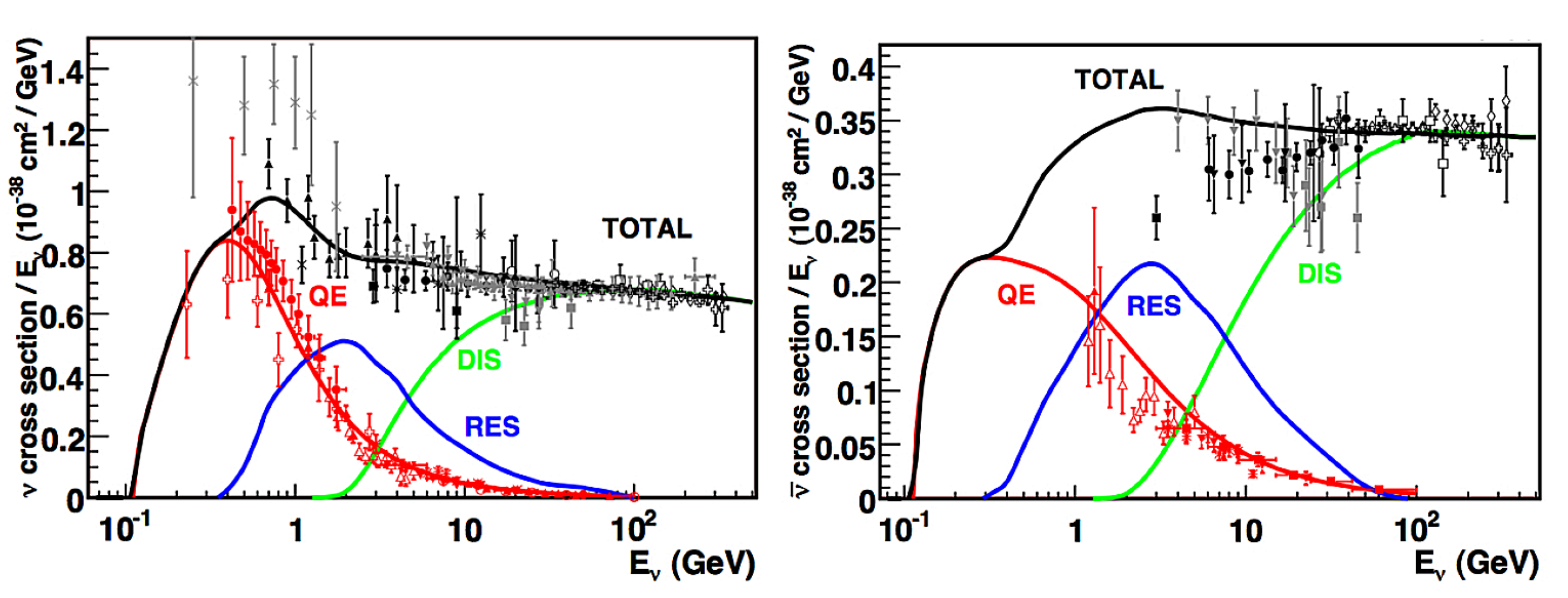}
  \end{center}
  \caption{
    The neutrino-nucleon (left panel) and antineutrino-nucleon (right
    panel) cross sections plotted as a function of (anti)neutrino
    energy \cite{FormaggioZeller:2012:zz}. 
    The data are compared to the expectations of the models described
    in \cite{Casper:2002sd}.
    The processes that contribute to the total cross section (shown by
    the black lines) are: quasi-elastic (QE, red lines) scattering;
    resonance production (RES, blue lines); and deep inelastic
    scattering (DIS, green lines).
    The uncertainties in the energy range of interest are typically 
    $10-40$\%.
    Figure taken from \cite{Hewett:2012ns}.
  }
  \label{Fig:XSect}
\end{figure}

Fig. \ref{Fig:XSect} shows the present data on the charged-current
neutrino-scattering cross sections in the relevant energy range.  
The neutrino flux that will be generated by the 3.8\,GeV stored muon
beam proposed here will allow cross section measurements in the
neutrino-energy range $1-3$\,GeV, the region in which the $\nu_\mu N$
data shown in Fig. \ref{Fig:XSect} is sparse.
Moreover, $\nu_e$ appearance searches rely on $\nu_e N$ cross sections
for which there is essentially no data. 
At present, estimates of the electron-neutrino cross sections are made
by extrapolation of the muon neutrino cross sections.
Such extrapolations suffer from substantial uncertainties arising from
non-perturbative hadronic corrections and it is therefore essential
that detailed measurements of the $\nu_e N$ and $\nu_\mu N$ scattering
cross sections and hadron-production rates are performed.
The $\nu$STORM facility, therefore, has a unique opportunity.  
The flavor composition of the beam and the neutrino energy
spectrum are both known precisely.
In addition, the storage ring instrumentation combined with
measurements at the near detector will allow the neutrino flux to be
measured with a precision of 1\%.
Substantial event rates may be obtained in a fine-grained detector
placed between 20\,m and 50\,m from the storage ring.
Therefore, the objective is to measure the $\nu_e N$ and 
$\nu_\mu N$ scattering cross sections for neutrino energies in the
range $1-3$\,GeV with a precision approaching 1\%.
This will be a critical contribution to the search for sterile
neutrinos and will be of fundamental importance to the present and
next generation of long-baseline neutrino oscillation experiments.%
%
%

\section{Facility}
\label{sec:Facility}
The basic concept for the facility is presented in Fig.~\ref{fig:STORM}.  A high-intensity proton 
source places beam on a target, 
producing a large spectrum of secondary pions.  Forward pions are focused by a collection element into a transport channel. 
Pions decay within the first straight of the decay ring and a fraction of the resulting muons are stored in the ring.
Muon decay within the straight 
sections will produce $\nu$ beams of known flux and flavor via:
$\mu^+ \rightarrow e^+$ + $\bar{\nu}_\mu$ + $\nu_e$ or $\mu^- \rightarrow e^-$ + $\nu_\mu$ + $\bar{\nu}_e$.
For the implementation which is described here, we choose a 3.8 GeV/c storage ring to obtain the desired spectrum 
of $\simeq$ 2 GeV neutrinos (see Fig.~\ref{fig:nuflux}). This 
means that we must capture pions at a momentum of approximately 5 GeV/c.
\subsection{Targeting and capture}
\label{sec:TnC}
\begin{figure}
  \begin{center}$
    \begin{array}{cc}
	  \includegraphics[width=0.48\textwidth]{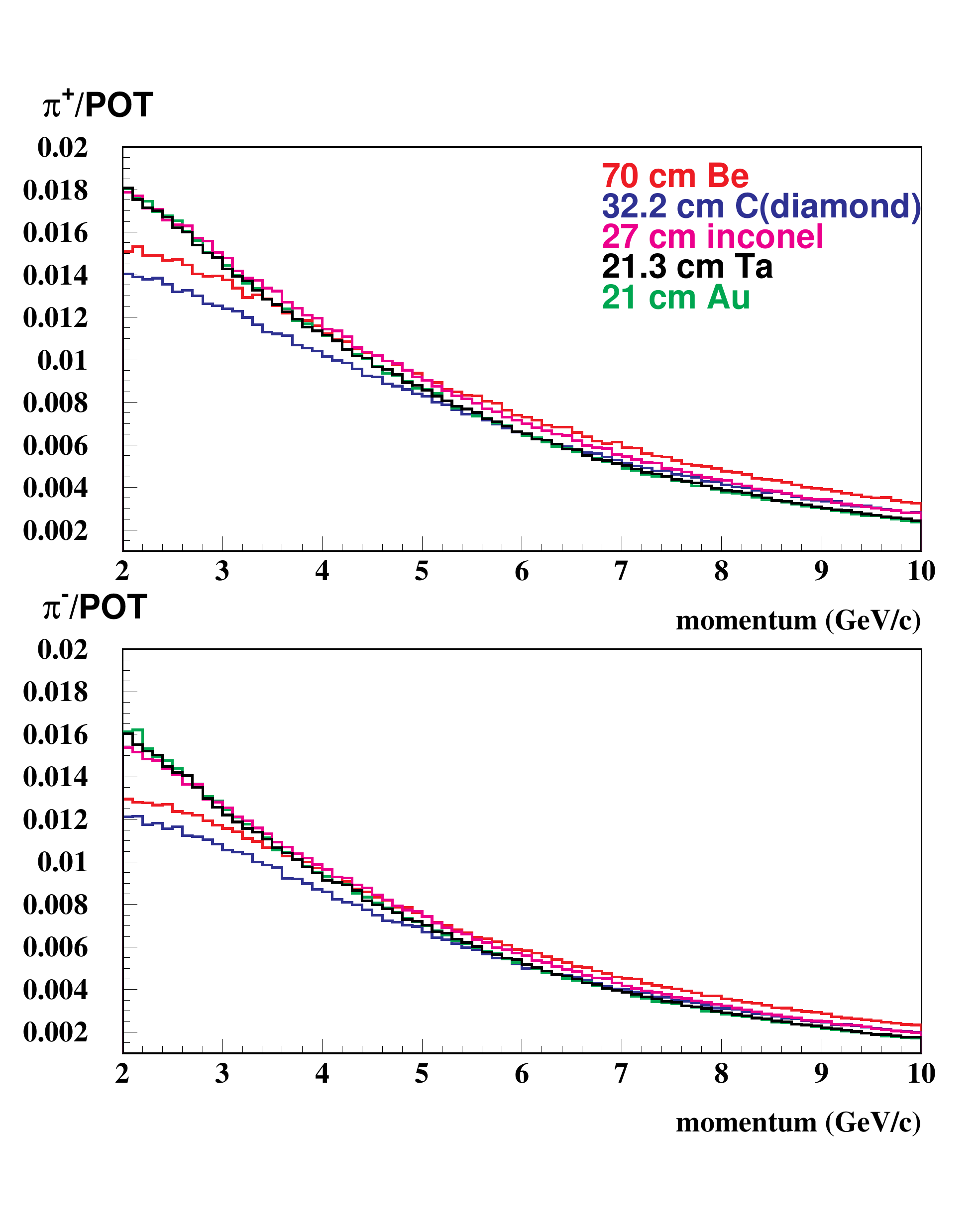} &
	  \includegraphics[width=0.48\textwidth]{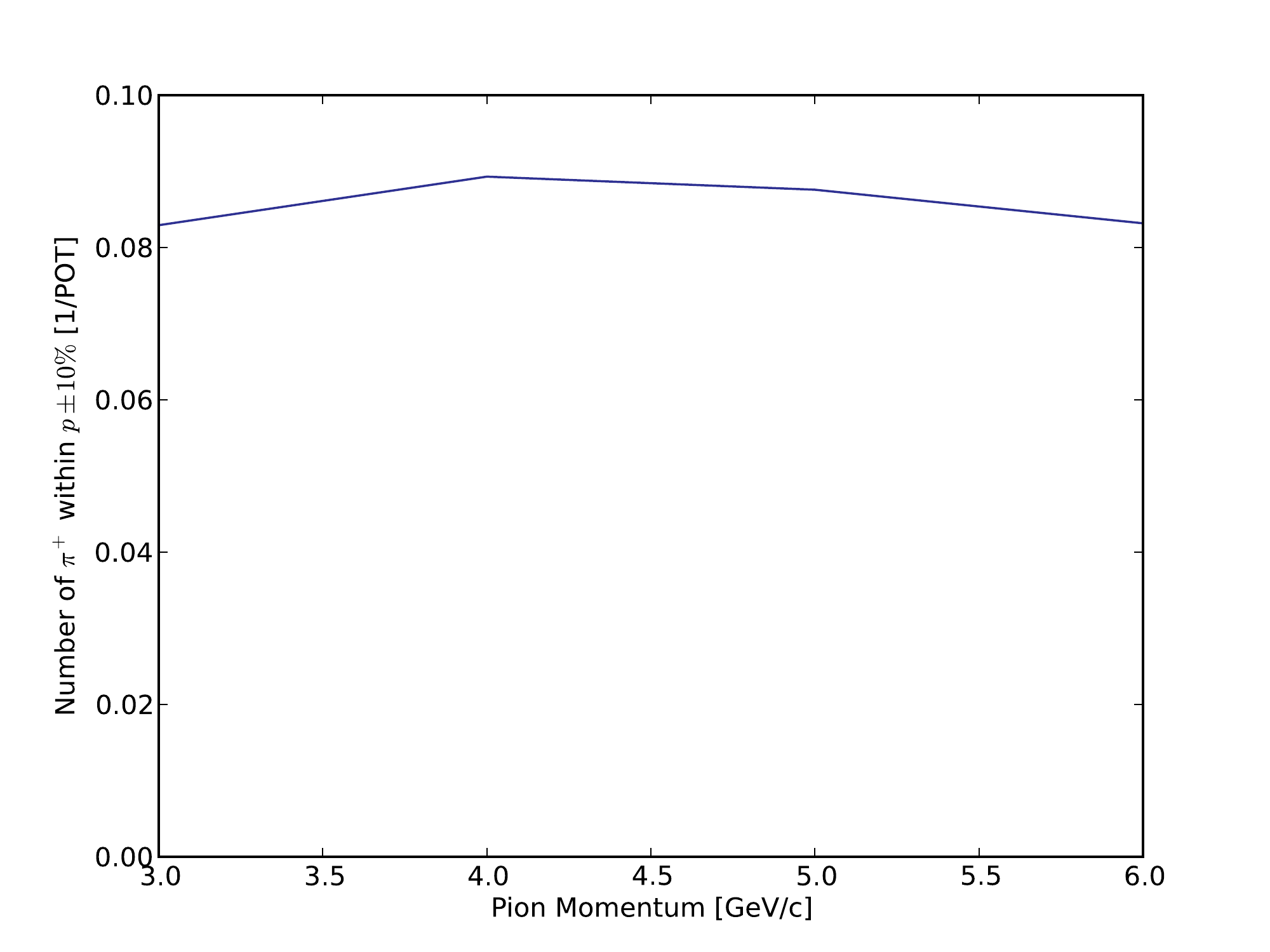}
	\end{array}$
  \end{center}	
\caption{Left: $\pi$ production off various targets into a forward cone of 120 mrad per 100 MeV bin. 
Upper $\pi^+$, lower $\pi^-$. Right: Integrated Production for the case of 70 cm Be target.}
\label{fig:piprod}
\end{figure}

The number of pions produced off various targets by 60 GeV/c protons has been simulated 
with the MARS code \cite{Mokhov:2000en}.  The results of 
this analysis on a number of different targets are shown in Fig.~\ref{fig:piprod} (left) where the number 
in a foward cone of 120 mrad, per 
proton on target, as a function of energy is given.

In Fig.~\ref{fig:piprod} (left) we see that the pion production decreases monotonically with increasing momentum.
Fig.~\ref{fig:piprod} (right) shows number of pions produced off a 70 cm Be target as a function of energy where a
linear interpolation is used to integrate $\pi$(p) in $\pm$10\% momentum bins. We see that that yield is relatively flat
in energy.  Since the integration range is relative (the range increases with increasing momentum), this compensates for
the monotonic decrease shown in Fig.~\ref{fig:piprod} (left).
We have also performed a target optimization based on a conservative estimate for the decay-ring acceptance of 2 mm-radian. This corresponds 
to a decay ring with 11 cm internal radius and a $\beta$ function of 600 cm.  Measurements of positive pion production at 
70 GeV \cite{Bozhko:1980,Barkov:1981} are in $\sim$ 30\% agreement with the MARS predictions for production of pions in
the momentum range of 3-5 GeV/c and at small angles. It is well known that the maximum yield can be achieved with 
a target radius of $\sim 3\times$ the 
proton beam RMS size . The optimal target length depends on the target material and the secondary pion momentum. 
Results of the optimization study are presented in Table~\ref{tab:yield}. We see that
approximately 0.11 $\pi^+$/POT can be collected into a $\pm$ 10\% momentum acceptance off medium/heavy 
targets assuming ideal capture.
\begin{table}
\centering
\caption {$\pi^+$ yield/POT with 60 GeV/c protons, into 2 mm radian acceptance. }
\label{tab:yield}
\begin{tabular}{|l|c|c|c|c|c|c|}
\hline
material & momentum (GeV/c)& $\pm 15\%$ & $\pm 10\%$    &$\pm 5\%$  & target length (cm)  & density (g/cm$^3$)\\
\hline\noalign{\smallskip}
Carbon   & 3   & 0.085       & 0.056  & 0.028 & 27.3 & 3.52   \\
Carbon   & 5   & 0.099       & 0.067  & 0.033 & 32.2 & 3.52   \\
Inconel  & 3   & 0.131       & 0.087  & 0.044 & 19.2 & 8.43   \\
Inconel  & 5   & 0.136       & 0.091  & 0.045 & 27.0 & 8.43   \\
Tantalum  & 3   & 0.164       & 0.109  & 0.054 & 15.3 & 16.6   \\
Tantalum  & 5   & 0.161       & 0.107  & 0.053 & 21.3 & 16.6   \\
Gold      & 3   & 0.177       & 0.118  & 0.059 & 18.0 & 19.32   \\
Gold      & 5   & 0.171       & 0.112  & 0.056 & 21.0 & 19.32   \\
\noalign{\smallskip}\hline
\end{tabular}
\end{table} 

\begin{figure}[h]
  \begin{center}$
    \begin{array}{cc}
	  \includegraphics[width=0.48\textwidth]{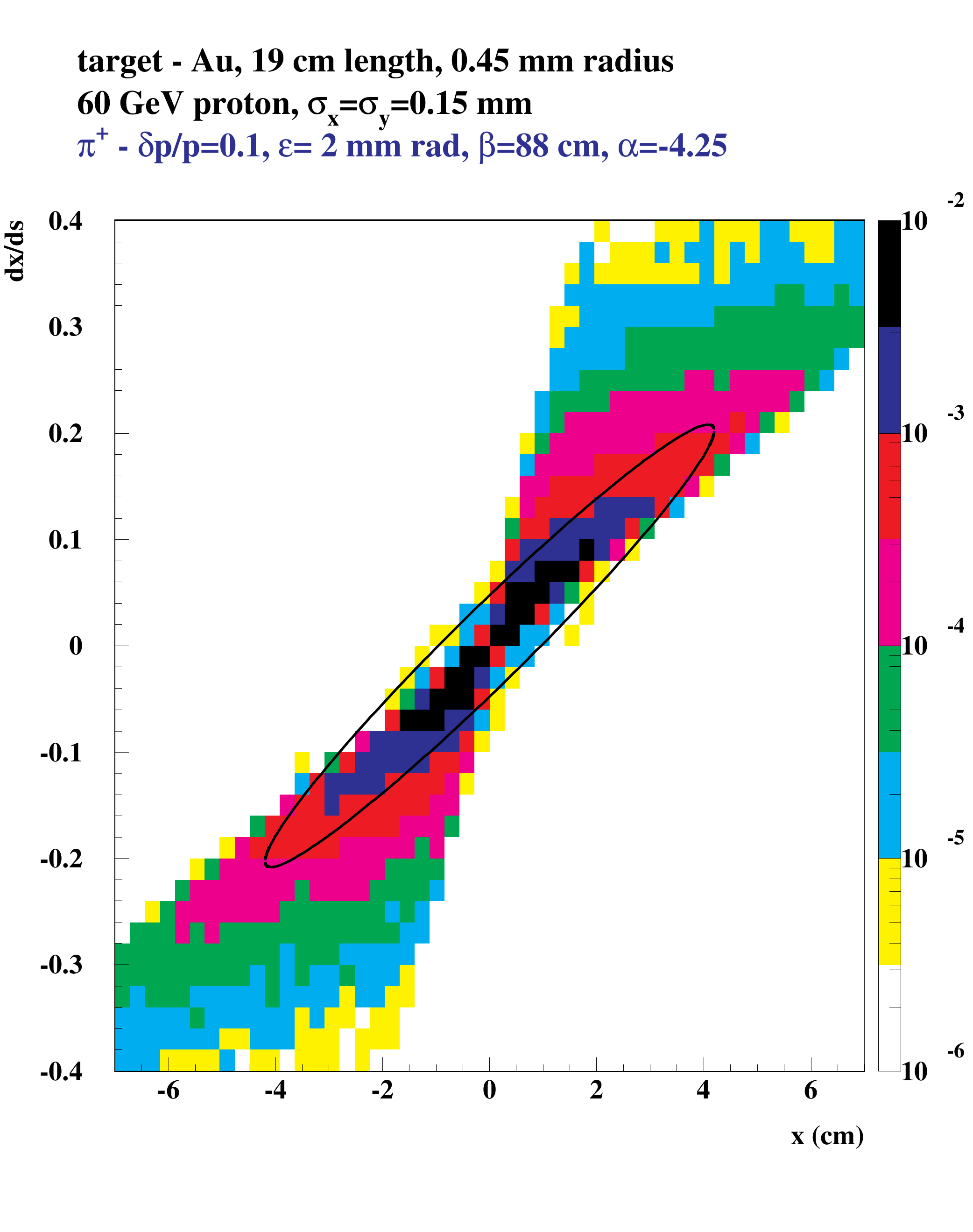} &
	  \includegraphics[width=0.48\textwidth]{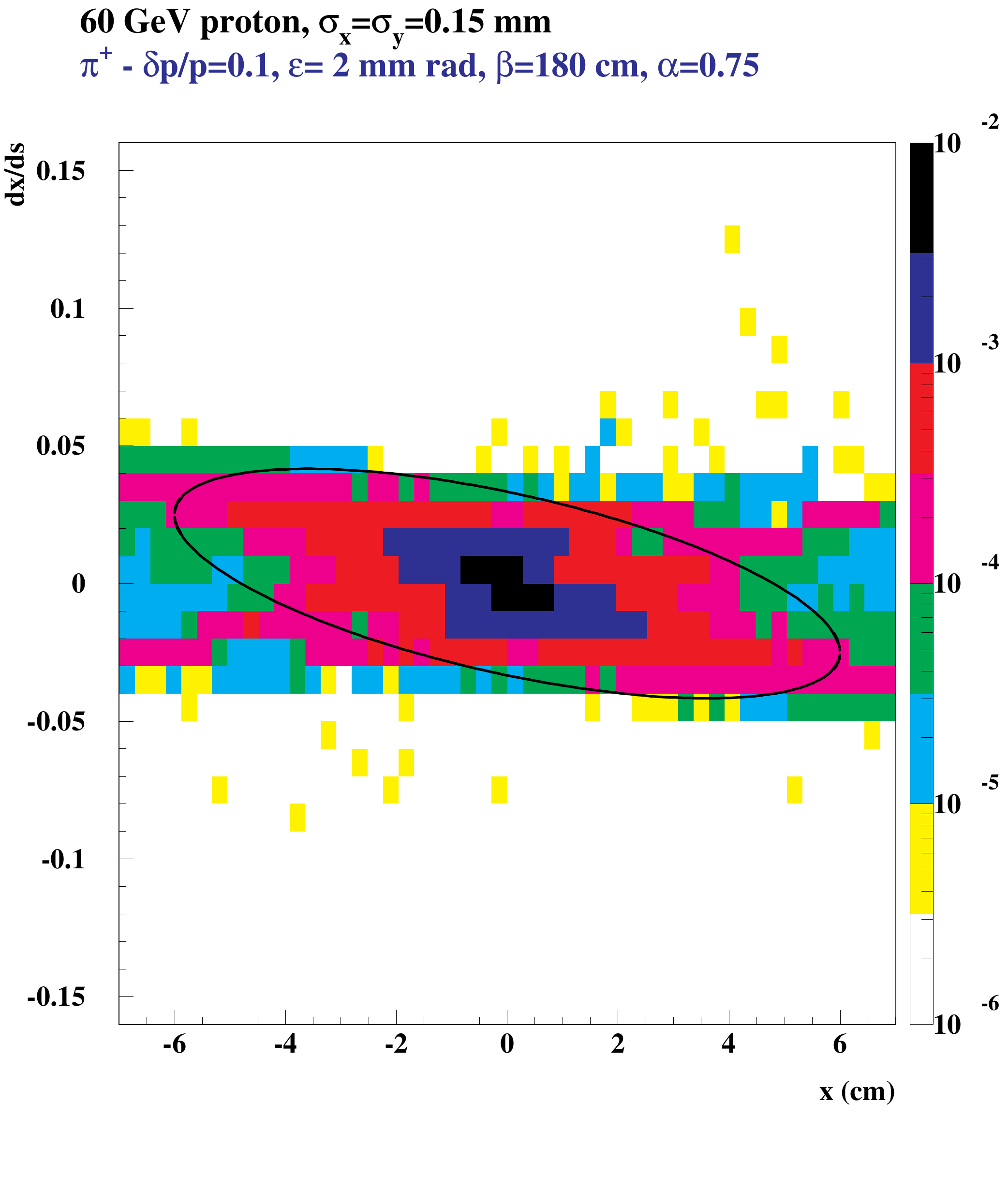}
	\end{array}$
  \end{center}	
\caption{Left: Pion spatial distribution just downstream of the target. Right: Pion spatial 
		distribution just downstream of the horn described above}
\label{fig:pion-target}
\end{figure}

Regarding capture/collection, we have looked into two options, a lithium lens and a horn.  
The existing Fermilab lithium lens has a working gradient of 2.6 Tesla/cm at 15 Hz. The optimal
distance between the target and lens center is about 25 cm. Pions produced into a 2 mm-radian
acceptance have a wide radial distribution, however.  Fig.~\ref{fig:pion-target} (Left) shows the
$\pi$ radial distribution 5 cm downstream of the target.
The current Fermilab lens with its 1 cm radius would
capture only 40\% of the pions in a $\pm$ 10\% momentum bin. With a 2 cm lens, the transmission 
factor increases up to 60\%. Further improvement could be achieved by increasing the lens
gradient, but increasing the gradient reduces the focal length. Maximal transmission could reach 80\%
with a 4 Tesla/m gradient and a 2 cm lens radius.  But this is beyond the current state-of-the-art for 
an operating lens and the target downstream 
end would then need to be very close to the lens.  With a NuMI-like horn operating at 300 kA and using a
22 cm gold target, it is possible to collect  0.088 $\pi^+$/POT within a momentum band of $5 \pm 0.5$ GeV/c.
The $\beta$ function of the pion beam after the horn is about 200 cm in this case. Note that shape of the
NuMI horn inner conductor was chosen to maximize the yield of neutrinos with energy $\le$ 12 GeV.
Optimization of the horn inner shape could increase the number of collected pions.  The spatial 
distribution of the pions just downstream of the horn is given in Fig.~\ref{fig:pion-target} (Right).
For our muon flux calculations we use a 20\% loss of pions during the collection phase (from the 0.88 above
and the numbers in Table~\ref{tab:yield} for gold, 5 GeV/c and $\pm$ 10\% capture.  The transport efficiency is assumed to be $\sim1$ and the injection efficiency is assumed to be 90\%.
\vspace{-4mm}
\subsection{Injection options}
\begin{figure}[h]
  \centering{
    \includegraphics[width=0.65\textwidth]{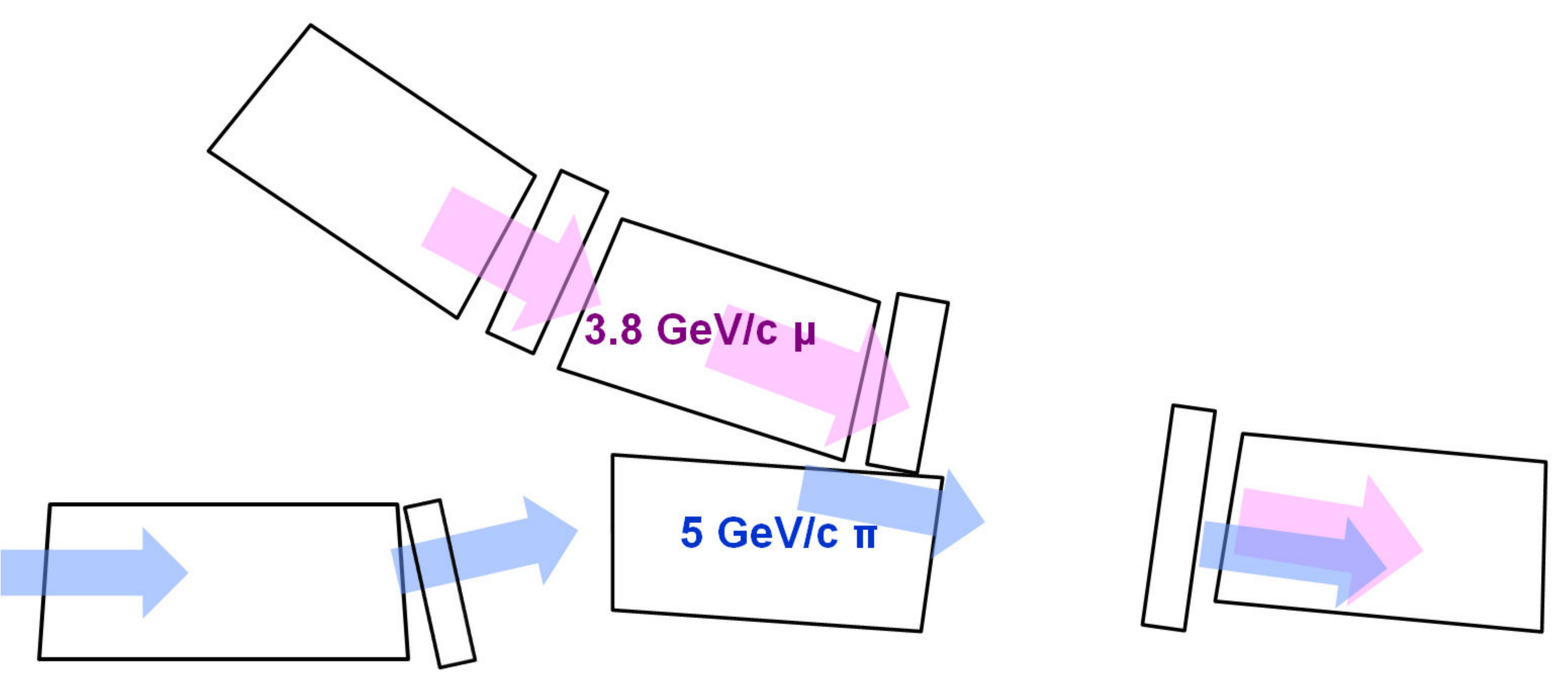}}
  \caption{Stochastic injection concept}
  \label{fig:SI}
\end{figure}

An obvious goal for the facility is to collect as many pions as possible (within the limits of available beam power),
inject them into the decay ring and capture as many muons as possible from the $\pi \rightarrow \mu$ decays.
With pion decay within the ring, non-Liouvillean ``stochastic injection" is possible. In stochastic injection, 
the $\simeq$ 5 GeV/c pion beam is transported from the target into the storage ring and dispersion-matched into a long 
straight section. (Circulating and injection orbits are separated by momentum.)  Decays within that straight 
section provide muons that are within the $\simeq$ 3.8 GeV/c ring momentum acceptance. With stochastic injection, muons 
from a beam pulse as long as the Main Injector circumference (3000m) can be accumulated, and no injection kickers 
are needed, see Fig.~\ref{fig:SI}.  Note: for 5.0 GeV/c pions, the decay length is $\simeq$ 
280m; $\simeq$ 42\% decay within the 150m decay ring straight.

As mentioned in section \ref{sec:Overview}, decay before injection requires a separate decay transport 
line and full-aperture fast kickers matching the pion beam pulse to the ring. 
The decay channel could be based on the conventional FODO channel focused by normal conducting quadrupoles. A preliminary design consisting of 36 cells with the total length of 165.6 m has been done. The quadrupoles are 0.8 m long with the  full aperture of about 30 cm and the gradient of ~9.7 T/m. The phase advance could be adjusted to provide stable focusing for the full pion momentum range decaying both backward and forward into the useful muon momentum range sets by the final ring acceptance (currently up to $\pm$16\%) and to the muon beam being formed simultaneously. The decay channel would need to be followed by a dedicated broad momentum matching section to couple the decay channel with the ring while keeping high transmission. The muon injection into the storage ring requires full-aperture fast kickers and septum magnets, matching the $\mu$ beam pulse to the ring. A preliminary considerations suggests that such kickers and the septum can be constructed based on the existing technology, subject to verification in future studies.  Developing a scenario for extraction from the Main Injection would also have to be included in any future studies.
At this point (and in the rest of this document), we are assuming pion decay in the ring.

\subsection{Muon decay ring}
We have investigated both a FODO racetrack and a FFAG racetrack for the muon decay.  The FODO ring that is described in detail below uses both normal and superconducting magnets.
A FODO lattice using only normal-conducting magnets (B $\lesssim$~2T) is also being developed. In this case, the arcs are twice as long ($\simeq$~50m), but the straight sections would be similar.  The racetrack FFAG (RFFAG) described below uses normal-conducting magnets, but a preliminary investigation with the use of super-ferric magnets for this lattice has been done.  In this case, the ring circumference would be reduced from $\sim$~600m to $\sim$~450m and the operating costs would be drastically reduced.
Table~\ref{tab:rings} gives a comparison (at our current level of understanding) between the FODO and the RFFAG with regard to the ratio of the total number of useful muons stored per POT assuming that capture off the target and injection into the rings are the same for both.  Acceptance for all the decay ring options we are considering will be studied and compared in order to obtain a cost/performance optimum.
\begin{table}[h]
\centering
\caption {Relative $\mu$ yield for FODO vs. RFFAG rings}
\label{tab:rings}
\begin{tabular}{|l|l|l|}
\hline
Parameter  & FODO \hspace{5mm} & RFFAG \\
\hline\noalign{\smallskip}
L$_{straight}$ (m) & 150 & 240 \\
Circumference (m) & 350 & 606 \\
Dynamic aperture A$_{dyn}$  & 0.7 & 0.95   \\
Momentum acceptance   & $\pm~10\%$ & $\pm~16\%$ \\
$\pi$/POT within momentum acceptance \hspace{10mm} & 0.112 & 0.171 \\
Fraction of $\pi$ decaying in straight (F$_s$) & 0.41 & 0.57 \\
Ratio of L$_{straight}$ to ring circumference ($\Omega$) & .43 & .40  \\
Relative factor (A$_{dyn} \times \pi$/POT $\times$ F$_s \times \Omega$) & 0.014 & 0.037 \\
\noalign{\smallskip}\hline
\end{tabular}
\end{table} 
\subsubsection{Separate element FODO racetrack}
\begin{figure}[h]
  \centering{
    \includegraphics[width=0.9\textwidth]{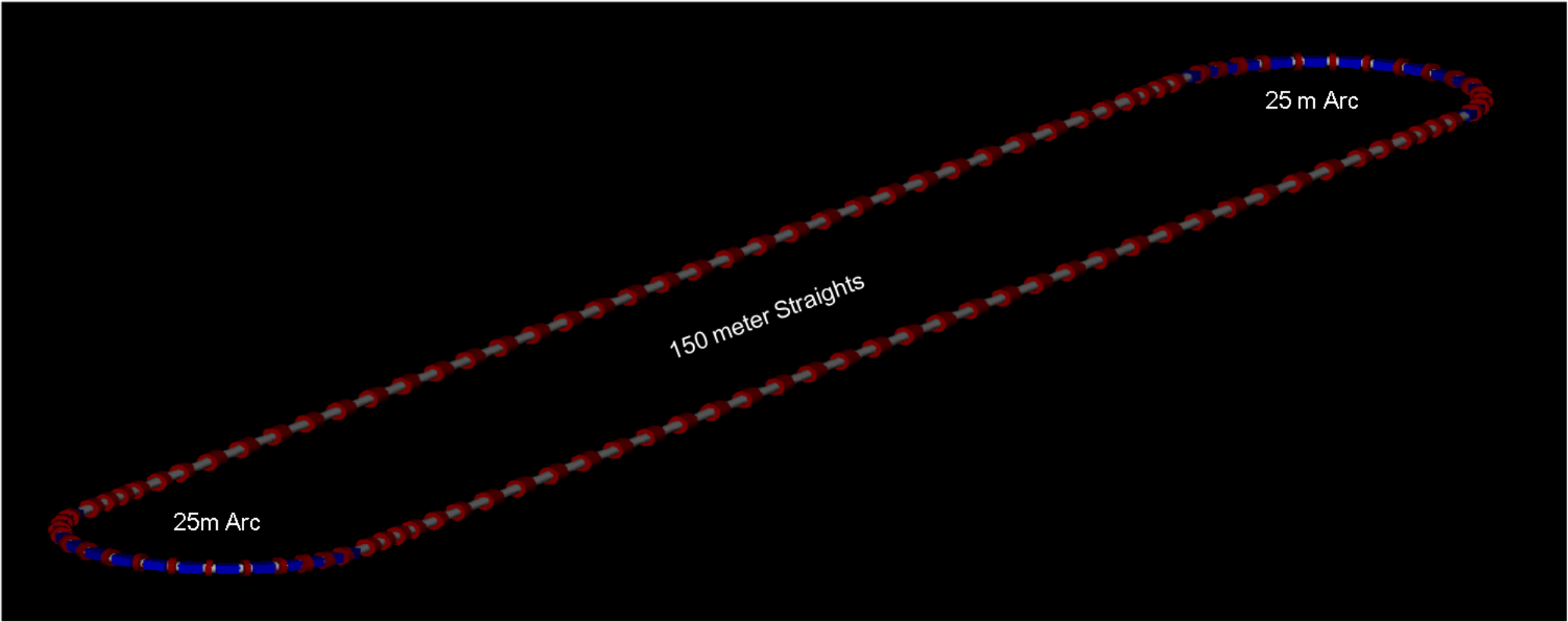}}
  \caption{Racetrack ring layout: 150 m straights and 25 m 180 deg. arcs}
  \label{fig:01}
\end{figure}

\begin{figure}[h]
  \centering{
    \includegraphics[width=0.9\textwidth]{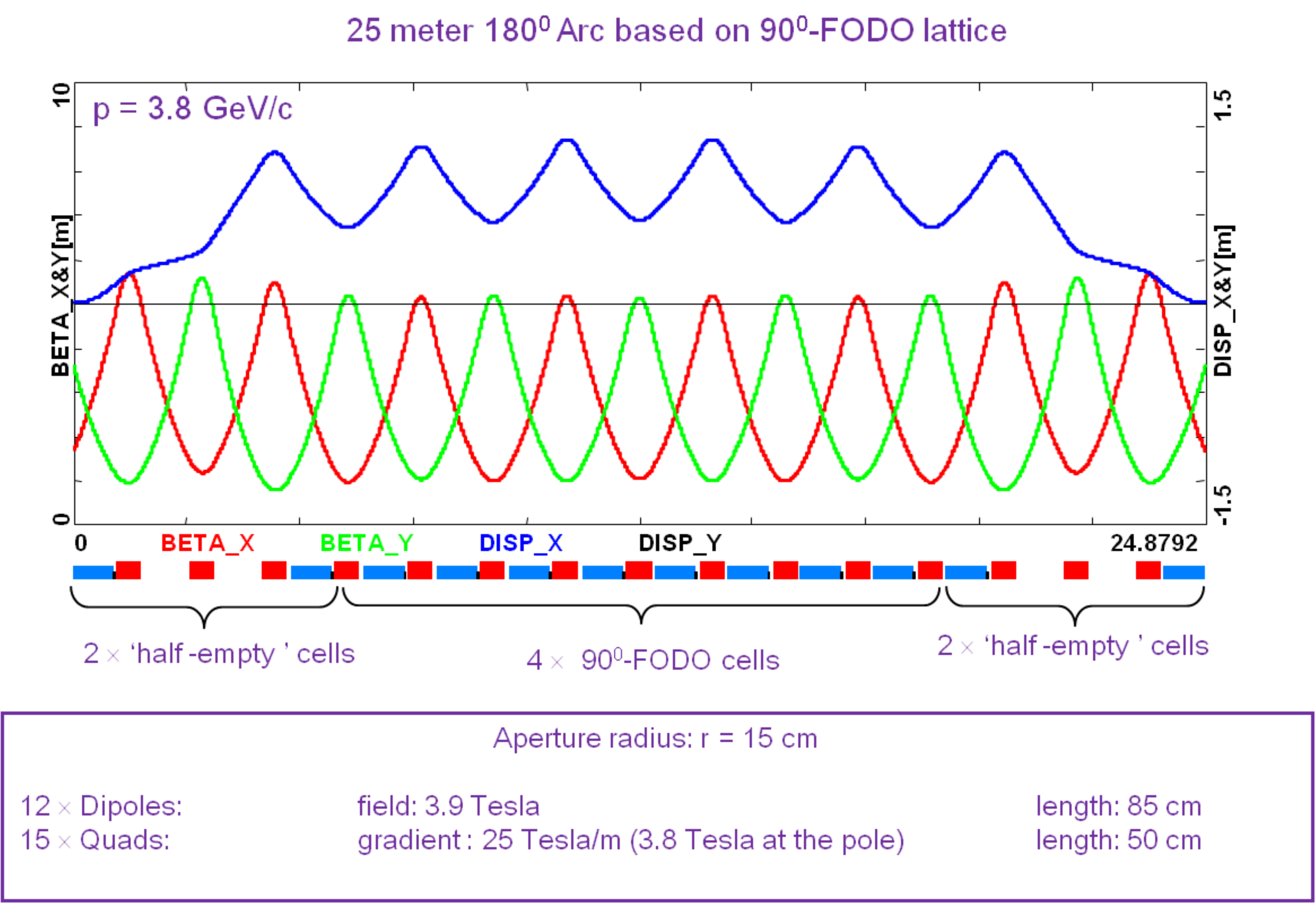}}
  \caption{Arc optics with dispersion suppression via missing dipoles with the so called 'half empty' cells; two of them at both arc ends.}
  \label{fig:02}
\end{figure}

\begin{figure}[h]
  \centering{
    \includegraphics[width=0.9\textwidth]{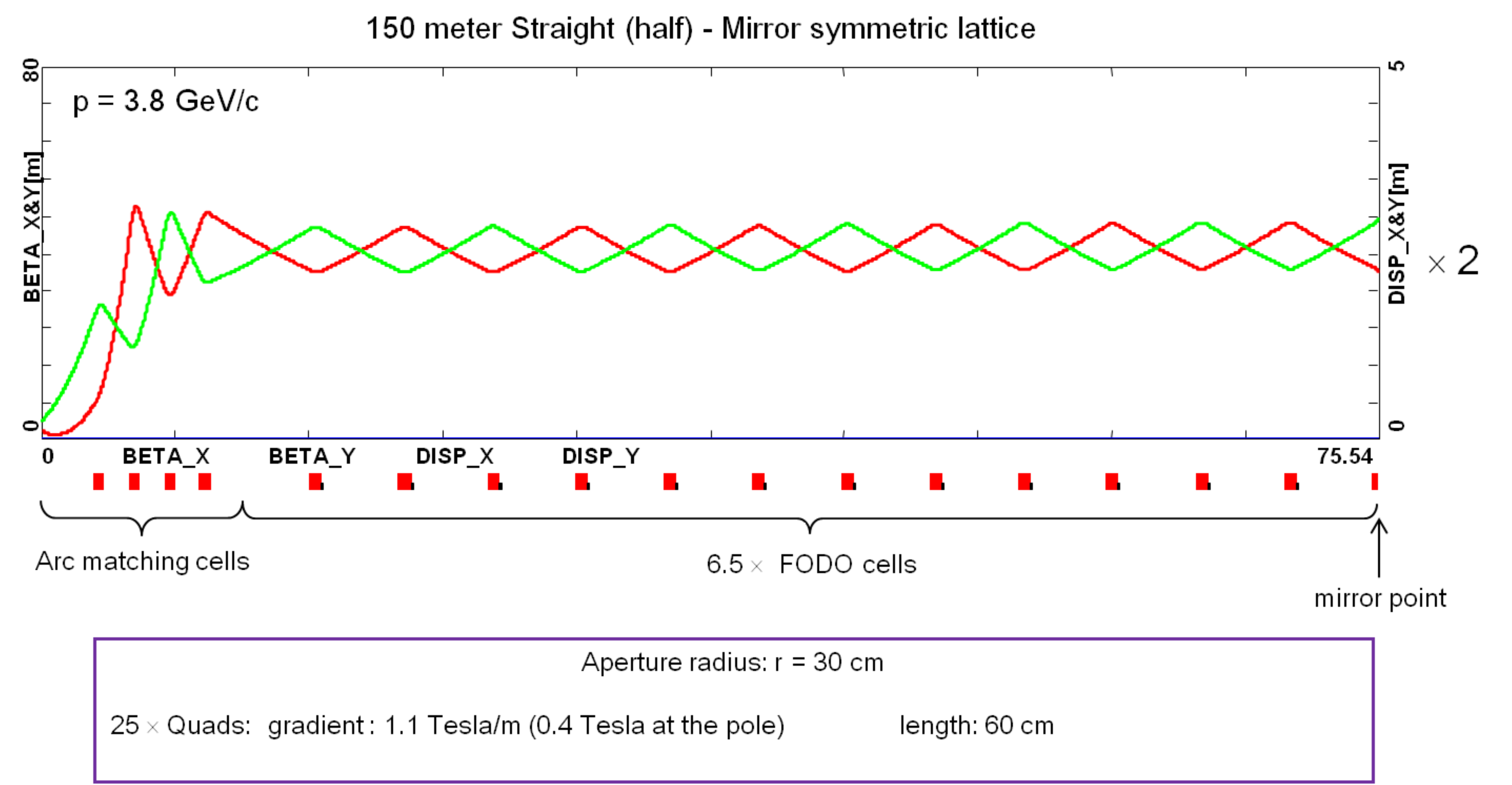}}
  \caption{Decay straight optics configured with ``high-$\beta$" ($\simeq$ 40 m) weakly focusing FODO cells smoothly matched to the arc. Only half of the 150 meter long straight is shown, with the mirror symmetry point indicated on the left end.}
  \label{fig:03}
\end{figure}

\begin{figure}[htpb]
  \centering{
    \includegraphics[width=0.9\textwidth]{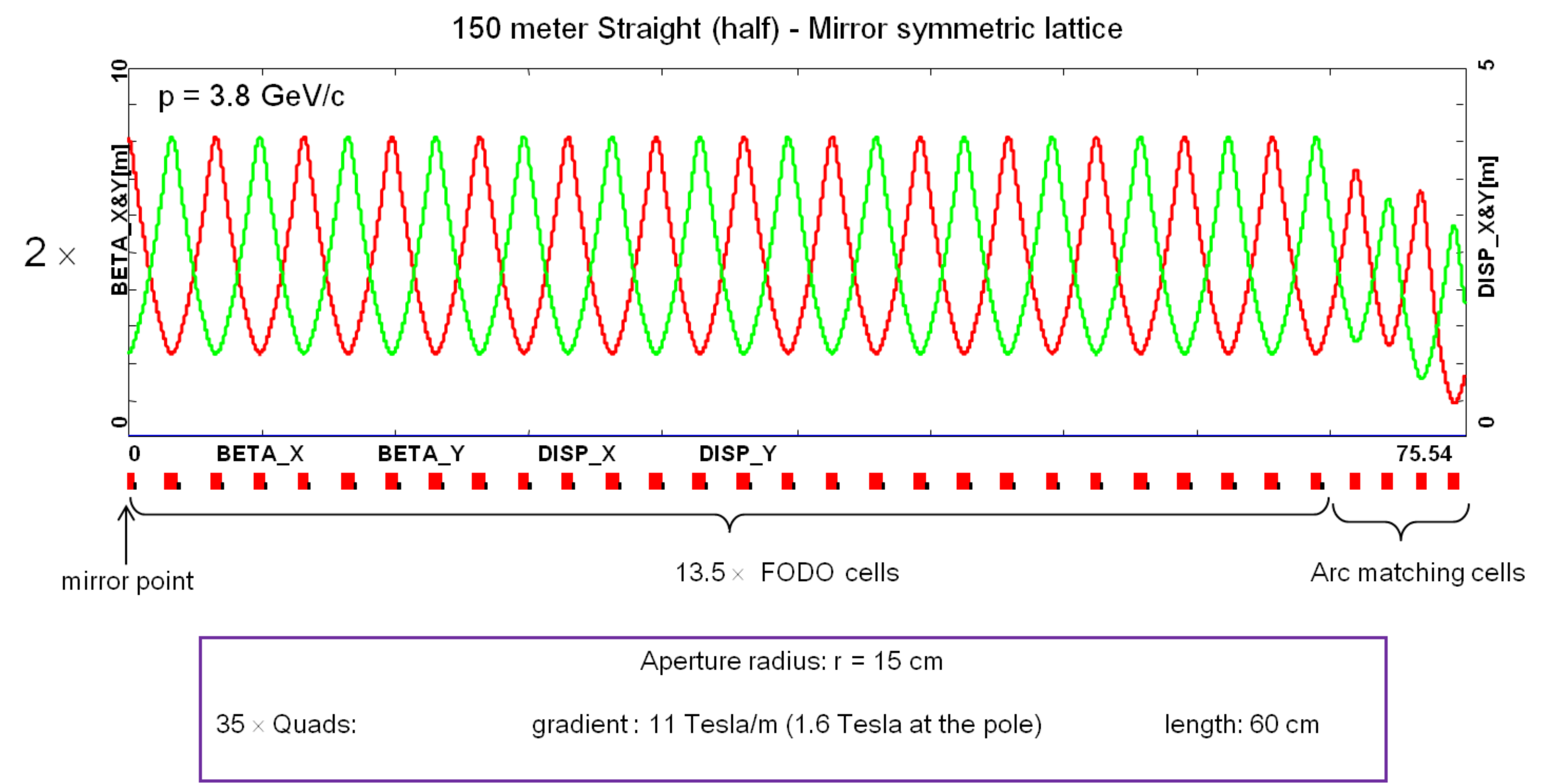}}
  \caption{The other straight optics configured with ``low-$\beta$" ($\simeq$ 5 m) weakly focusing FODO cells smoothly matched to the arc. Only half of the 150 meter long straight is shown, with the mirror symmetry point indicated on the right end.}
  \label{fig:04}
\end{figure}

\begin{figure}[htpb]
  \centering{
    \includegraphics[width=0.9\textwidth]{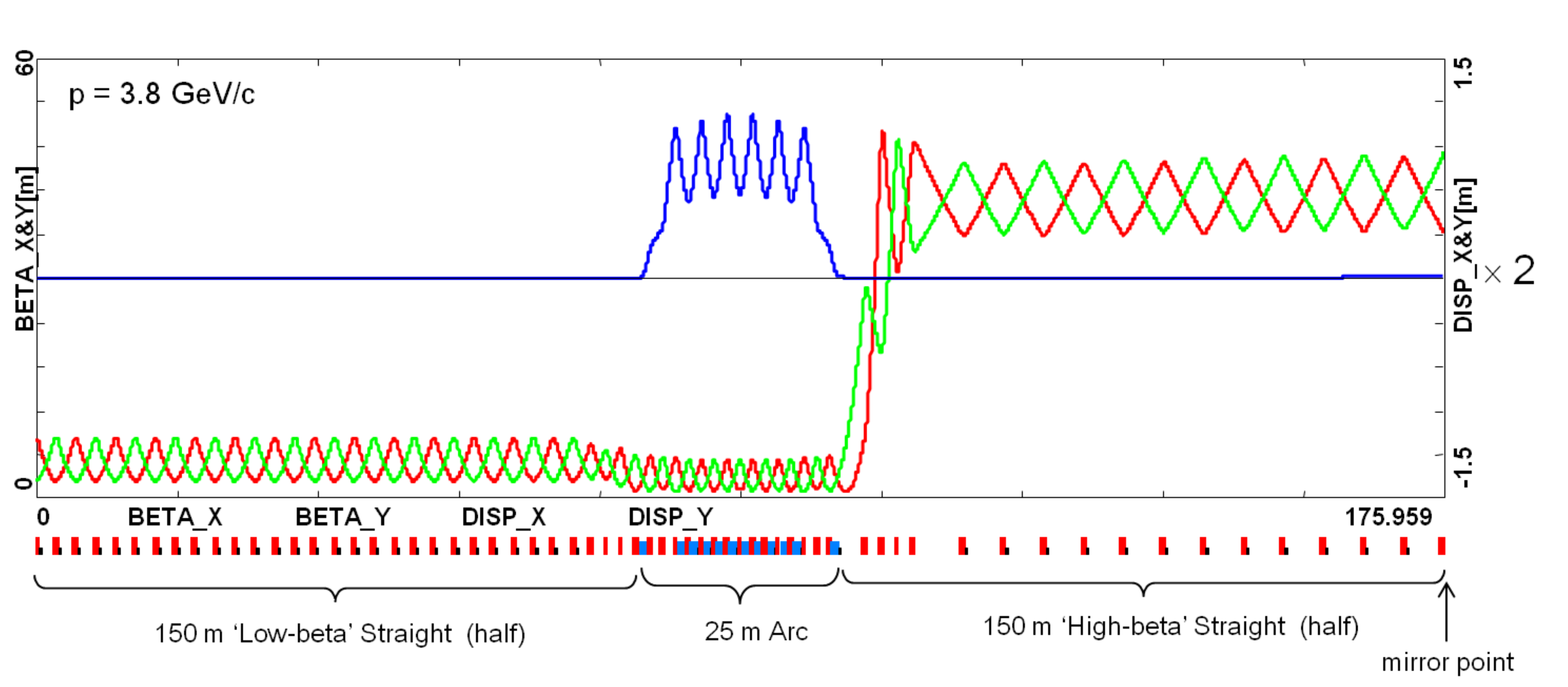}}
  \caption{Complete racetrack ring lattice. Only half of the ring is shown, with the mirror symmetry point indicated on the right end.}
  \label{fig:05}
\end{figure}

Here we propose a compact racetrack ring design based on separate function magnets (bends and quadrupoles only) configured with various flavors of FODO lattice. The ring layout is illustrated in Fig.~\ref{fig:01}.
The design goal for the ring was to maximize both the transverse and momentum acceptance (around 3.8 GeV/c central momentum), while maintaining reasonable physical apertures for the magnets in order to keep the cost down. This was accomplished by employing strongly focusing optics in the arcs (90 deg. phase advance per cell FODO); featuring small $\beta$ functions ($\simeq$ 3 m average) and low dispersion ($\simeq$ 0.8 m average). The linear optics for one of the 180 deg. arcs is illustrated in Fig.~\ref{fig:02}.
The current FODO lattice design incorporates a missing-magnet dispersion suppressor.
The missing-magnet dispersion suppressor provides an ideal location for the implementation of 
stochastic injection, see Fig.~\ref{fig:SI}.  With a dispersion of $\eta$ $\simeq$ 1.2m at the drift, the 5 and 3.8 
GeV/c orbits are 
separated by $\simeq$ 30 cm; an  aperture of $\simeq \pm $ 15cm is available for both the 5 GeV/c 
$\pi$ and 3.8 GeV/c $\mu$ orbits. 
To maintain high compactness of the arc, while accommodating adequate drift space for the injection chicane to merge, two special ``half empty" cells with only one dipole per cell were inserted at both arc ends to suppress the horizontal dispersion. This solution allowed us to limit the overall arc length to about 25 m, while keeping the dipole fields below 4 Tesla. The arc magnets assume a relatively small physical aperture radius of 15 cm, which limits the maximum field at the quadrupole magnet pole tip to less than 4 Tesla. 
On the other hand, the decay straight requires much larger values of $\beta$ functions ($\simeq$ 40 m average)in order to maintain small beam divergence ($\simeq$ 7 mrad). The resulting muon beam divergence is a factor of 4 smaller than the characteristic decay cone of 1/$\gamma$ ($\simeq$ 0.028 at 3.8 GeV). As illustrated in Fig.~\ref{fig:03}, the decay straight is configured with a much weaker focusing FODO lattice (30 deg. phase advance per cell). It uses normal conducting large aperture (r = 30 cm) quads with a modest gradient of 1.1 Tesla/m (0.4 Tesla at the pole tip). Both the arc and the straight are smoothly matched via a compact telescope insert, as illustrated in Fig.~\ref{fig:03}.
The ``other" 150 meter straight, which is not used for neutrino production, can be designed using much tighter FODO lattice (60 deg. phase advance per cell), with rather small $\beta$ functions comparable to the one in the arc ($\simeq$ 5 m average). This way one can restrict the aperture radius of the straight to 15 cm.  Again, the second straight uses normal conducting, quads with a gradient of 11 Tesla/m (1.6 Tesla at the pole tip). Both the arc and the straight are smoothly matched, as illustrated in Fig.~\ref{fig:04}.
Finally, the complete racetrack ring architecture is illustrated in Fig.~\ref{fig:05}. It features the ``low-$\beta$" straight (half) matched to the 180 deg. arc and followed by the ``high-$\beta$" decay straight (half) connected to the arc with a compact telescope insert.
To summarize the magnet requirements, both 180 deg. arcs were configured with 3.9 Tesla dipoles and 25 Tesla/m quads (superconducting magnets with 15 cm aperture radius). Both straights use normal conducting magnets: the decay straight---1.1 Tesla/m quads with 30 cm aperture radius and the 'other' straight
---11 Tesla/m quads with 15 cm aperture radius.

\begin{figure}[htpb]
  \centering{
    \includegraphics[width=1.0\textwidth]{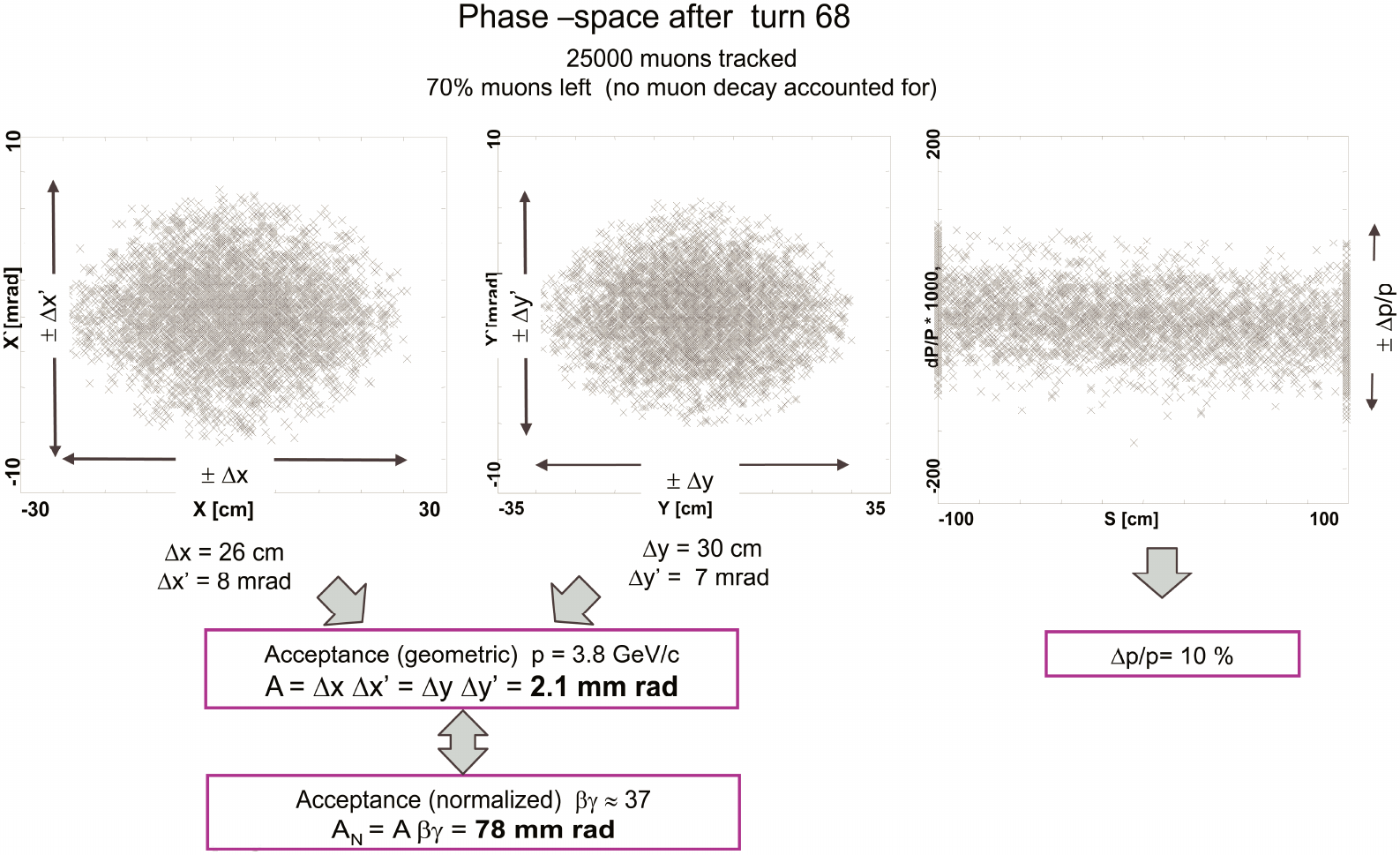}}
  \caption{Dynamic aperture study – resulting transverse and longitudinal phase-space acceptance after 68 turns of tracking. For illustration, the phase-space 'snapshots' were taken at the middle of the decay straight.}
  \label{fig:06}
\end{figure}

The ring acceptance was studied via symplectic tracking (with the OptiM code) of 25,000 muons through 68 turns (e-folding muon decay). The dynamic losses amounted to 30\% (70\% muons survived 68 turns without accounting for muon decay). The resulting acceptance is summarized in terms of the transverse and longitudinal phase-space projections resulting from multi-particle tracking as illustrated in Fig.~\ref{fig:06}.
In summary, the ring features transverse acceptance (normalized) of 78 mm rad both in x and y (or geometric acceptance of 2.1 mm rad) for the net momentum acceptance of $\pm 10\%$

%
%
%
%
%
%
\subsubsection{Advanced scaling FFAG}
\begin{figure}[h!]
	\begin{center}
		\includegraphics[width=\textwidth]{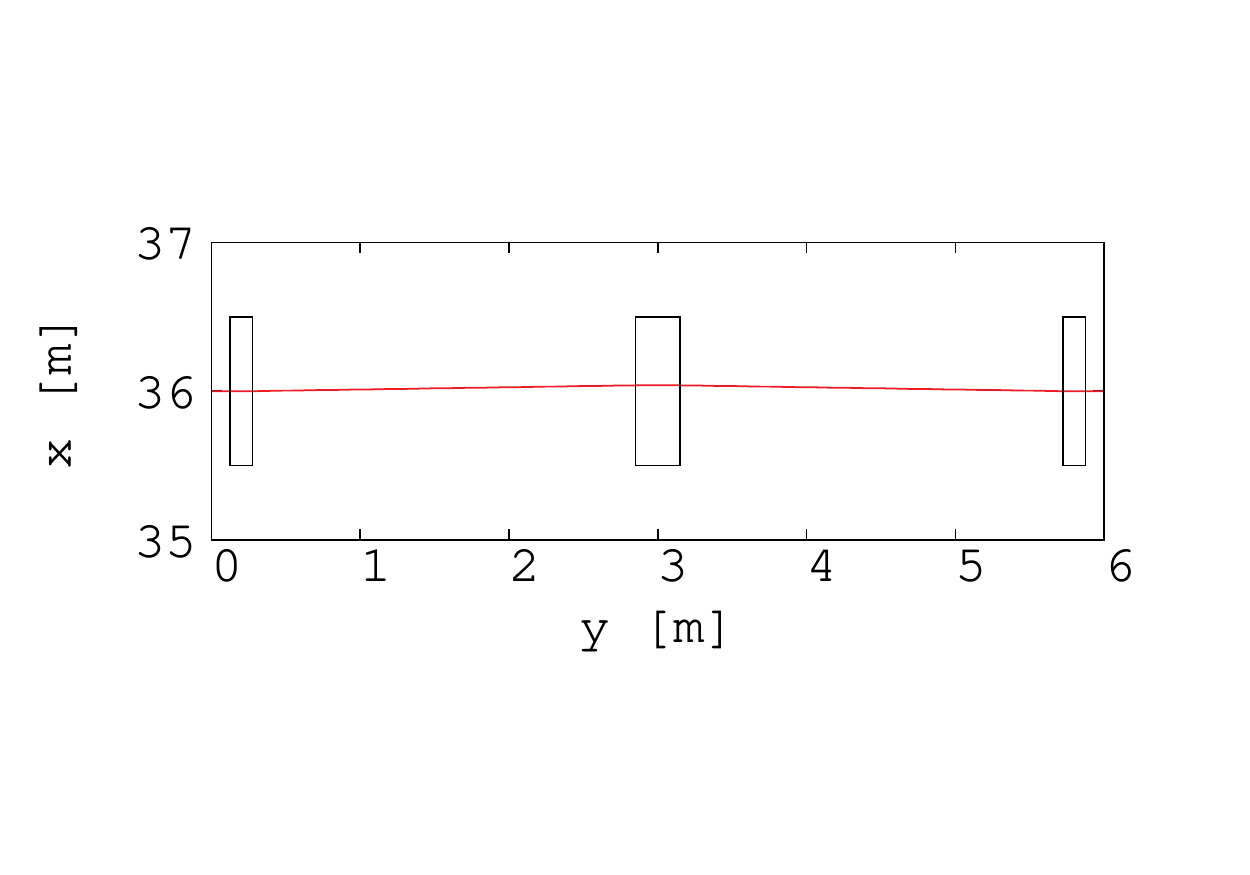}
		\caption{Top view of the straight scaling FFAG cell. The 3.8 GeV/c muon reference trajectory is shown in red. 
		Effective field boundaries with collimators are shown in black.}
		\label{str-traj}
	\end{center}
\end{figure}

\begin{figure}[h!]
   \begin{minipage}[b]{.48\linewidth}
       \includegraphics[width=8.cm]{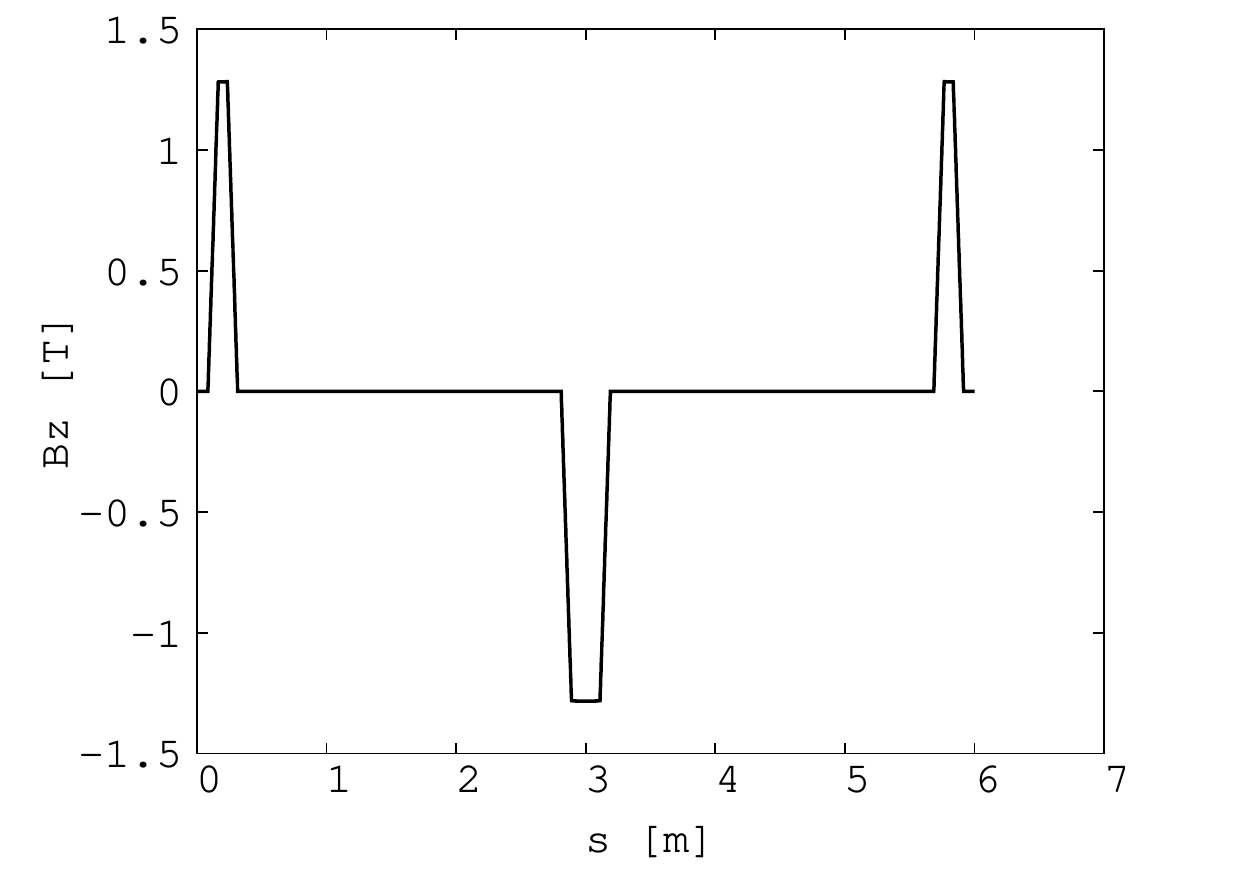}
	\caption{Vertical magnetic field for 3.8 GeV/c muon reference trajectory in the straight scaling FFAG cell.}
	\label{str-bz}
	 \end{minipage} \hfill
   \begin{minipage}[b]{.48\linewidth}
    	\includegraphics[width=8cm]{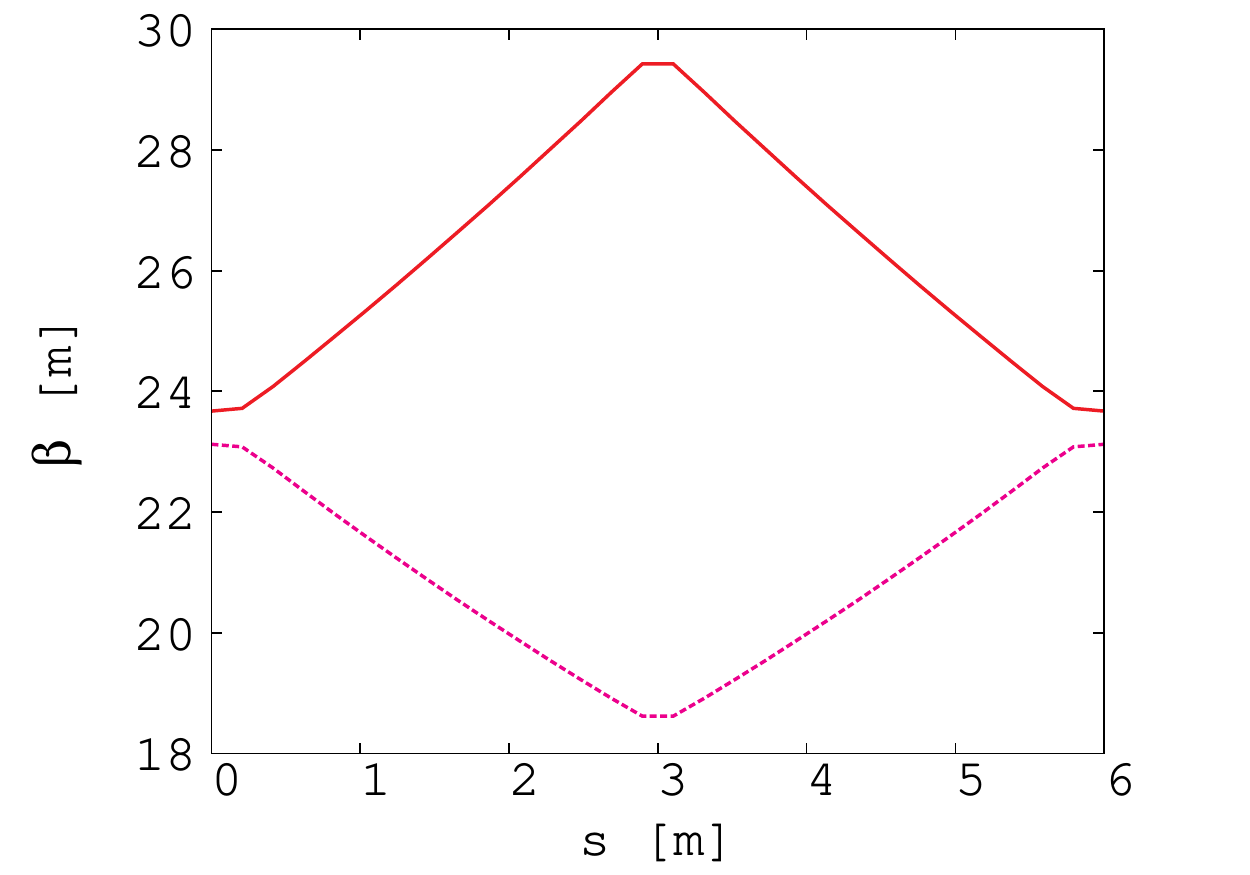}
	\caption{Horizontal (plain red) and vertical (dotted purple) periodic $\beta$ functions of the straight scaling FFAG cell.}
	\label{str-beta}
   \end{minipage}
\end{figure}

\begin{figure}[h!]
	\begin{center}
		\includegraphics[width=10cm]{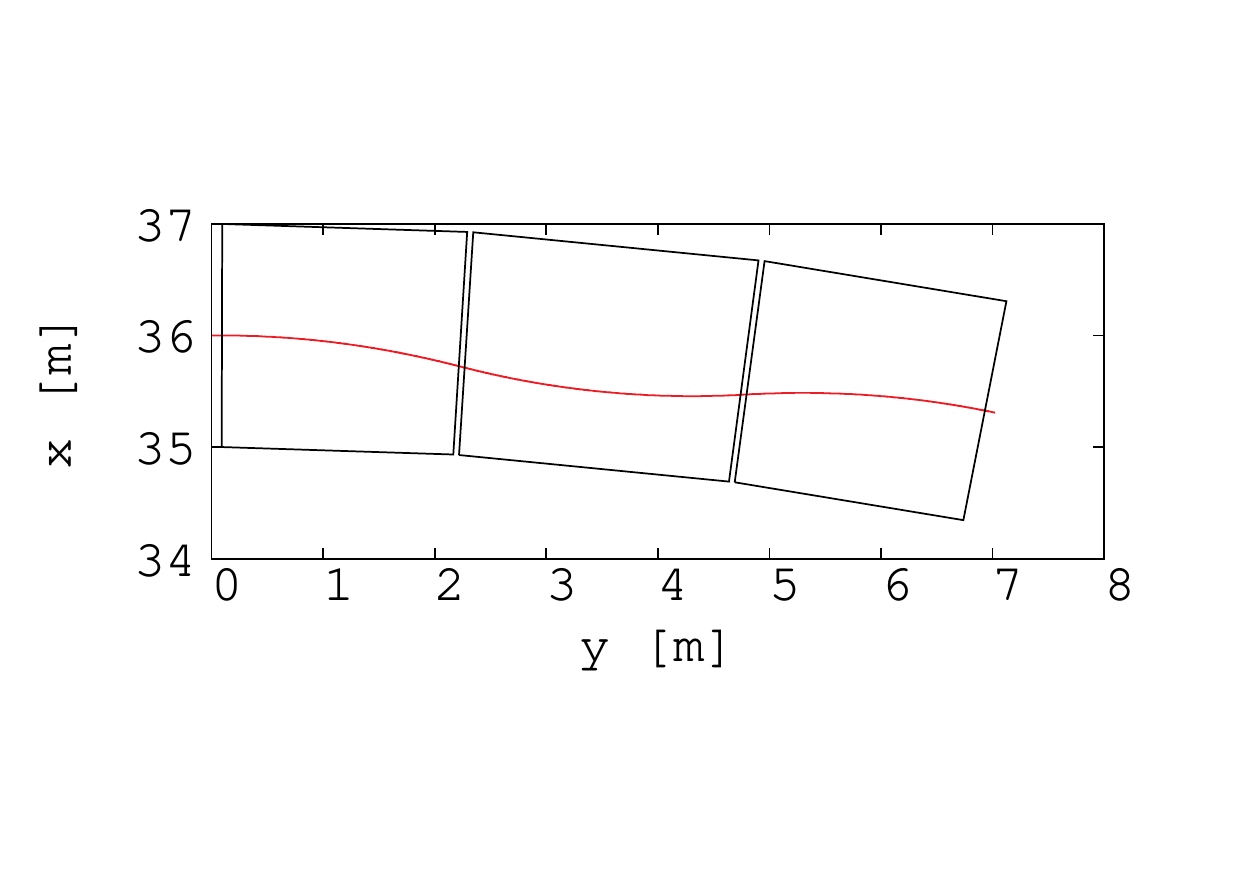}
		\caption{Top view of the circular scaling FFAG cell. The 3.8~GeV/c muon reference trajectory is shown in red. 
		Effective field boundaries with collimators are shown in black.}
		\label{circ-traj}
	\end{center}
\end{figure}

\begin{figure}[h!]
   \begin{minipage}[b]{.49\linewidth}
       \includegraphics[width=8.cm]{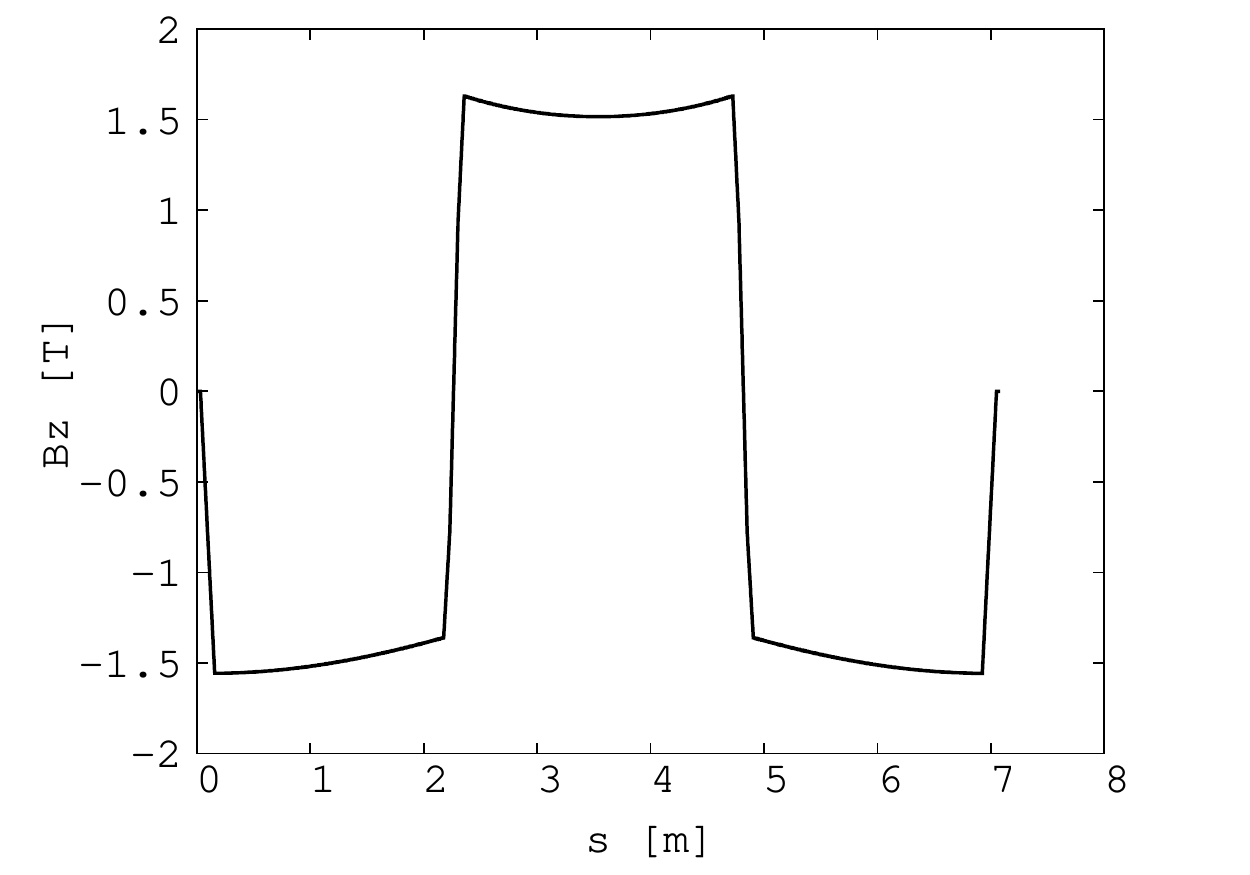}
	\caption{Vertical magnetic field for the 3.8~GeV/c muon reference trajectory in the circular scaling FFAG cell.}
	\label{circ-bz}
	 \end{minipage} \hfill
   \begin{minipage}[b]{.49\linewidth}
    	\includegraphics[width=8cm]{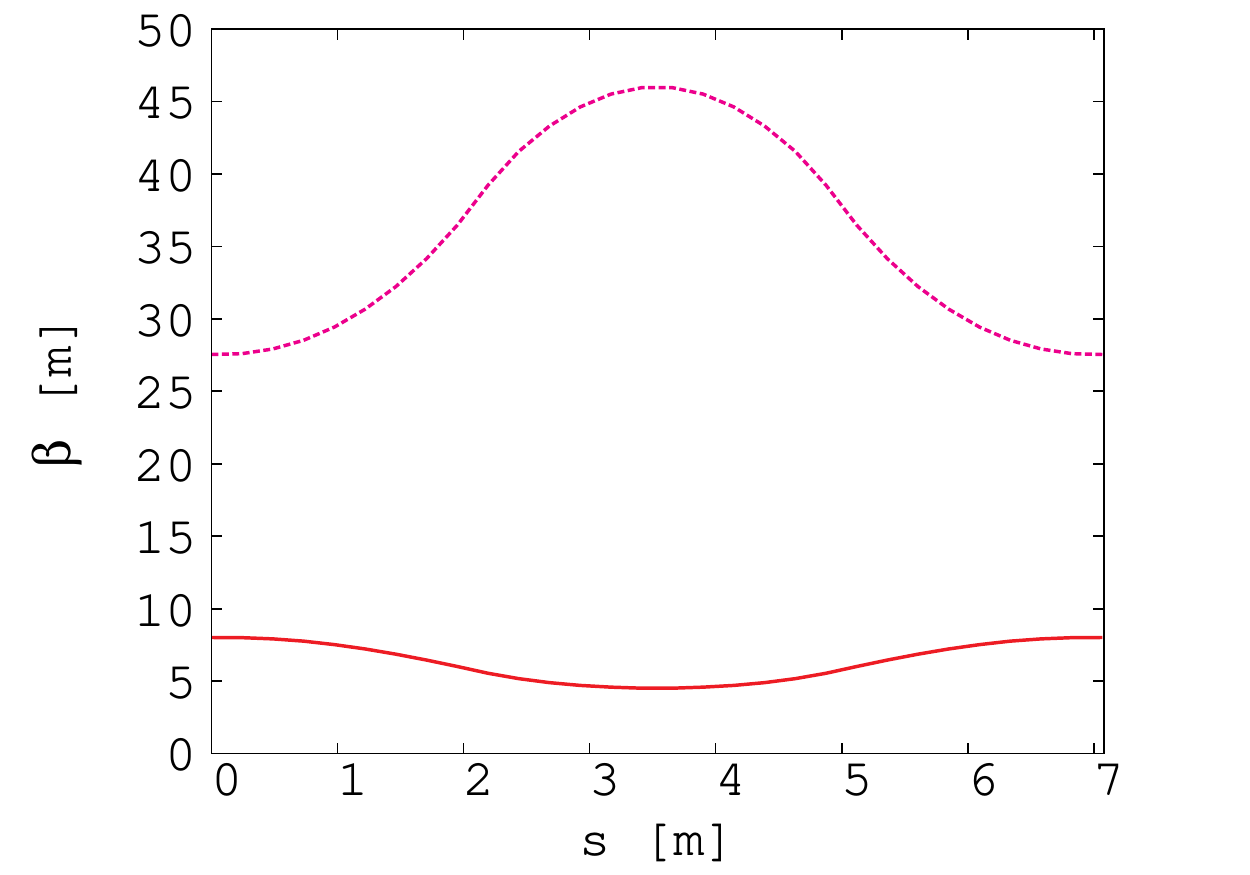}
	\caption{Horizontal (plain red) and vertical (dotted purple) periodic $\beta$ functions of the circular scaling FFAG cell.}
	\label{circ-beta}
   \end{minipage}
\end{figure}

The racetrack FFAG ring is composed of two cell types: a) a straight scaling FFAG cell and b) a
circular scaling FFAG cell.  There are 40~straight FFAG cells in each long straight section 
(80 for the  whole ring) and 16~circular FFAG cells in each of the arc sections.
\paragraph{Straight scaling FFAG cell parameters}
In the straight scaling FFAG cell, the vertical magnetic field $B_{sz}$ in the median plane follows:
\begin{displaymath}
B_{sz}=B_{0sz} e^{m(x-x_0)}  \mathcal{F},
\end{displaymath}
with $x$ the horizontal Cartesian coordinate, $m$ the normalized field gradient, $ \mathcal{F}$ an 
arbitrary function and $B_{0sz}=B_{sz}(x_0)$.
The parameters of the straight scaling FFAG cell are summarized in Table~\ref{tab-straight}.
\begin{table}[!h]
    \centering
\begin{tabular}{lcc}
\hline
\hline
Cell type			&		& DFD triplet \\[-1.5mm]
Number of cells in the ring	 &	& 80 \\[-1.5mm]
Cell length	&	& 6~m\\[-1.5mm]
 $x_0$ &&	36~m\\[-1.5mm]
m-value      &         & 2.65\,m$^{-1}$         \\[-1.5mm]
         Packing factor & & 0.1 \\[-1mm]
         Collimators ($x_{min}, x_{max}, z_{max}$) && (35.5~m, 36.5~m, 0.3~m)\\[-1.5mm]
         Periodic cell dispersion  && 0.38~m\\[-1.5mm]
          Horizontal phase advance  && 13.1\,deg. \\[-1.5mm]
Vertical phase advance  && 16.7\,deg. \\[-1.mm]
\hline
D$_1$ magnet parameters &&\\[-1.5mm]
& Magnet center & 0.2~m\\[-1.5mm]
& Magnet length & 0.15~m\\[-1.5mm]
& Fringe field fall off & Linear (Length: 0.04~m)\\[-1.5mm]
& $B_0(x_0=36~m)$ & 1.28067~T\\[-1.mm]
\hline
F magnet parameters &&\\[-1.5mm]
& Magnet center & 3~m\\[-1.5mm]
& Magnet length & 0.3~m\\[-1.5mm]
& Fringe field fall off & Linear (Length: 0.04~m)\\[-1.5mm]
& $B_0(x_0=36~m)$ & -1.15037~T\\[-1.mm]
\hline
D$_2$ magnet parameters &&\\[-1.5mm]
& Magnet center & 5.8~m\\[-1.5mm]
& Magnet length & 0.15~m\\[-1.5mm]
& Fringe field fall off & Linear (Length: 0.04~m)\\[-1.5mm]
& $B_0(x_0=36~m)$ & 1.28067~T\\[-1.mm]
\hline
\hline
\end{tabular} 
 \caption{Parameters of the straight scaling FFAG cell.}
 \label{tab-straight}
\end{table}
The cell is shown in Fig.~\ref{str-traj}. The red line represents the $\simeq$ 3.8 GeV/c muon reference trajectory, 
and its corresponding magnetic field is shown in Fig.~\ref{str-bz}. Periodic $\beta$ functions are shown in 
Fig.~\ref{str-beta}.
%
\paragraph{Circular scaling FFAG cell parameters}
In the circular scaling FFAG cell, the vertical magnetic field $B_{cz}$ in the median plane follows
\begin{displaymath}
B_{cz}=B_{0cz} \left( \frac{r}{r_0}\right)^k \mathcal{F},
\end{displaymath}
with $r$ the radius in polar coordinates, $k$ the geometrical field index, $ \mathcal{F}$ an 
arbitrary function and $B_{0cz}=B_{cz}(r_0)$.
The parameters of the circular scaling FFAG cell are summarized in Table~\ref{tab-circ}.
\begin{table}[h!]
    \centering
\begin{tabular}{lcc}
\hline
\hline
Cell type			&		& FDF triplet \\[-1.5mm]
Number of cells in the ring	 &	& 32 \\[-1.5mm]
Cell opening angle	&	& 11.25~deg\\[-1.5mm]
 $r_0$ &&	36~m\\[-1.5mm]
k-value      &         & 10.85        \\[-1.5mm]
         Packing factor & & 0.96 \\[-1.5mm]
         Collimators ($r_{min}, r_{max}, z_{max}$) && (35~m, 37~m,  0.3~m)\\[-1.5mm]
         Periodic cell dispersion  && 1.39~m (at~ 3.8~GeV/c)\\[-1.5mm]
          Horizontal phase advance  && 67.5\,deg. \\[-1.5mm]
Vertical phase advance  && 11.25\,deg. \\[-1.mm]
\hline
F$_1$ magnet parameters &&\\[-1.5mm]
& Magnet center & 1.85~deg\\[-1.5mm]
& Magnet length & 3.4~deg\\[-1.5mm]
& Fringe field fall off & Linear (Length: 0.1~deg)\\[-1.5mm]
& $B_0(r_0=36~m)$ & -1.55684~T\\[-1.mm]
\hline
D magnet parameters &&\\[-1.5mm]
& Magnet center & 5.625~deg\\[-1.5mm]
& Magnet length & 4.0~deg\\[-1.5mm]
& Fringe field fall off & Linear (Length: 0.1~deg)\\[-1.5mm]
& $B_0(r_0=36~m)$ & 1.91025~T\\[-1.mm]
\hline
F$_2$ magnet parameters &&\\[-1.5mm]
& Magnet center & 9.4~deg\\[-1.5mm]
& Magnet length & 3.4~deg\\[-1.5mm]
& Fringe field fall off & Linear (Length: 0.1~deg)\\[-1.5mm]
& $B_0(r_0=36~m)$ & -1.55684~T\\[-1.mm]
\hline
\hline
\end{tabular} 
 \caption{Parameters of the circular scaling FFAG cell.}
 \label{tab-circ}
\end{table}
The cell is shown in Fig.~\ref{circ-traj}. The red line represents the 3.8~GeV/c muon reference trajectory, and its 
corresponding magnetic field is shown in Fig.~\ref{circ-bz}. Periodic $\beta$ functions are shown in Fig.~\ref{circ-beta}.
\paragraph{Single particle tracking}
Stepwise tracking using Runge Kutta integration in a field model with linear fringe fields has been performed where 
interpolation of the magnetic field away from the mid-plane has been done to first order. Only single particle tracking has 
been done so far. We used $\mu^+$ with a central momentum, $p_0$, of 3.8~GeV/c, a minimum momentum, 
$p_{min}$, of 3.14~GeV/c and a maximum momentum, $p_{max}$, of 4.41~GeV/c.  $\Delta p/p_0$ is thus $\pm 16\%$.
The tracking step size was 1~mm. The exit boundary of a cell is the entrance boundary of the next cell.

\begin{figure}[h!]
	\begin{center}
		\includegraphics[width=10cm]{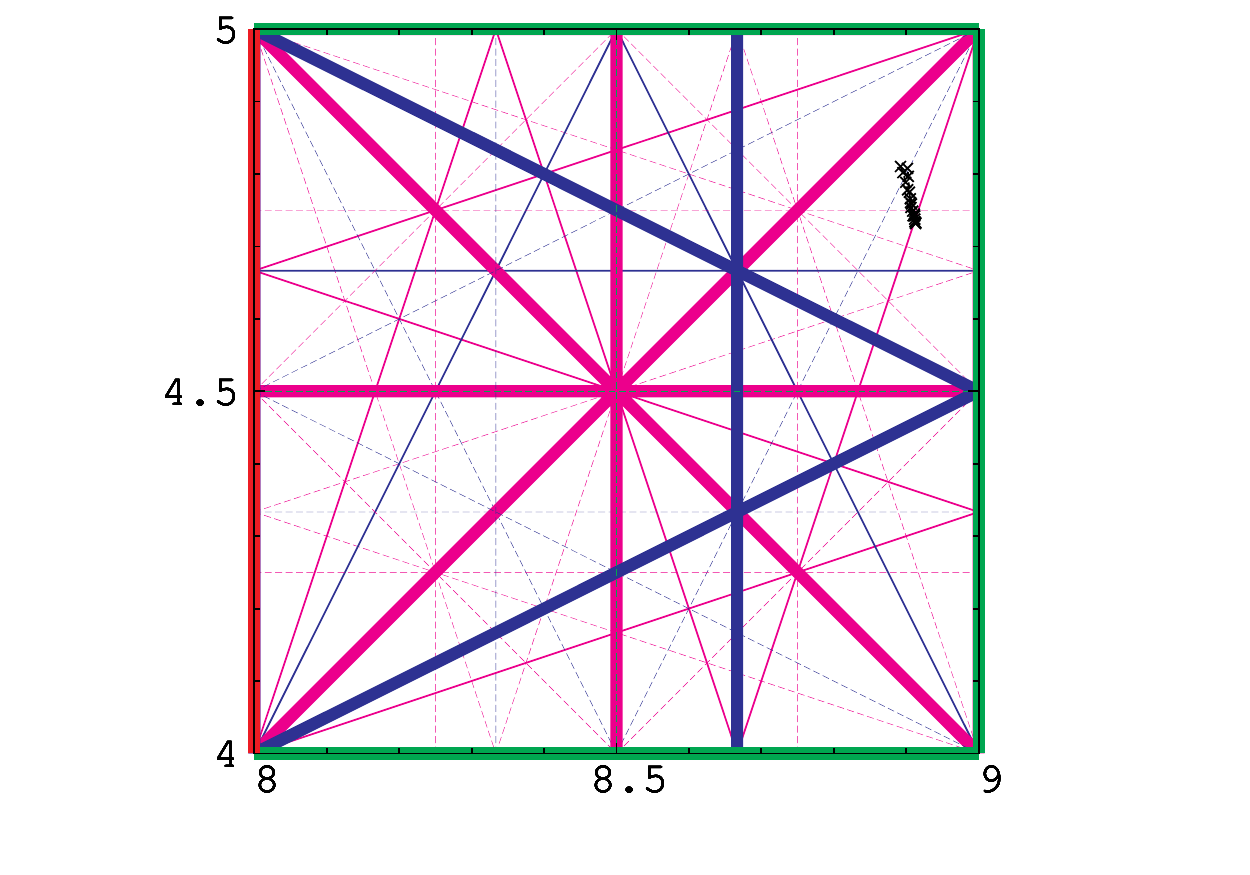}
		\caption{Tune diagram for muons from $p_{min}$ to $p_{max}$ ($\pm16\%$ in momentum around 3.8~GeV/c). Integer (red), 
		half-integer (green), third integer (blue) and fourth integer (purple) normal resonances are plotted. Structural 
		resonances are in bold.}
		\label{tunediag}
	\end{center}
\end{figure}

The ring tune point is (8.91,4.72) at $p_0$. Stability of the ring tune has been studied over the momentum range. The 
tune shift is presented in Fig.~\ref{tunediag}. The tune point stays within a 0.1 shift.
 
\begin{figure}[h!]
	\begin{center}
		\includegraphics[width=15cm]{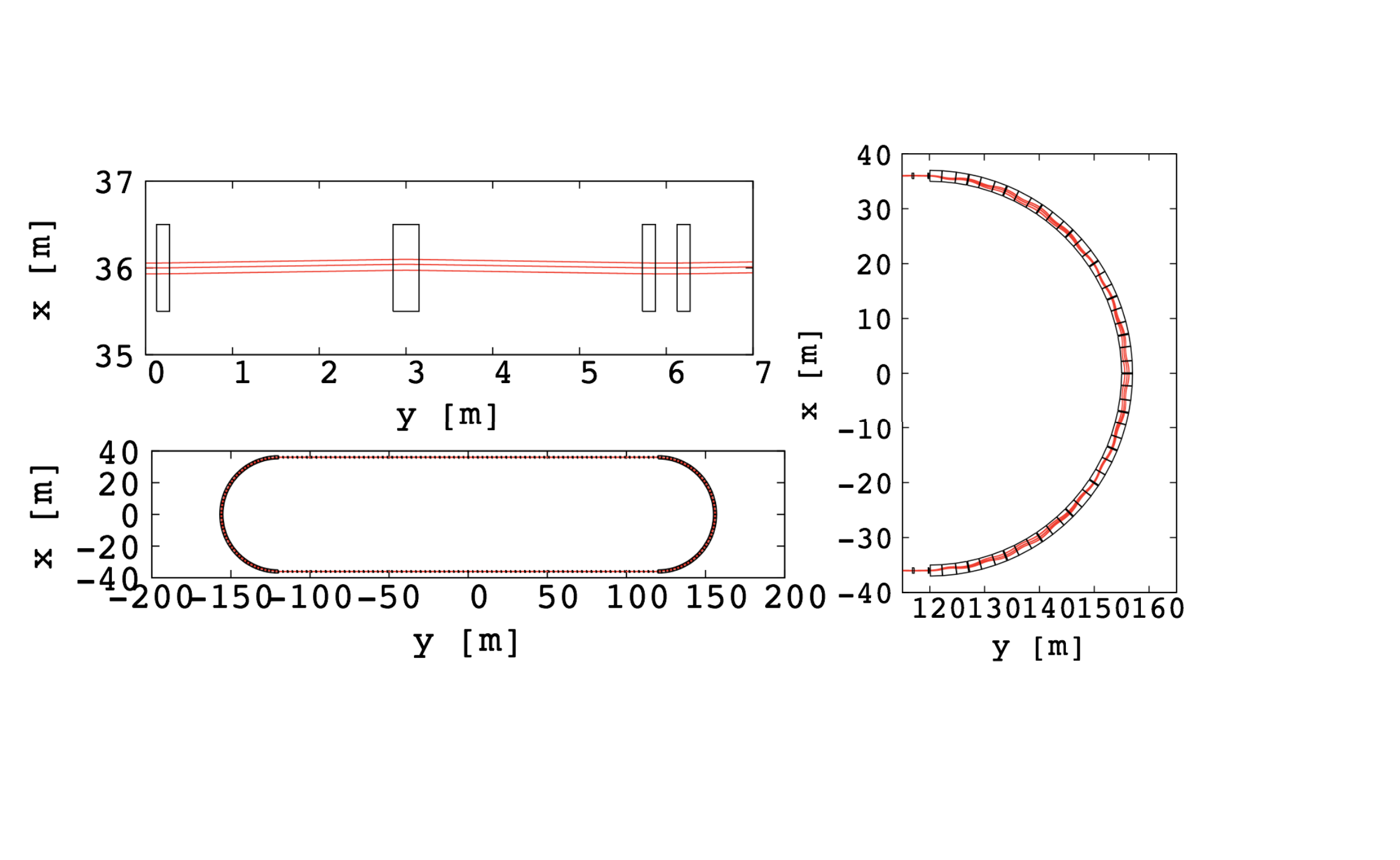}
		\caption{Top view of the racetrack FFAG lattice (bottom left scheme). The top left shows a zoom 
		of the straight section and on the right we show a zoom of the arc section. $p_0$, $p_{min}$, and $p_{max}$ 
		muon closed orbits are shown in red. Effective field boundaries with collimators are shown in black.}
		\label{single-traj}
	\end{center}
\end{figure}

\begin{figure}[h!]
	\begin{center}
		\includegraphics[width=10cm]{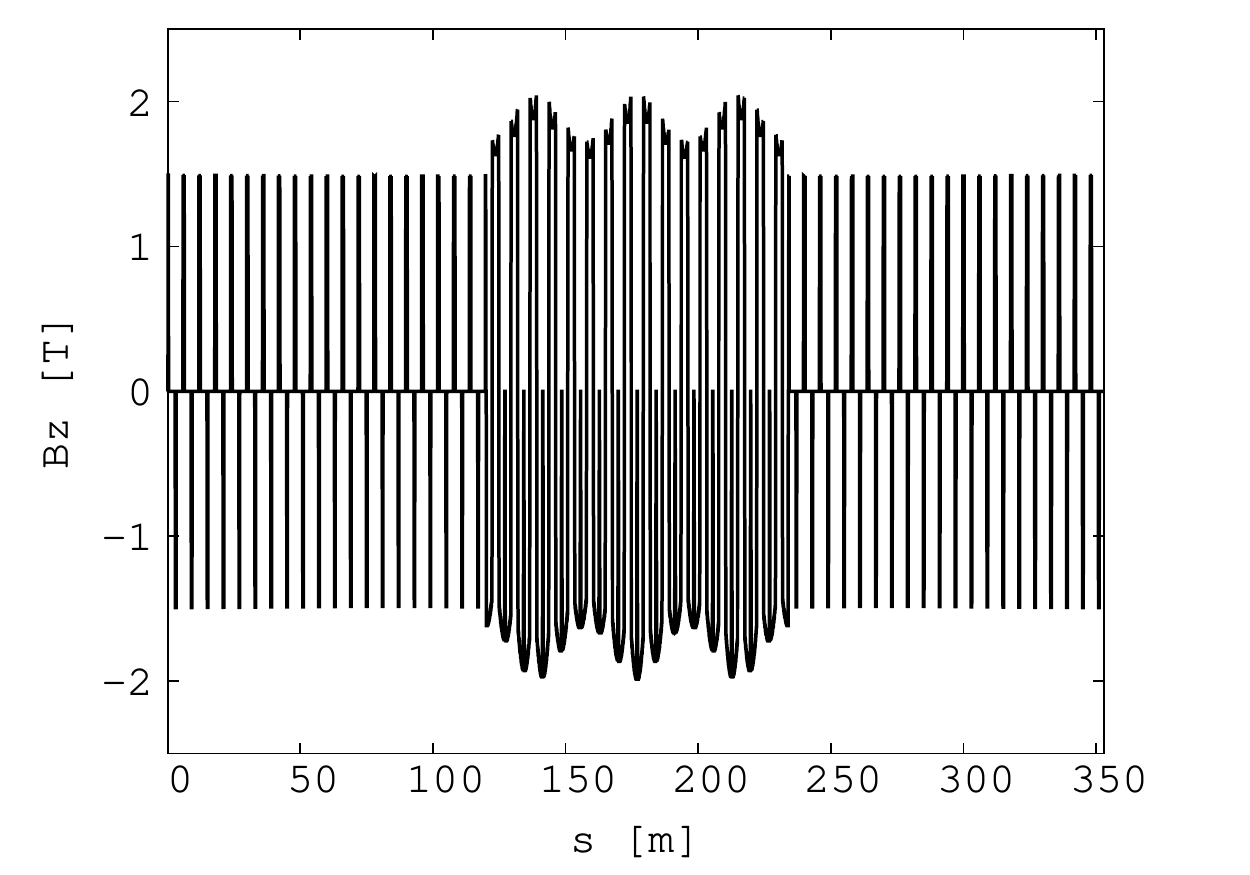}
		\caption{Vertical magnetic field for $p_{max}$ muon closed orbit in the racetrack FFAG ring.}
		\label{bz-single}
	\end{center}
\end{figure}

\begin{figure}[h!]
	\begin{center}
		\includegraphics[width=10cm]{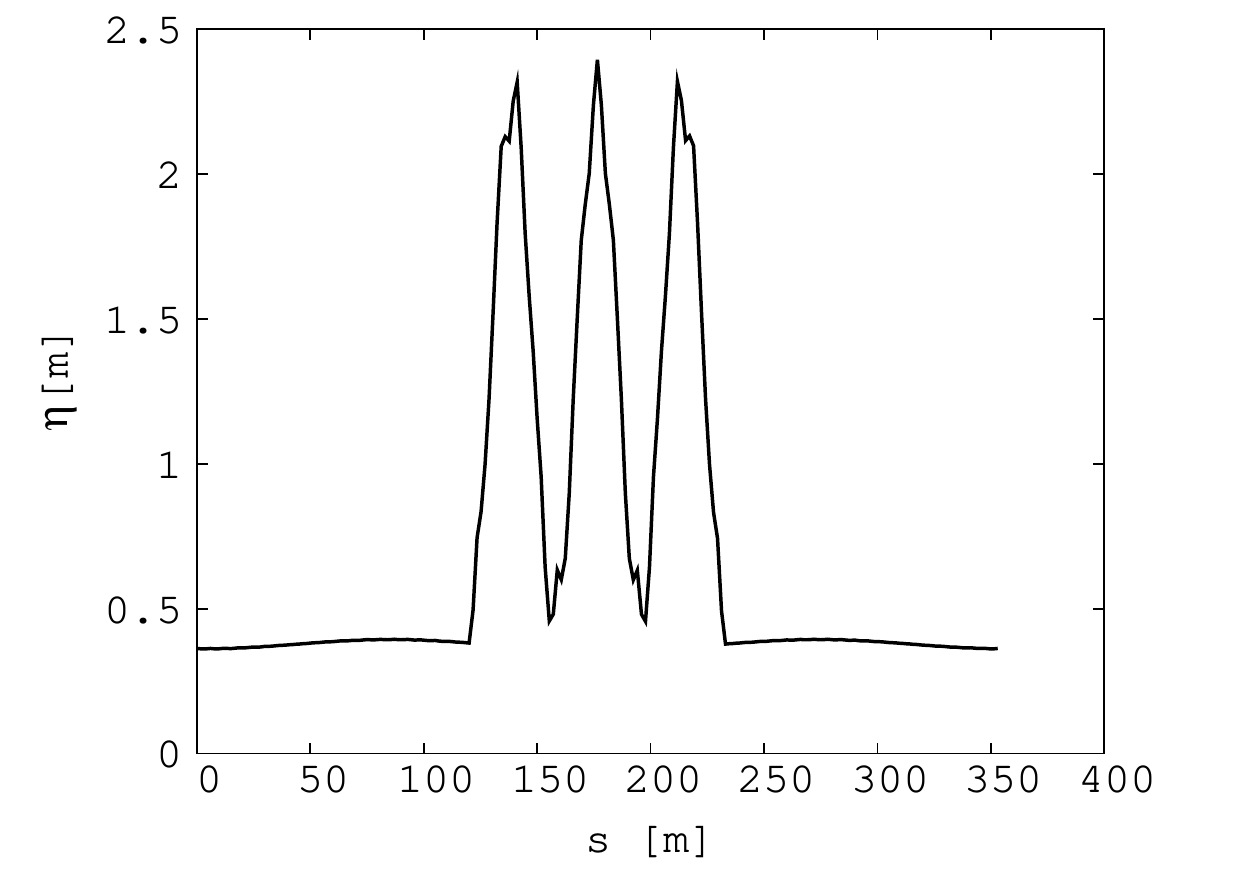}
		\caption{Dispersion function for $p_0$ in half of the ring. The plot is centered on the arc part.}
		\label{disp-single}
	\end{center}
\end{figure}

\begin{figure}[h!]
	\begin{center}
		\includegraphics[width=12cm]{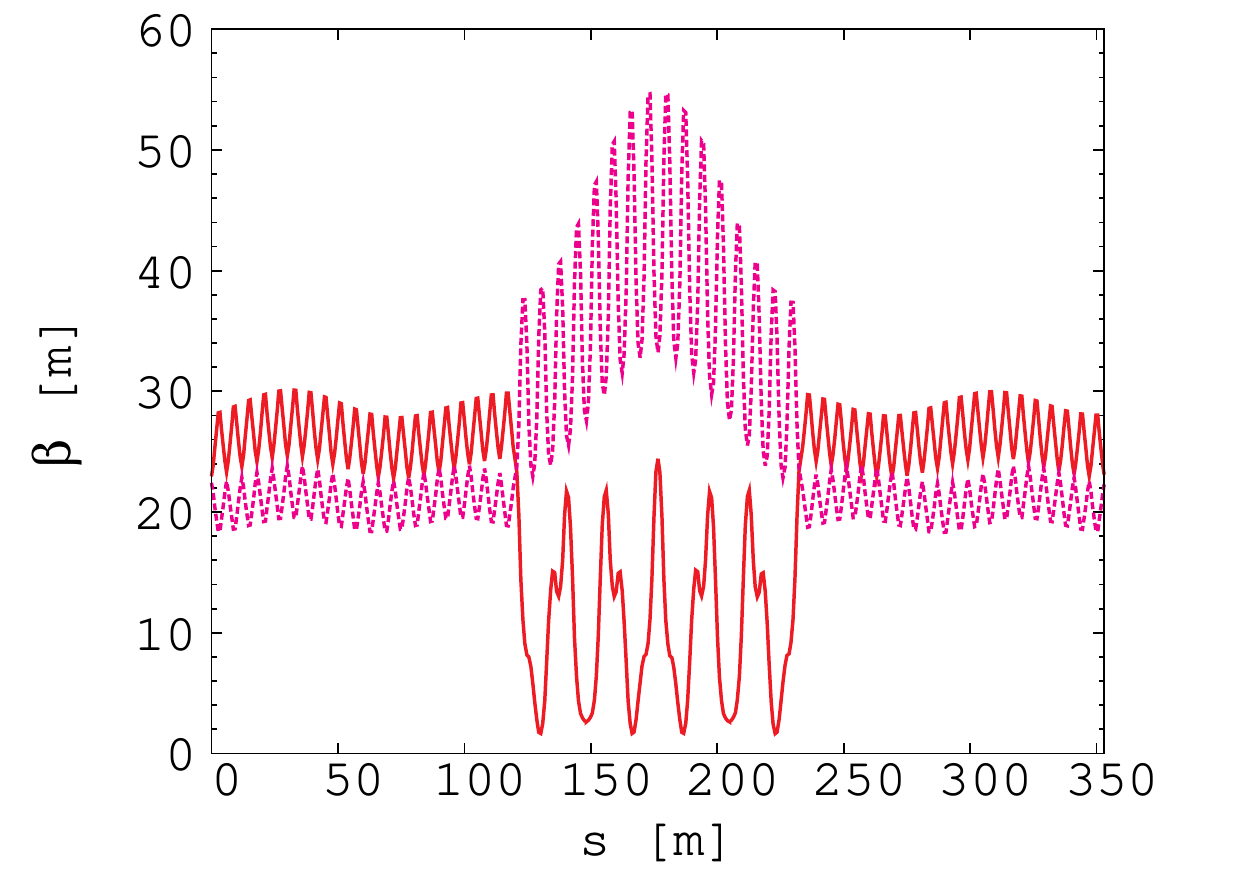}
		\caption{Horizontal (plain red) and vertical (dotted purple) periodic $\beta$ functions of half
		of the ring for $p_{0}$. The plot is centered on the arc part.}
		\label{beta-single}
	\end{center}
\end{figure}

Closed orbits of $p_0$, $p_{min}$, and $p_{max}$ particles are shown in Fig.~\ref{single-traj}. The magnetic field for the $p_{max}$ closed orbit is presented in Fig.~\ref{bz-single}. Dispersion at $p_{0}$ is shown in Fig.~\ref{disp-single}. $\beta$ functions for $p_0$, $p_{min}$, and $p_{max}$ are plotted in Fig.~\ref{beta-single}. 

\begin{figure}[h!]
   \begin{minipage}[b]{.49\linewidth}
   \centering
       \includegraphics[width=8.cm]{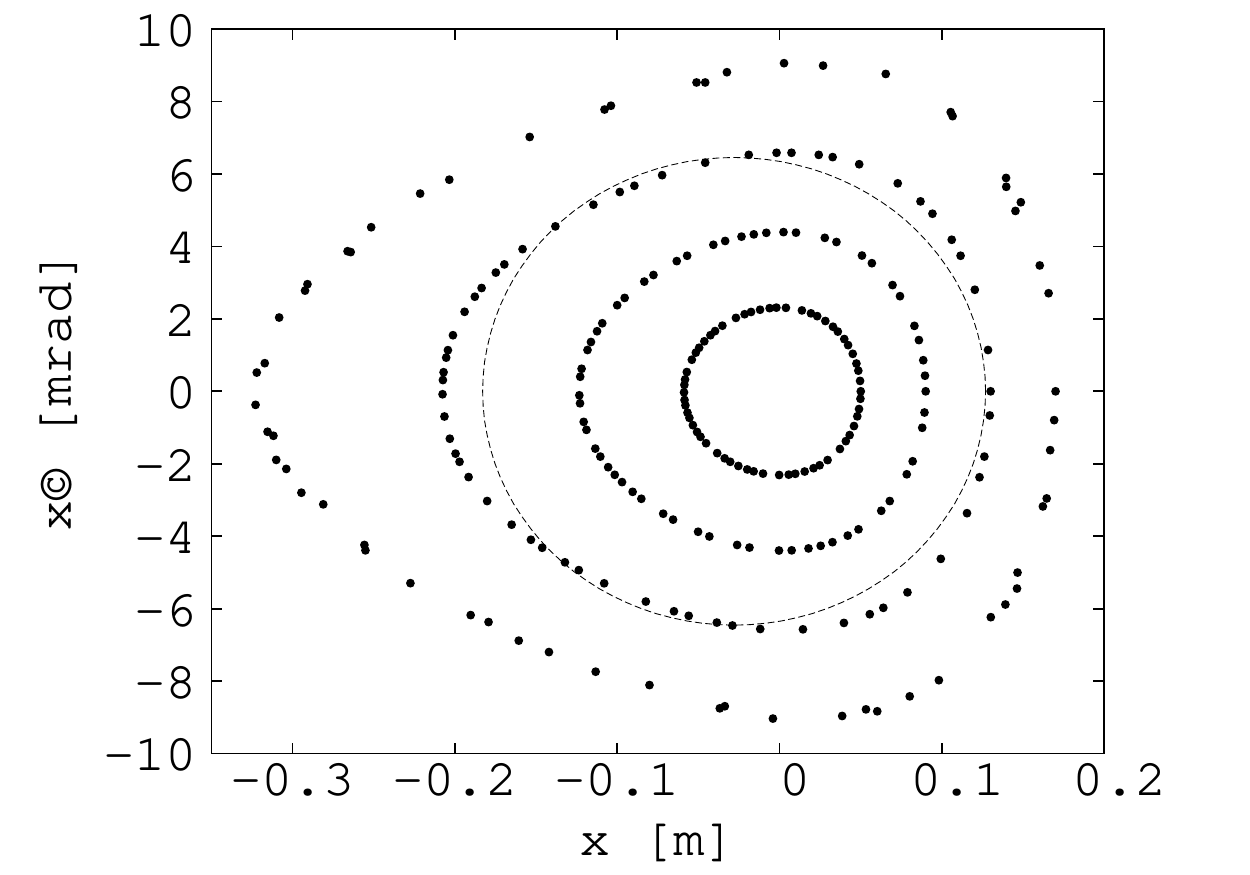}
	\caption{Stable motions in the horizontal Poincare map for different initial amplitudes (5~cm, 9~cm, 13~cm and 17~cm) over 30 turns for $p_0$. The ellipse shows a 1 mm-radian unnormalized emittance.}
	\label{poincarrex-single}
	 \end{minipage} \hfill
   \begin{minipage}[b]{.49\linewidth}
   \centering
    	\includegraphics[width=8.cm]{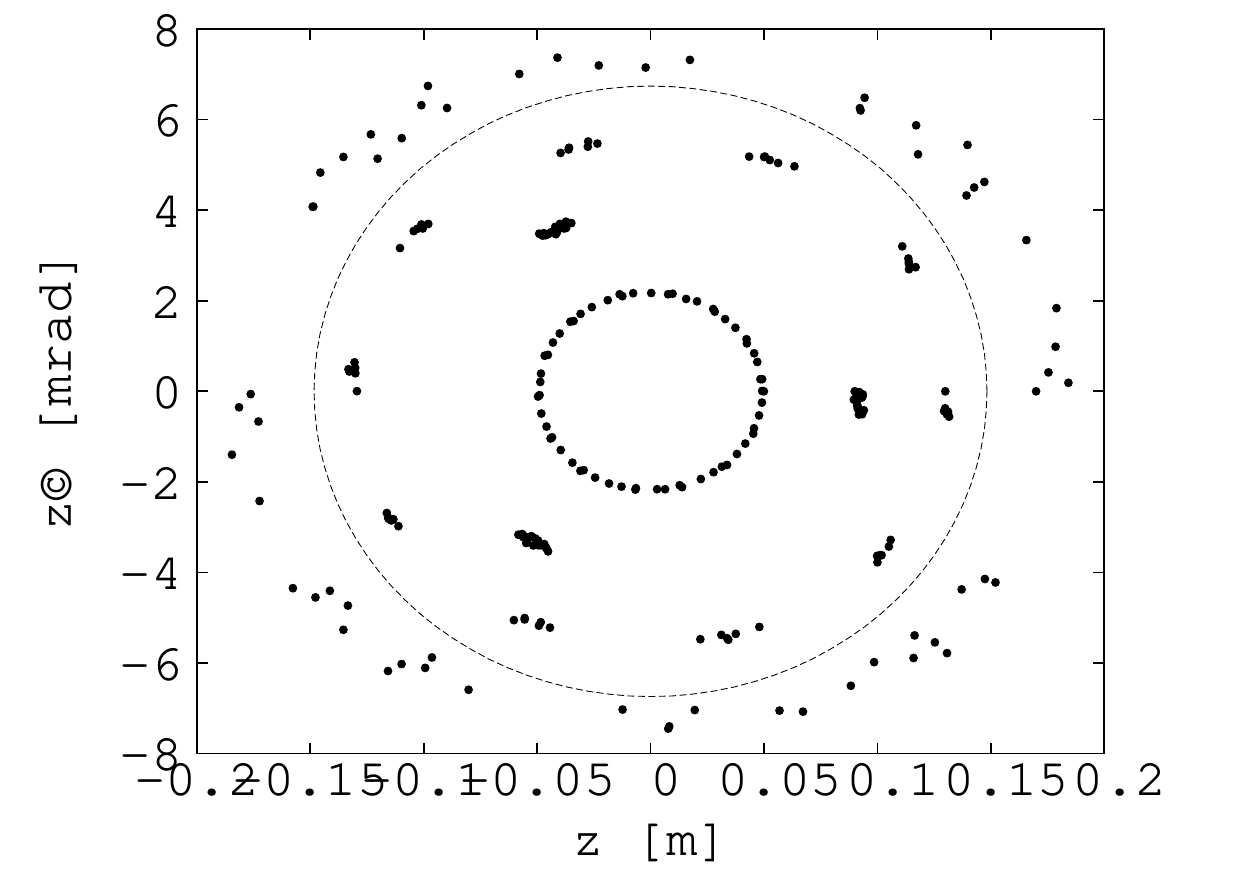}
	\caption{Stable motions in the vertical Poincare map for different initial amplitudes (5~cm, 9~cm, 13~cm and 17~cm) over 30 turns for $p_0$. The ellipse shows a 1 mm-radian unnormalized emittance.}
	\label{poincarrez-single}
   \end{minipage}
\end{figure}

\begin{figure}[h!]
   \begin{minipage}[b]{.49\linewidth}
   \centering
       \includegraphics[width=8.cm]{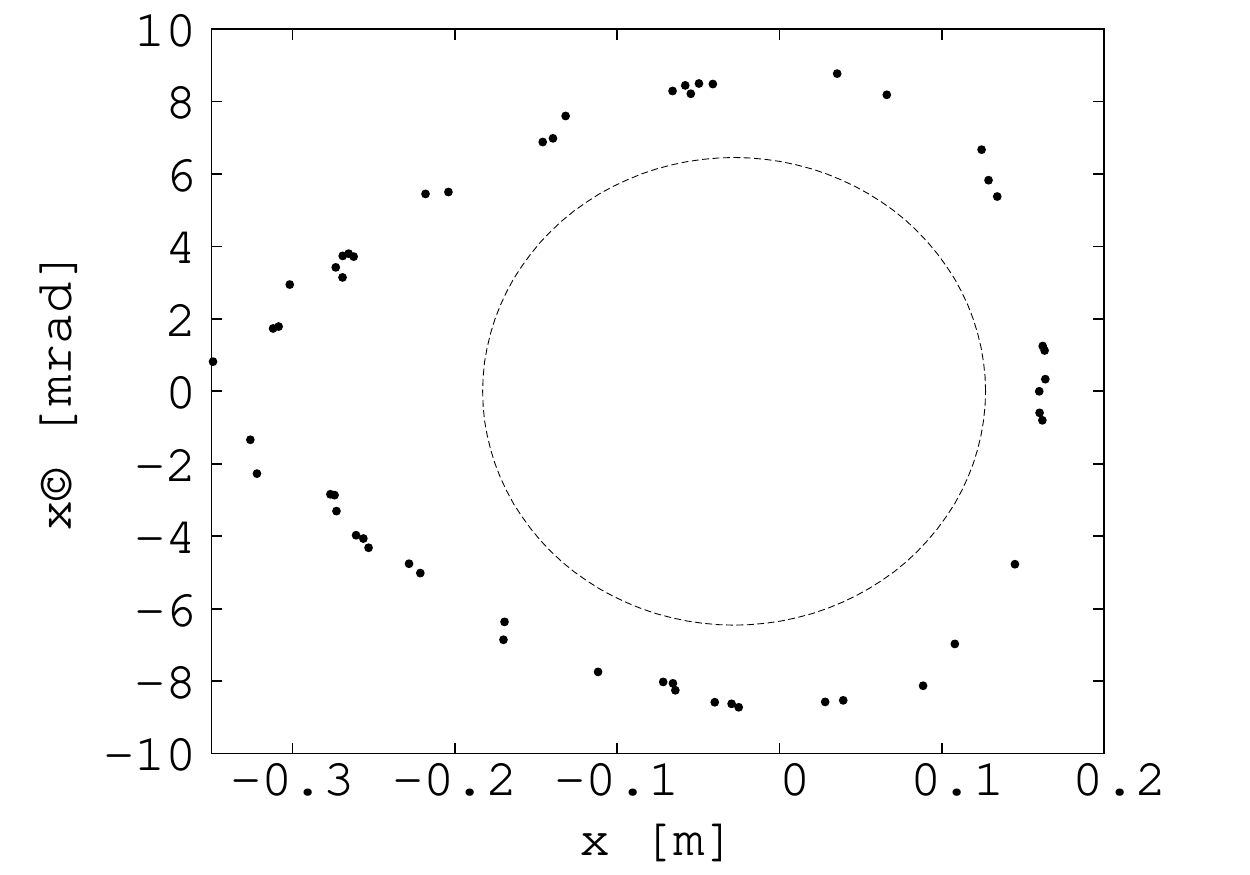}
	\caption{Horizontal Poincare map for maximum initial amplitude (16~cm) with stable motion over 30 turns for $p_{min}$. The ellipse shows a 1 mm-radian unnormalized emittance.}
	\label{poincarrex-single-pmin}
	 \end{minipage} \hfill
   \begin{minipage}[b]{.49\linewidth}
   \centering
    	\includegraphics[width=8.cm]{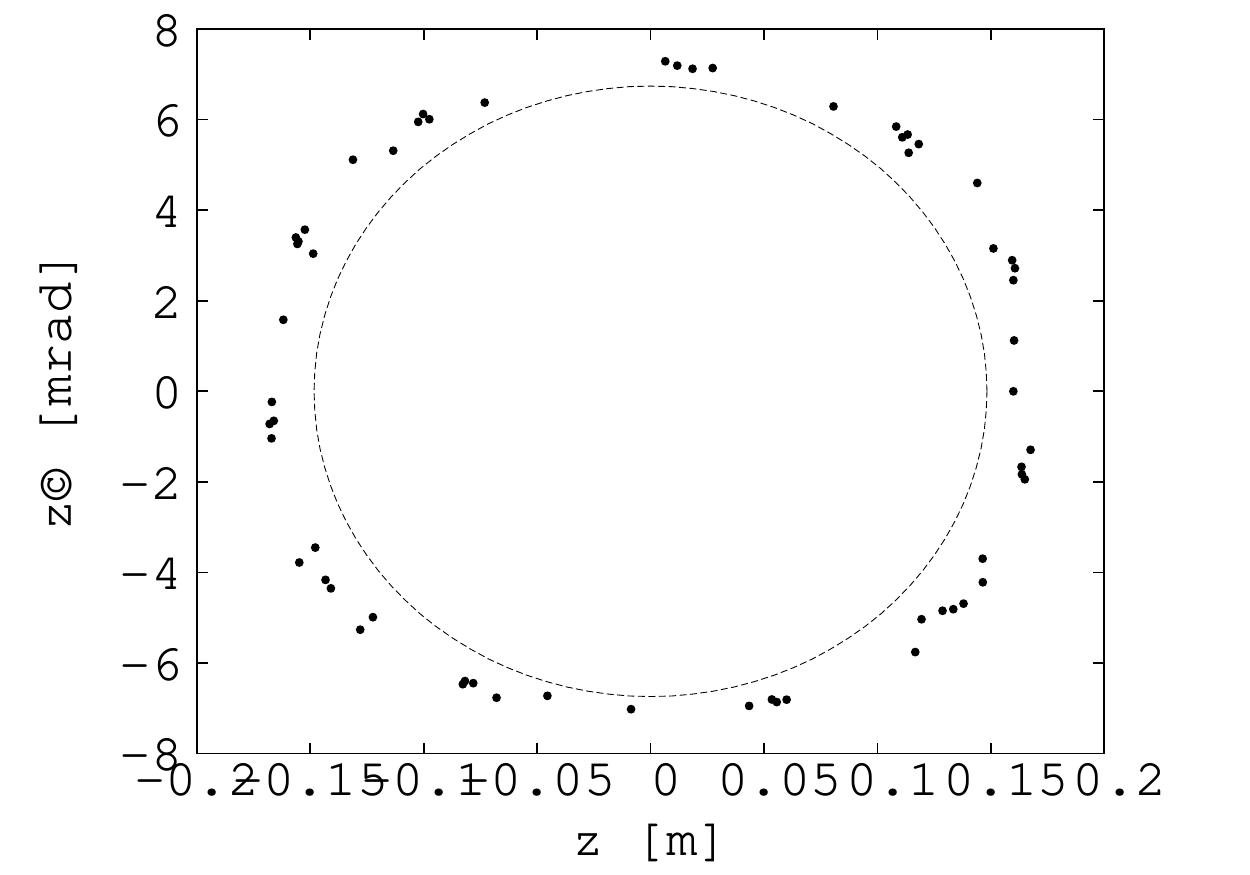}
	\caption{Vertical Poincare map for maximum initial amplitude (16~cm) with stable motion over 30 turns for $p_{min}$. The ellipse shows a 1 mm-radian unnormalized emittance.}
	\label{poincarrez-single-pmin}
   \end{minipage}
\end{figure}

\begin{figure}[h!]
   \begin{minipage}[b]{.49\linewidth}
   \centering
       \includegraphics[width=8.cm]{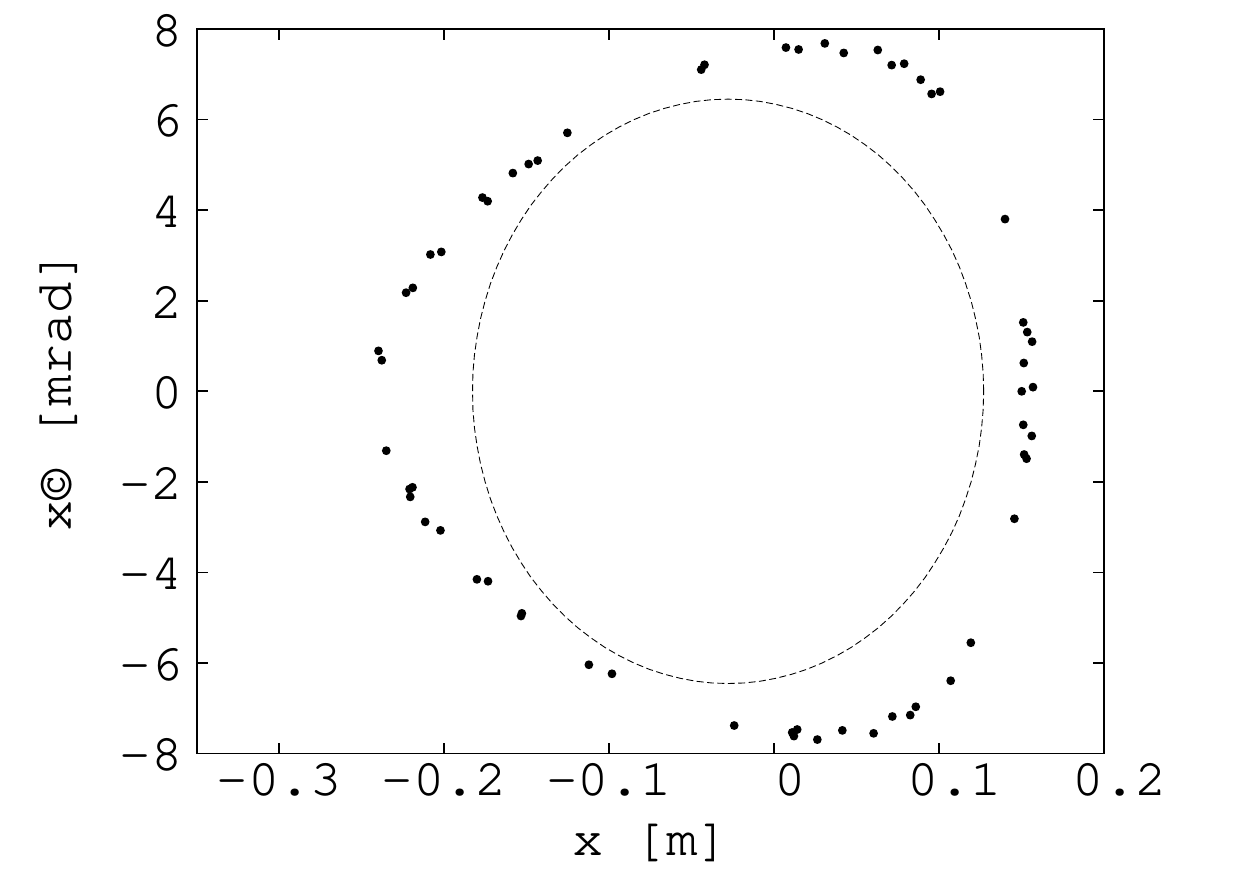}
	\caption{Horizontal Poincare map for maximum initial amplitude (15~cm) with a stable motion over 30 turns for $p_{max}$. The ellipse shows a 1 mm-radian unnormalized emittance.}
	\label{poincarrex-single-pmax}
	 \end{minipage} \hfill
   \begin{minipage}[b]{.49\linewidth}
   \centering
    	\includegraphics[width=8.cm]{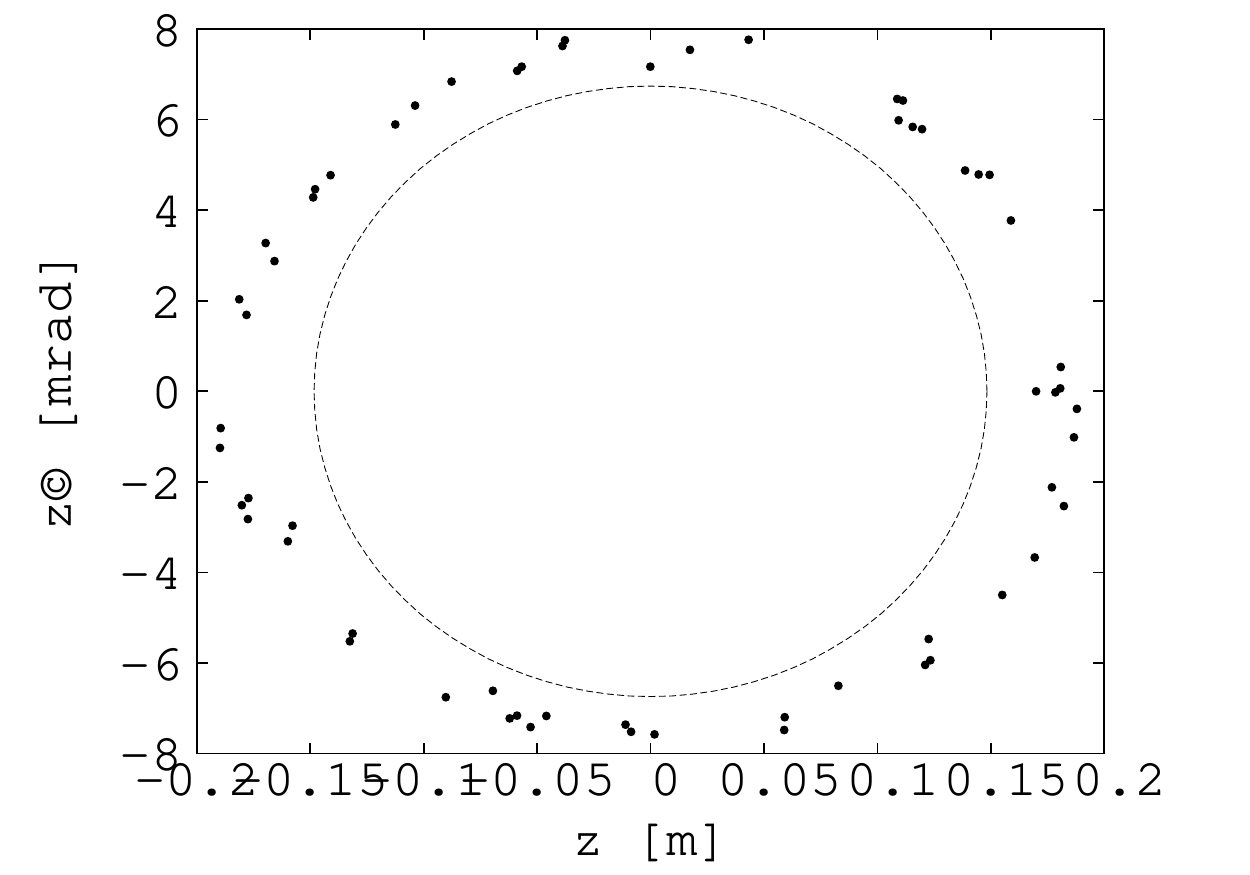}
	\caption{Vertical Poincare map for maximum initial amplitude (17~cm) with a stable motion over 30 turns for $p_{max}$. The ellipse shows a 1 mm-radian unnormalized emittance.}
	\label{poincarrez-single-pmax}
   \end{minipage}
\end{figure}

An acceptance study at fixed energy has also been done. The maximum amplitudes with stable motion at $p_0$ over 30 turns 
are shown for horizontal and vertical motion in Fig.~\ref{poincarrex-single} (left) and in Fig.~\ref{poincarrez-single} 
(right), respectively. 
The same procedure has been done for $p_{min}$ (see Fig.~\ref{poincarrex-single-pmin})
and $p_{max}$ (see Fig.~\ref{poincarrex-single-pmax}). The results are comparable. The 
unnormalized maximum emittance is more than 1 mm-radian.

\begin{figure}[h!]
   \begin{minipage}[b]{.49\linewidth}
   \centering
       \includegraphics[width=8.cm]{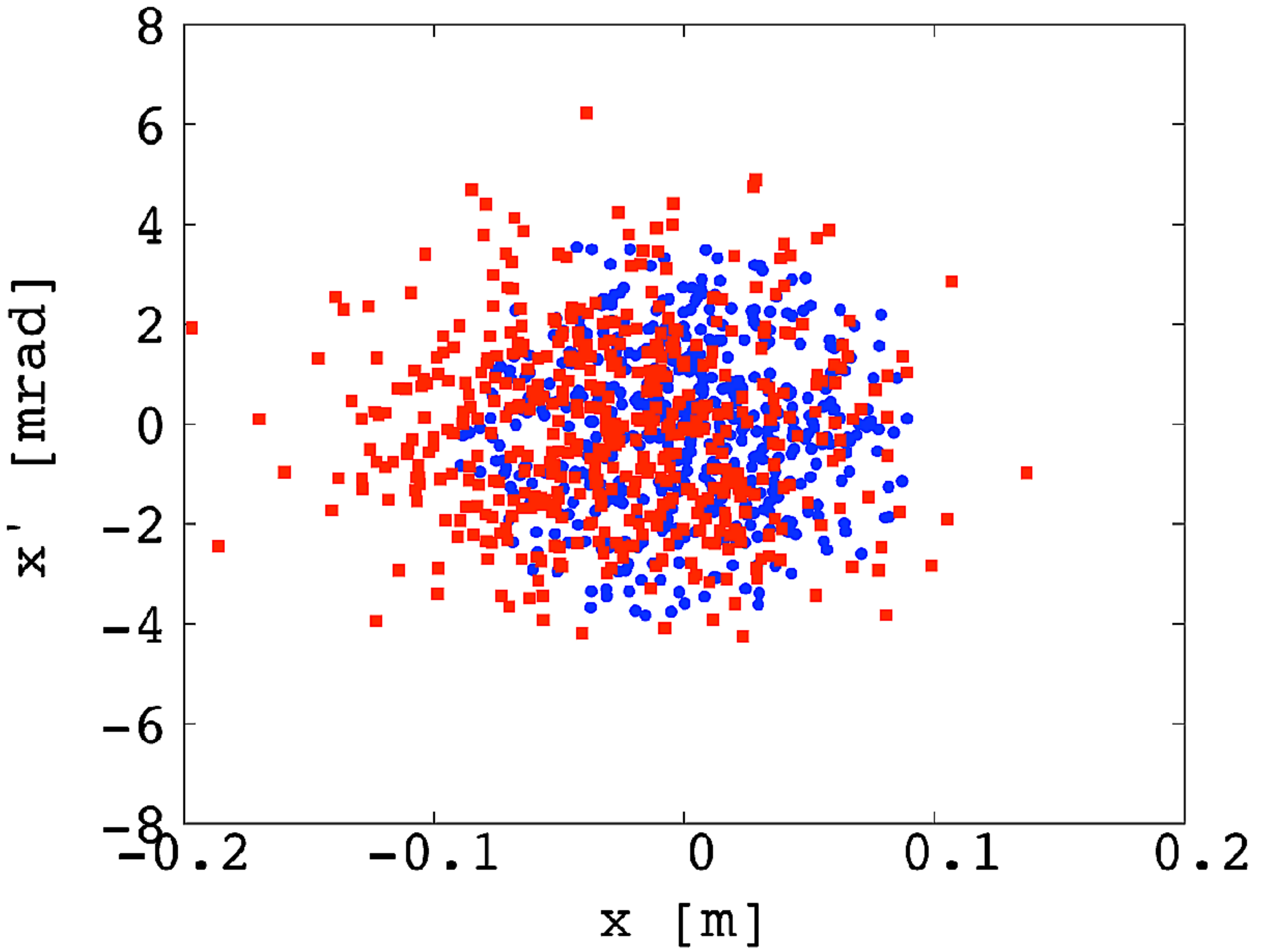}
	\caption{Beam tracking results in the horizontal phase space for a beam with $\Delta p/p_0=\pm 16\%$.   The blue shows the initial particle distribution and the red the final distribution after 60 turns.}
	\label{fig:Bthp}
	 \end{minipage} \hfill
   \begin{minipage}[b]{.49\linewidth}
   \centering
    	\includegraphics[width=8.cm]{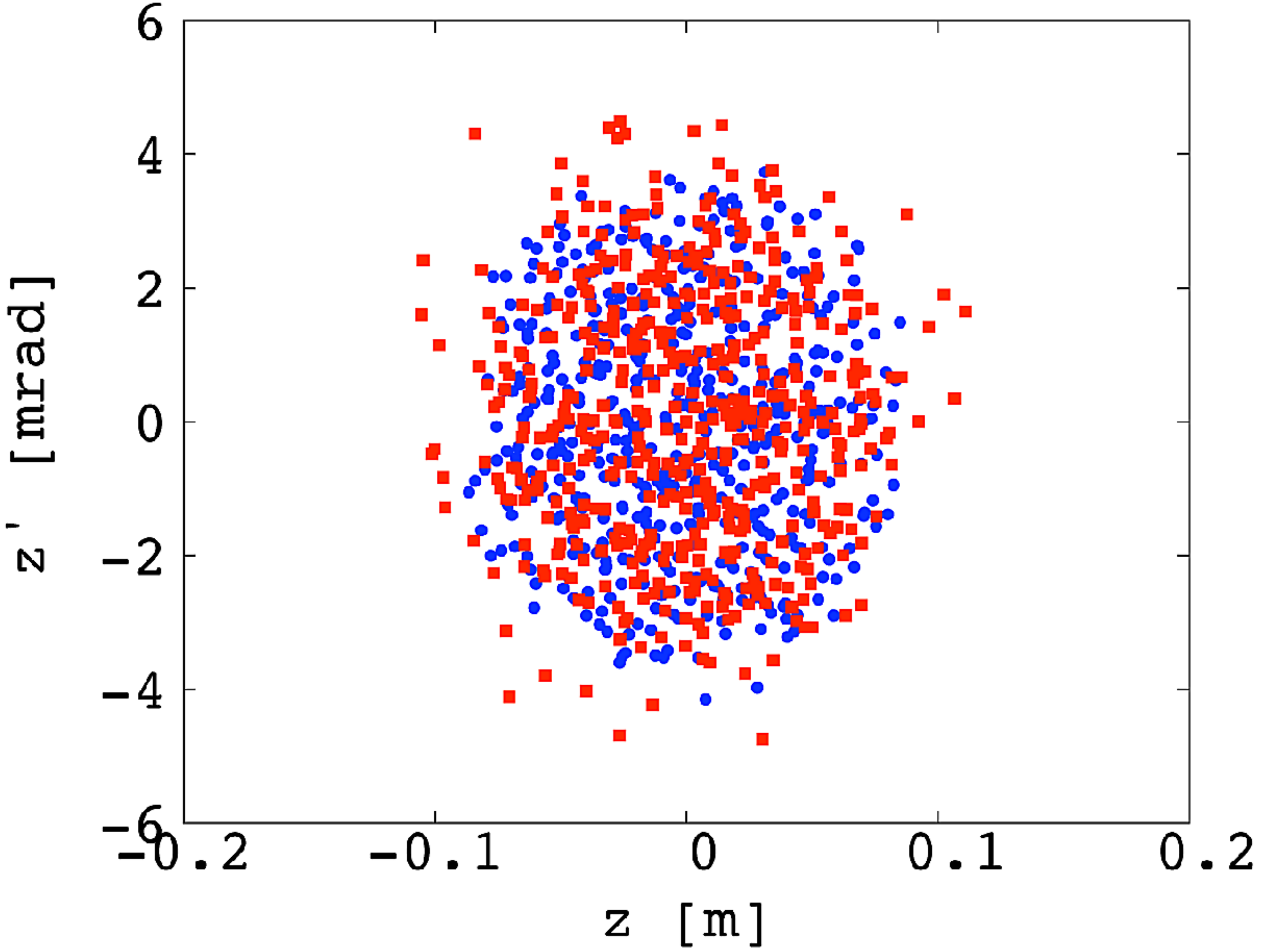}
	\caption{Beam tracking results in the vertical phase space for a beam with $\Delta p/p_0=\pm 16\%$.   The blue shows the initial particle distribution and the red the final distribution after 60 turns. }
	\label{fig:Btvp}
   \end{minipage}
\end{figure}

\paragraph{Multi-particle tracking}
 Multi-particle beam tracking in 6-D phase space has been carried out for the beam with $\Delta p/p_0=\pm 16\%$.   Fig.~\ref{fig:Bthp} and \ref{fig:Btvp} show the results of the beam tracking simulation in the horizontal and vertical directions, respectively.  A normalized emittance of 14 mm-radian in the transverse direction is assumed.    In these figures, the blue dots show the initial particle distribution and the red ones are after 60 turns.  No beam loss is observed in 60 turns.
 
\clearpage
\section{Far Detector - SuperBIND}
\label{sec:far}
The Super B Iron Neutrino Detector (SuperBIND) is an iron and scintillator sampling 
calorimeter which is similar in concept to the MINOS detectors \cite{Michael:2008bc}.
We have chosen a cross section of approximately 5 m in order to maximize the ratio of 
the fiducial mass to total mass.  The magnetic field will be toroidal as in MINOS and 
SuperBIND will also use extruded scintillator for the readout planes.  Details on the 
iron plates, magnetization, scintillator, photodetector and electronics are given below.
Fig.~\ref{fig:SuperBIND} gives an overall schematic of the detector.
\begin{figure}[hbtp]
  \centering{
    \includegraphics[width=0.9\textwidth]{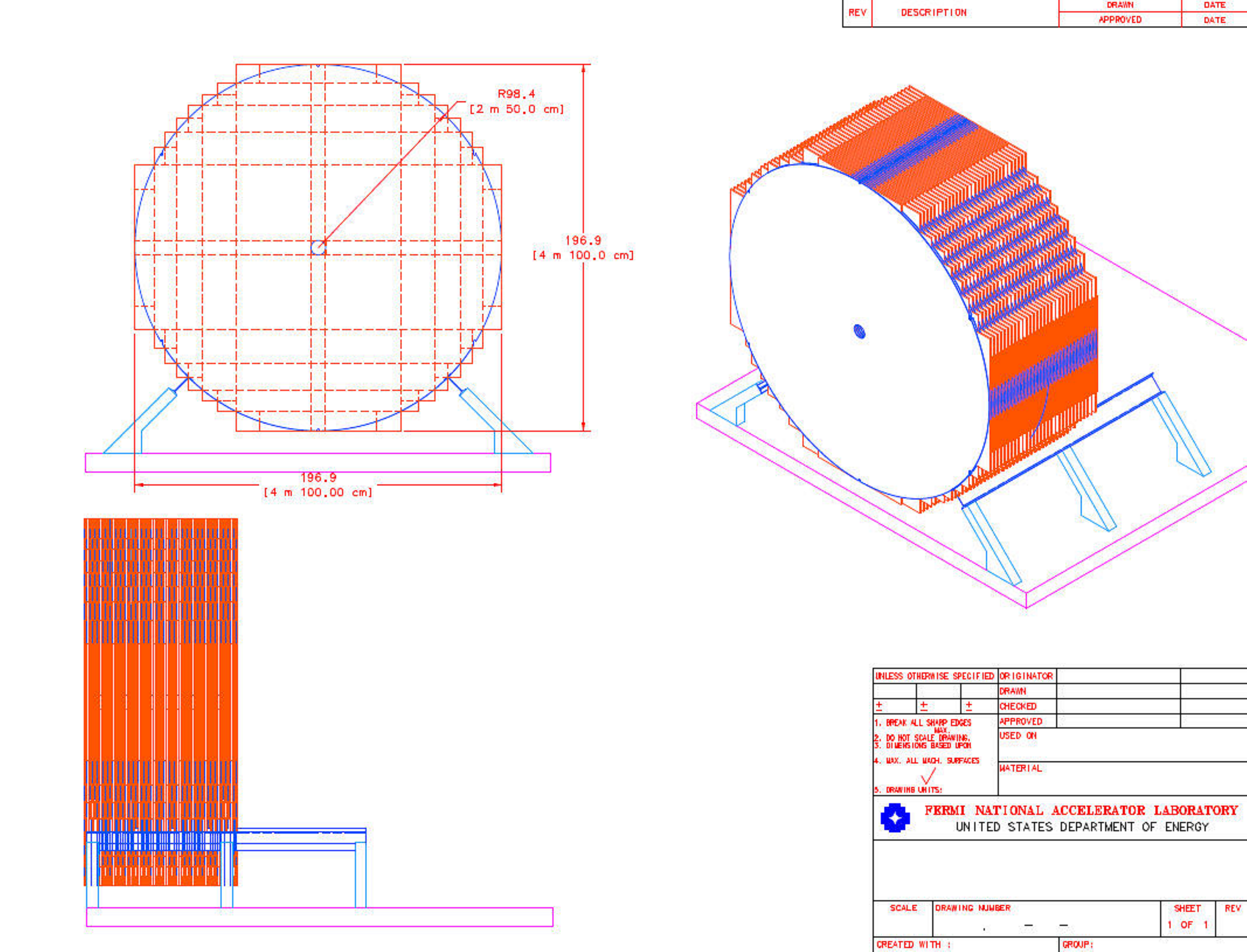}}
  \caption{Far Detector concept}
  \label{fig:SuperBIND}
\end{figure}
We note that within the Advanced European Infrastructures for Detectors at Accelerators (AIDA)
project, whose time line runs from 2011 to 2015, detectors similar to those planned for
$\nu$STORM will be built and characterized at CERN. The motivation is to test the 
capabilities for charge identification of $\le$~5GeV/c electrons in a Totally Active 
Scintillator Detector and $\le$5~GeV/c muons in a Magnetized Iron Neutrino Detector (MIND). 
These detector prototypes will provide further experience in the use of STL technology, 
and SiPMs and associated electronics, to complement the already large body of knowledge 
gained through past and current operation of this type of detector.
\subsection{Iron Plates}
For the Iron plates in SuperBIND, we are pursuing the following design strategy.  
The plates are cylinders with an overall diameter of 5 m and and depth of 1-2 cm.
Our original engineering design uses 1 cm plates, but we have simulated the detector
performance for both 1 cm and 2 cm thick plates.  They are fabricated 
from two semicircles that are skip welded together.  Instead of hanging the plates on ears 
(as was done in MINOS), we plan to stack in a cradle using a strong-back when starting the 
stacking.  We envision that no R\&D on the iron plates will be needed.  Final specification 
of the plate structure would be determined once a plate fabricator is chosen.
\subsection{Magnetization}
\begin{figure}[hbtp]
  \centering{
    \includegraphics[width=0.6\textwidth]{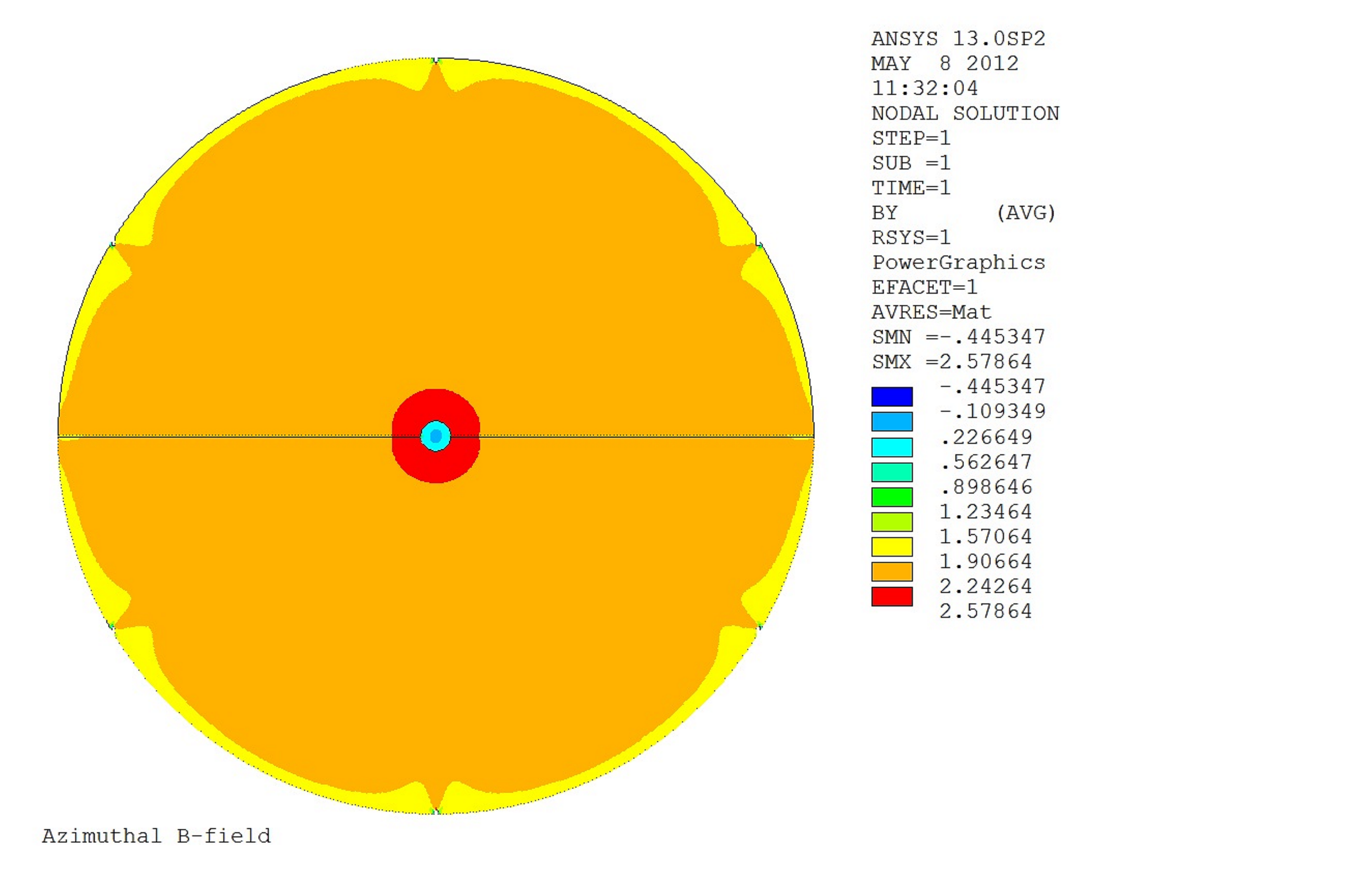}}
  \caption{Toroidal Field Map}
  \label{fig:b-Field}
\end{figure}

As was mentioned above, MIND will have a toroidal magnetic field like that of MINOS.  
For excitation, however, we plan to use the concept of the Superconducting Transmission 
Line (STL) developed for the Design Study for a Staged Very Large Hadron 
Collider \cite{Ambrosio:2001ej}.  Minimization of the muon charge mis-identification rate requires the highest field
possible in the iron plates.  SuperBIND requires 
a much large excitation current per turn than that of the MINOS near detector 
(40 kA-turns).  We have simulated 3 turns of the STL (20 cm hole).  The STL is described
in Appendix~\ref{sec:Appen} and shown in Fig.~\ref{fig:STL}.
%
%
%
%
%
Utilizing the SuperBIND plate geometry shown in Fig.~\ref{fig:SuperBIND}, a 2-d finite 
element magnetic field analysis for the plate was performed.  Fig.~\ref{fig:b-Field} shows the 
results of those calculations.  For this analysis, a 20 cm diameter hole for the STL was assumed, the CMS steel 
\cite{Smith:2004uf} BH curve was used and an excitation current of 250 
kA-turn was assumed.  This current represents approximately 80\% of the critical current achieved at 
6.5K in the STL test stand assembled for the VLHC proof-of-principle tests.  

\subsection{Detector planes}

\subsubsection{Scintillator}

Particle detection using extruded scintillator and optical fibers is a mature technology.\hfill  
MINOS has shown that co-extruded solid scintillator with embedded wavelength shifting (WLS) 
fibers and PMT readout produces adequate light for MIP tracking and that it can be 
manufactured with excellent quality control and uniformity in an industrial setting.  Many 
experiments use this same technology for the active elements of their detectors, such as 
the K2K Scibar \cite{Maesaka:2003jt}, the T2K INGRID, P0D, and 
ECAL \cite{Kudenko:2008ia} and the Double-Chooz cosmic-ray veto detectors \cite{Greiner:2007zzd}.

Our initial concept for the readout planes for SuperBIND is to have both an $x$ and a $y$ 
view between each plate. The simulations  done to date have assumed a scintillator extrusion 
profile that is 1.0 $\times$ 1.0 cm$^2$.  This gives both the required point resolution and 
light yield.  

\subsubsection{Scintillator extrusions}

The existing SuperBIND simulations have assumed that the readout planes will use an 
extrusion that is 1.0 $\times$ 1.0 cm$^2$.  A 1 mm hole down the centre 
of the extrusion is provided for insertion of the wavelength shifting fiber.  This is 
a relatively simple part to manufacture and has already been fabricated in a similar 
form for a number of small-scale applications.  The scintillator strips will consist 
of an extruded polystyrene core doped with blue-emitting fluorescent compounds, a 
co-extruded TiO$_2$ outer layer for reflectivity, and a hole in the middle for a WLS 
fiber.  Dow Styron 665 W polystyrene pellets are doped with PPO (1\% by weight) and 
POPOP (0.03\% by weight). The strips have a white, co-extruded, 0.25 mm thick TiO$_2$ 
reflective coating.  This layer is introduced in a single step as part of a 
co-extrusion process.  The composition of this coating is 15\% TiO$_2$ 
in polystyrene.  In addition to its reflectivity properties, the layer facilitates 
the assembly of the scintillator strips into modules. The ruggedness of this coating 
enables the direct gluing of the strips to each other and to the module skins which 
results in labour and time savings.  This process has now been 
used in a number of experiments.

\subsection{Photo-detector}

Given the rapid development in recent years of solid-state photodetectors 
based on Geiger mode operation of silicon avalanche photodiodes, we have 
chosen this technology for SuperBIND.  Although various names are used for this 
technology, we will use silicon photomultiplier or SiPM.

\subsubsection{SiPM Overview}

\begin{figure}[hbtp]
	\begin{center}$
	  \begin{array}{cc}
		\includegraphics[width=0.4\textwidth]{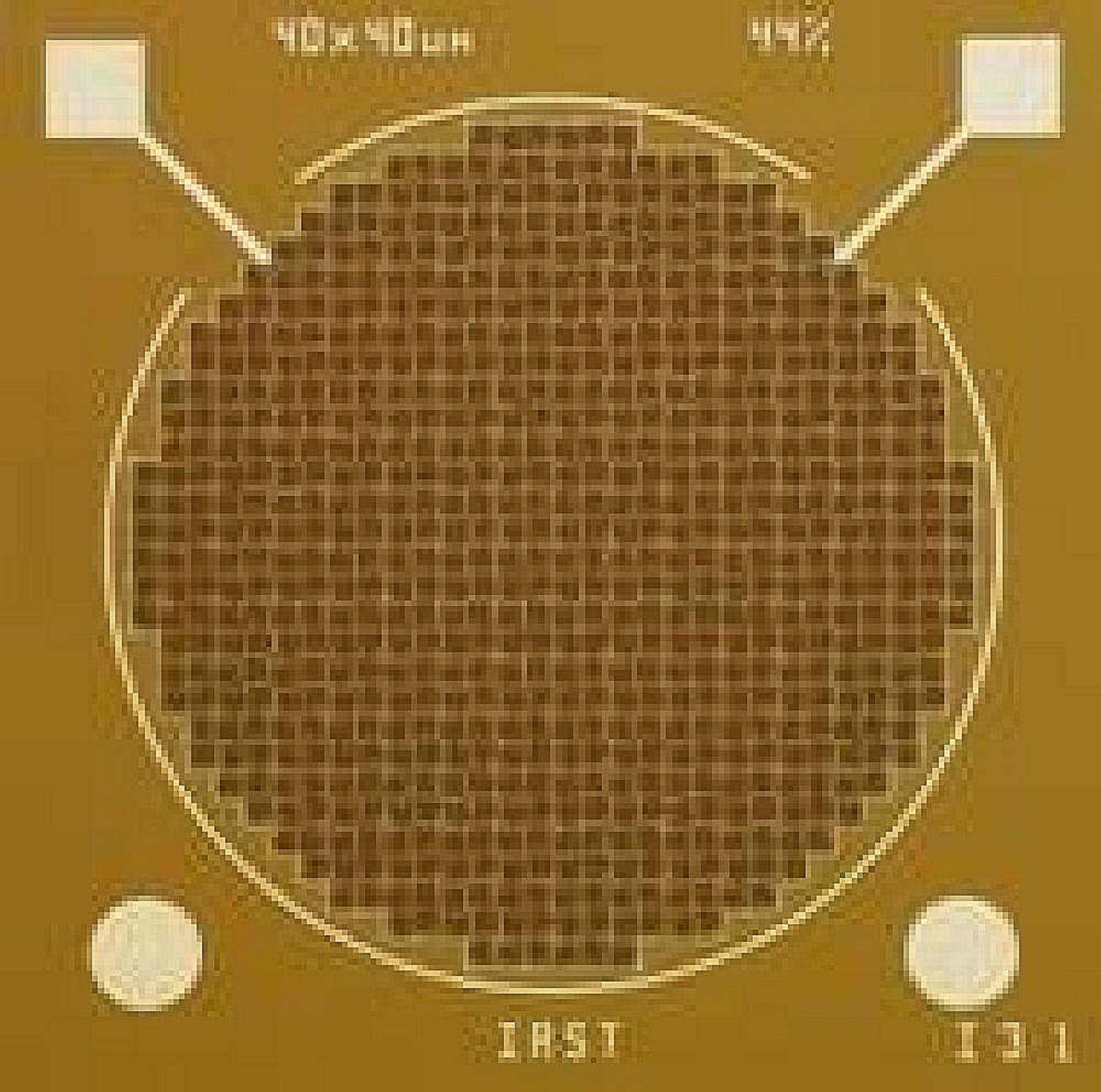} &
		\includegraphics[width=0.6\textwidth]{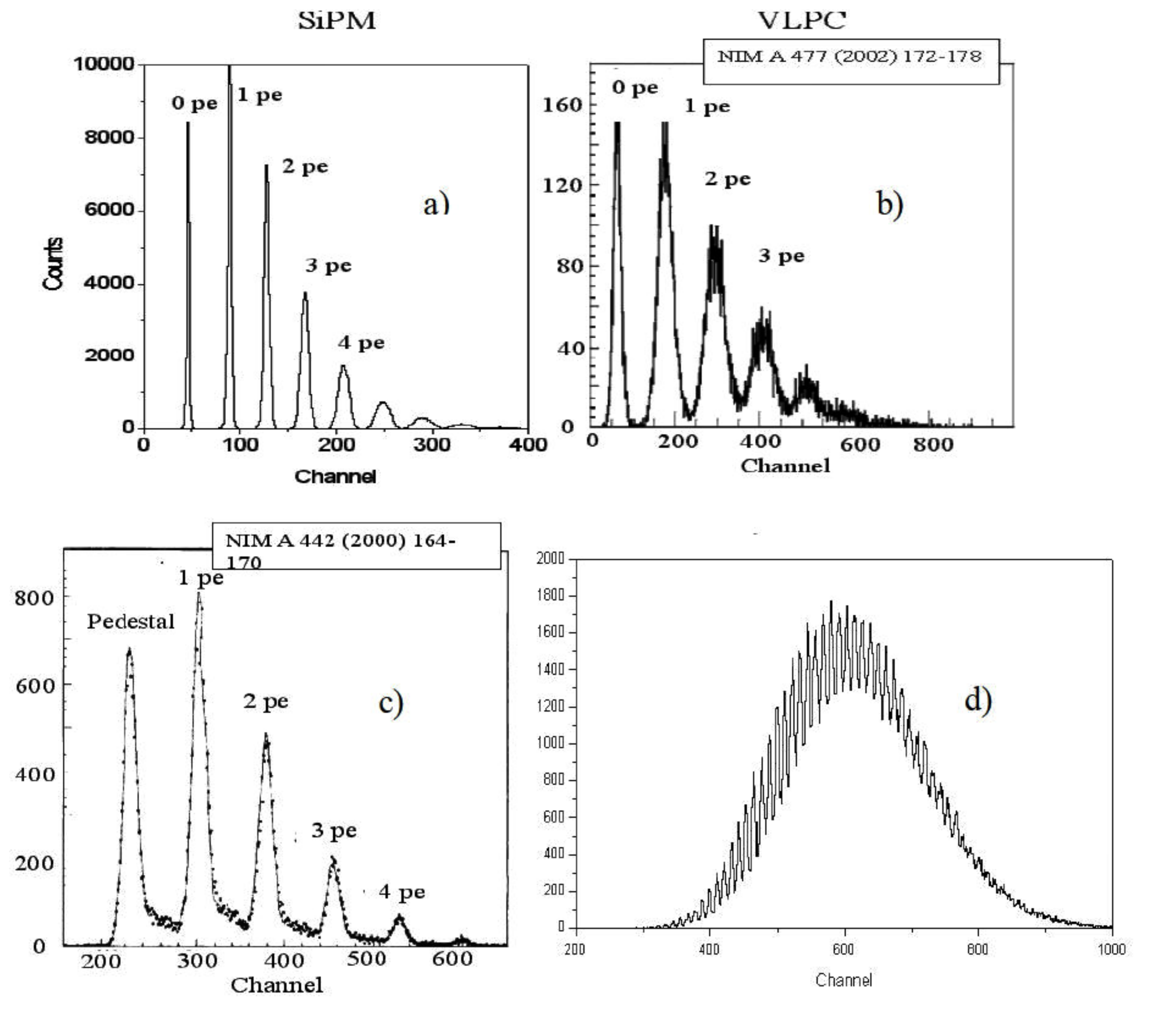}
	  \end{array}$
    \end{center}
  \caption{Photograph of SiPM (left) and SiPM photon counting capability (a) compared to 
    VLPC (b) and HPD (c) . The SiPM pulse height spectrum (d) for an intense 
	light burst with a mean photoelectron number of 46 is also shown.}
  \label{Fig:SiPM}
\end{figure}

SiPM is the often-used name for a type of photo detector formed by
combining many small avalanche photodiodes operated in the Geiger mode
to form a single detector \cite{Sadygov:1996, Bacchetta:1996dc}.
Detailed information and basic principles of operation of these
``multi-pixel"  photodiodes can be found  in a recent review paper and
the references therein \cite{Renker:2009zza}. The first generation of
these detectors use a polysilicon resistor connected to each avalanche
photodiode forming  a pixel. Pixels usually vary in size from 10
$\times 10 \, \mu$m$^2$ to 100 $\times~100~\mu$m$^2$ (see Fig.~\ref{Fig:SiPM}, left). 
All the diodes are connected to a common electrical point on one side, typically 
through the substrate, and all the resistors are connected to a common grid with metal 
traces on the other side to form a two node device.  A typical SiPM will have from 100 
to 10,000 of these pixels in a single device, with the total area from 1 to 10 mm$^2$.  
Because all the diodes and the individual quenching resistors are connected in parallel, 
the SiPM device as a whole appears as a single diode. In operation, the device appears 
to act somewhat like a conventional APD, but in detail it is radically different. 
Because the diodes are operated in the Geiger mode, and because every pixel of the 
SiPM device is nearly identical, the sum of the fired pixels gives the illusion of 
an analog signal that is proportional to the incident light, but it is an 
essentially digital device.  The photo counting capabilities of the SiPM are unmatched, as 
can be seen in Fig.~\ref{Fig:SiPM} (right) from \cite{Dolgoshein:2003nt}.
SiPMs have a number of advantages over conventional photo multiplier tubes, 
including high photon detection efficiency, complete immunity to magnetic fields, 
excellent timing characteristics, compact size and physical robustness. They are 
immune to nuclear counter effect and do not age. They are particularly well suited 
to applications where optical fibers are used, as the natural size of the SiPM is 
comparable to that of fibers. But the most important single feature of the SiPM is 
that it can be manufactured in standard microelectronics facilities using well 
established processing. This means that huge numbers of devices can be produced 
without any manual labor, making the SiPMs very economical as the number of 
devices grows. Furthermore, it is possible to integrate the electronics into 
the SiPM itself, which reduces cost and improves performance. Initial steps 
have been taken in this direction, though most current SiPMs do not have integrated 
electronics. But it is widely recognized that this is the approach that makes sense 
in the long run for many applications. It improves performance and reduces cost, 
and can be tailored to a specific application. As the use of SiPMs spreads, so 
will the use of custom SiPM with integrated electronics, just as ASICs have 
superseded standard logic in micro electronics. 

\begin{figure}[hbtp]
\centering\includegraphics[width=0.7\textwidth]{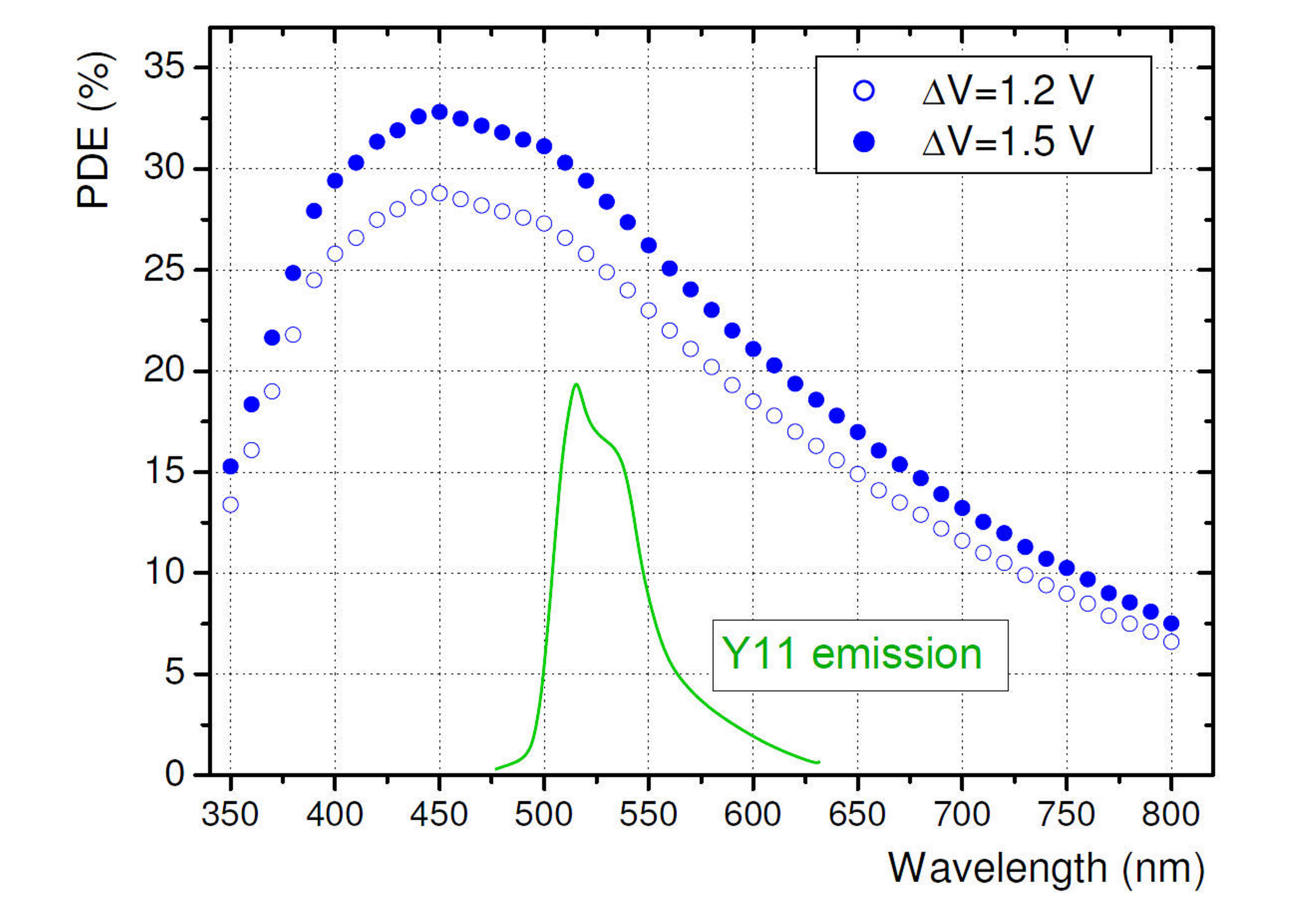}
\caption{Photon detection efficiency  of a Hamamatsu MPPC as a function of wavelength 
of the detected light at  $\Delta V$ of 1.2 and 1.5~V at 25$^{\circ}$C.  The Y11(150 ppm) 
Kuraray fiber emission spectrum for a fiber length of 150 cm (from Kuraray 
specification) is also shown.}
\label{fig:pde_mppc}
\end{figure}

The photon detection efficiency (PDE) of a SiPM is the product of 3  factors:
\begin{center}
\begin{equation}
{\rm PDE} = QE\cdot\varepsilon_{Geiger}\cdot\varepsilon_{pixel},
\label{eq:pde}
\end{equation}
\end{center}
where $QE$ is the wavelength-dependent quantum efficiency, $\varepsilon_{Geiger}$ is the 
probability  to initiate the Geiger discharge by a photoelectron, and $\varepsilon_{pixel}$ 
is the fraction of the total photodiode  area occupied by sensitive pixels. The  bias voltage 
affects one parameter in the expression~(\ref{eq:pde}), $\varepsilon_{Geiger}$.   The 
geometrical factor $\varepsilon_{pixel}$  is  completely 
determined by the photodiode topology, and is in the range 50-70\%.    The PDE of a 
device manufactured by Hamamatsu (Hamamatsu uses the name multi-pixel photon counter, 
MPPC) as function of wavelength of detected light is shown in Fig.~\ref{fig:pde_mppc}.

\subsubsection{Readout Electronics}
Currently, a number of companies are working on integrating electronics and SiPM 
detectors on the same device, on the same wafer. The first such device was announced 
by Philips in 2009 and a complete system for evaluation of this technology is 
commercially available. The system features a fully digital SiPM with active 
quenching and it is reasonable to expect that this technology will continue to 
advance and new devices with lower costs and better performance will appear.
However, one important disadvantage of integrating electronics with the photodetector 
is that the SiPM becomes an ASIC, an Application Specific Integrated Circuit, and it 
is much more likely that additional R\&D will be required to develop the system. The 
question then becomes what level of investment in research and development is 
justified in order to optimize the detector for the particular application described 
here. Clearly, it is much too early to answer this question, but generally we can 
outline three possible approaches, given the current state of SiPM development.

\begin{figure}[hbtp]
\centering\includegraphics[width=0.7\textwidth]{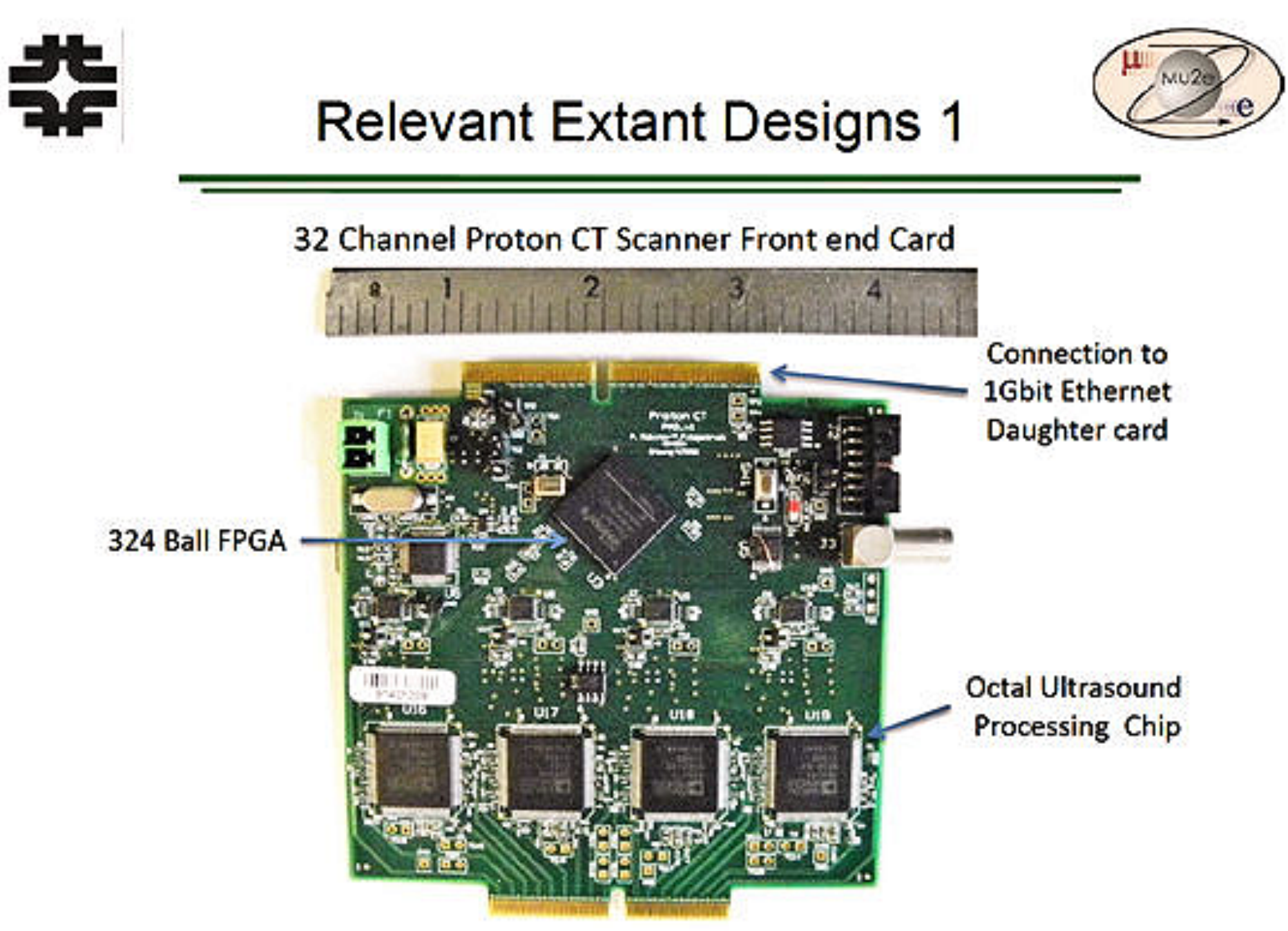}
\caption{32 channel SiPM readout card based on commercially available electronics.}
\label{fig:32CHOC}
\end{figure}

\begin{figure}[hbtp]
\centering\includegraphics[width=0.6\textwidth]{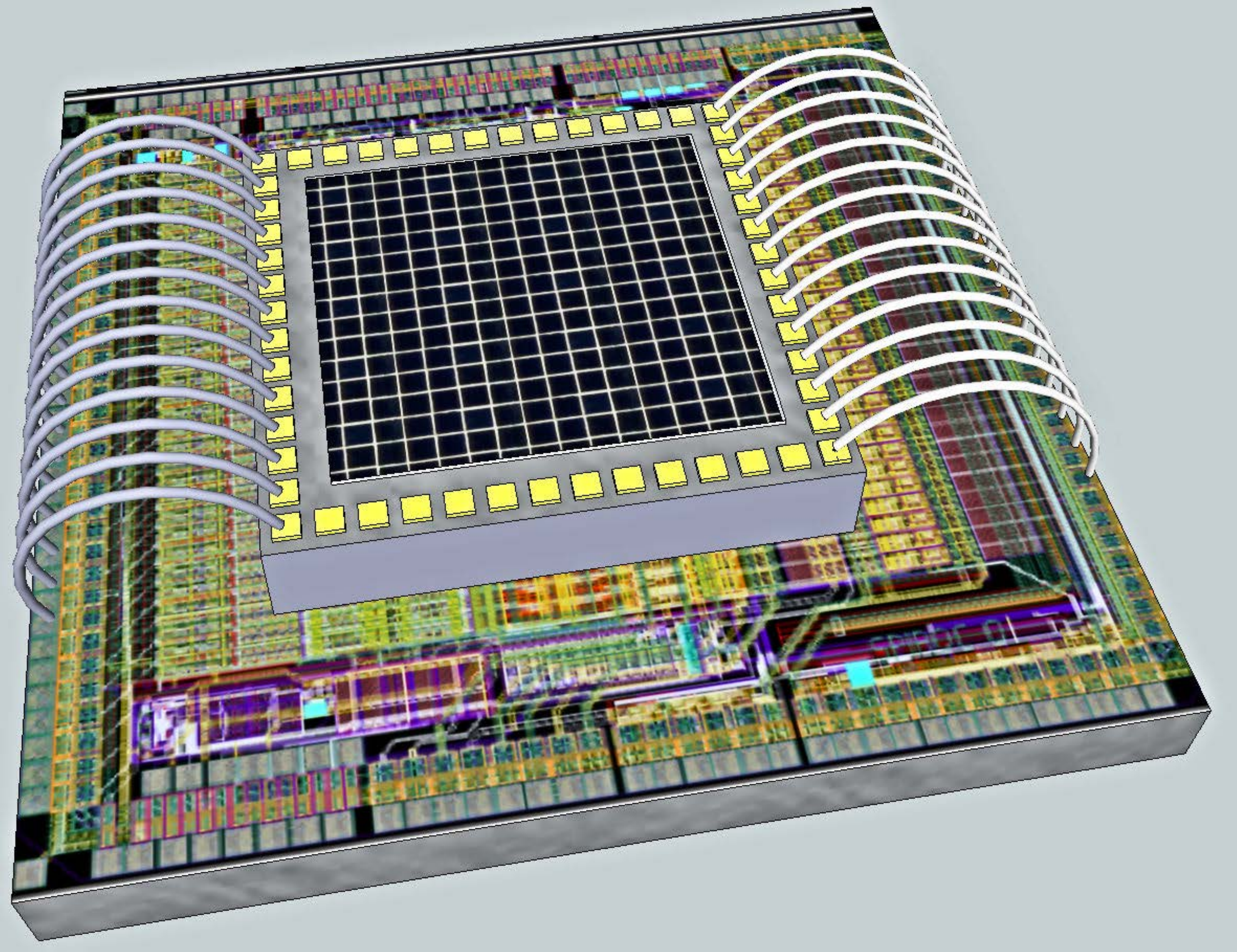}
\caption{A possible configuration for a hybrid approach is shown. The top chip 
is a SiPM, wire bonded to a readout chip on the bottom.}
\label{fig:SiPMHybrid}
\end{figure}

The first approach is to pursue commercially available ``analog" SiPMs coupled to 
commercially available, ``off the shelf" electronics. This approach is often 
referred to as ``COTS". This is the approach taken so far by existing experiments and those 
planned for the near future. This includes T2K, mu2e and CALICE. This has the 
advantage of low technical risk and has a well understood cost. A typical implementation 
of the electronics might be based on commercial AFE (analog front end) chips 
and FPGAs, with Ethernet readout. An example of a preliminary prototype for 
mu2e is shown in Fig.~\ref{fig:32CHOC}.
Another approach would be to adopt existing SiPMs to an existing ASIC designed 
specifically for SiPMs. This is not the same as developing a custom ASIC, as 
these devices already exist for some other experiments. There are many 
similarities between different experiments in high energy physics and the 
popularity and interest in SiPMs is driving development for various applications. 
Some examples of ASICs that have been used (or are being developed for use) with 
SiPMs are the TriP-t (developed at Fermilab for Dzero, now used by T2K for SiPM 
readout), TARGET (developed for Cherenkov Telescope Array) \cite{Bechtol:2011tr} as well 
as the EASIROC, the SPIROC and their derivative chips that were developed by the 
Omega group at IN2P3 in Orsay.
The third approach is to develop a custom solution, using either analog or digital 
SiPMs. This approach could potentially significantly reduce the per channel cost of 
both the photodetector and electronics, but involves higher technical risk and 
requires larger initial investment. This is clearly the best approach for a 
sufficiently large detector system, but more resources would need to be devoted 
to make a specific proposal for a custom SiPM development. One possible approach 
would be to slightly modify an existing SiPM to allow many connections between 
the SiPM and the readout ASIC.  This is essentially a hybrid solution with a ``near digital" 
SiPM, where a few SiPM pixels are wire bonded to an electronics channel. This would 
provide most of the benefits of digital SiPMs, but with a much shorter and simpler 
development effort.  A conceptual design is show in Fig.~\ref{fig:SiPMHybrid}

\clearpage
\section{Near Detectors}
\label{sec:near}
\providecommand{\nm}{\mbox{\boldmath $\nu_\mu$}}
\providecommand{\anm}{\mbox{\boldmath $\bar\nu_\mu$}} 
\providecommand{\nue}{\mbox{\boldmath $\nu_e$}} 
\providecommand{\ane}{\mbox{\boldmath $\bar\nu_e$}} 

\providecommand{\nmne}{\mbox{\boldmath $\nu_{\mu}\rightarrow\nu_e$}} 
\providecommand{\anmne}{\mbox{\boldmath $\bar\nu_{\mu} \rightarrow \bar\nu_e$}} 
\providecommand{\nmnt}{\mbox{\boldmath $\nu_{\mu}\rightarrow\nu_\tau$}} 
\providecommand{\nmnx}{\mbox{\boldmath $\nu_\mu\rightarrow\nu_x$}} 
\providecommand{\nent}{\mbox{\boldmath $\nu_e\rightarrow\nu_\tau$}} 
\providecommand{\nenx}{\mbox{\boldmath $\nu_e\rightarrow\nu_x$}} 

\providecommand{\thtwothree}{\mbox{\boldmath $\Theta_{23}$}}
\providecommand{\dmtwothree}{\mbox{\boldmath $\Delta m^2_{23}$}}
\providecommand{\sthtwothree}{\mbox{\boldmath $\sin^2 2\Theta_{23}$}} 
\providecommand{\wam}{\mbox{\boldmath $\sin^2 \theta_W$}} 

\providecommand{\enu}{\mbox{\boldmath $E_\nu$}}
\providecommand{\em}{\mbox{\boldmath $E_\mu$}}
\providecommand{\eh}{\mbox{\boldmath $E_{Had}$}}
\providecommand{\ne}{\mbox{\boldmath $\nu_e$}}

\begin{figure}[htbp]
  \begin{center}$
    \begin{array}{cc}
      \includegraphics[width=0.49\textwidth]{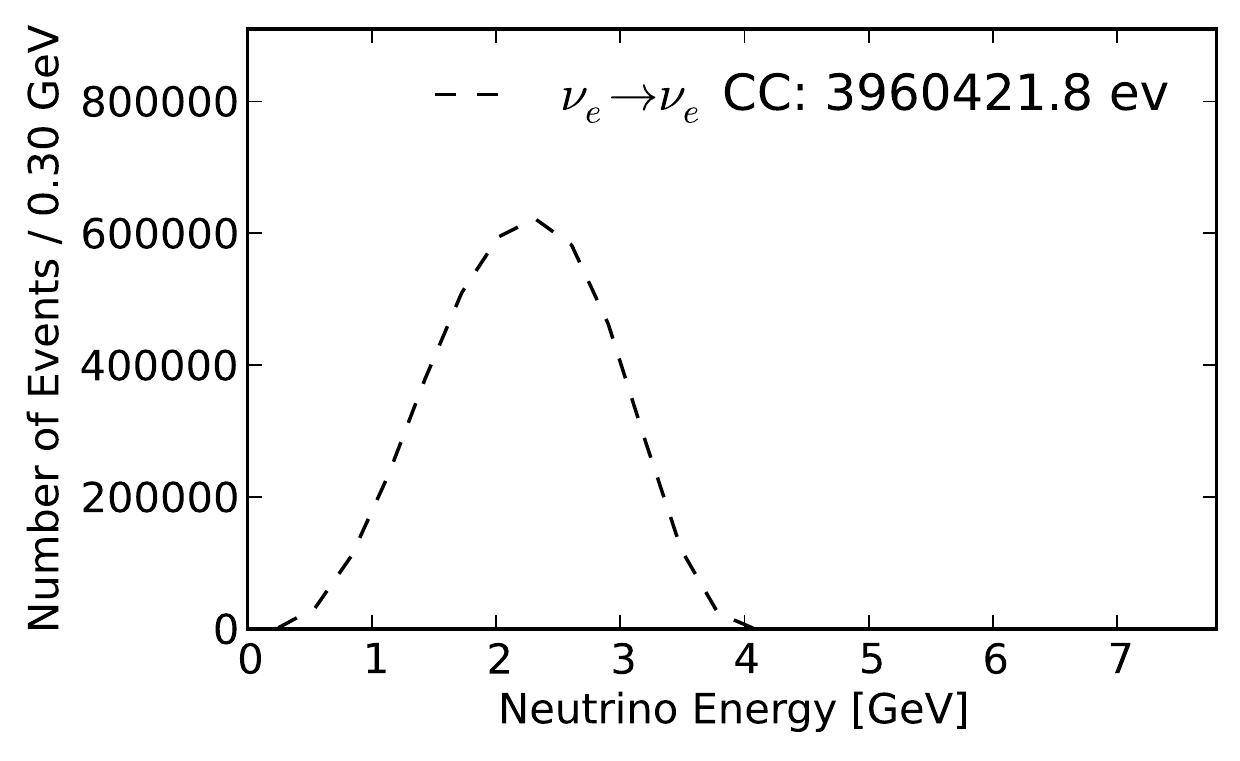}  &
      \includegraphics[width=0.49\textwidth]{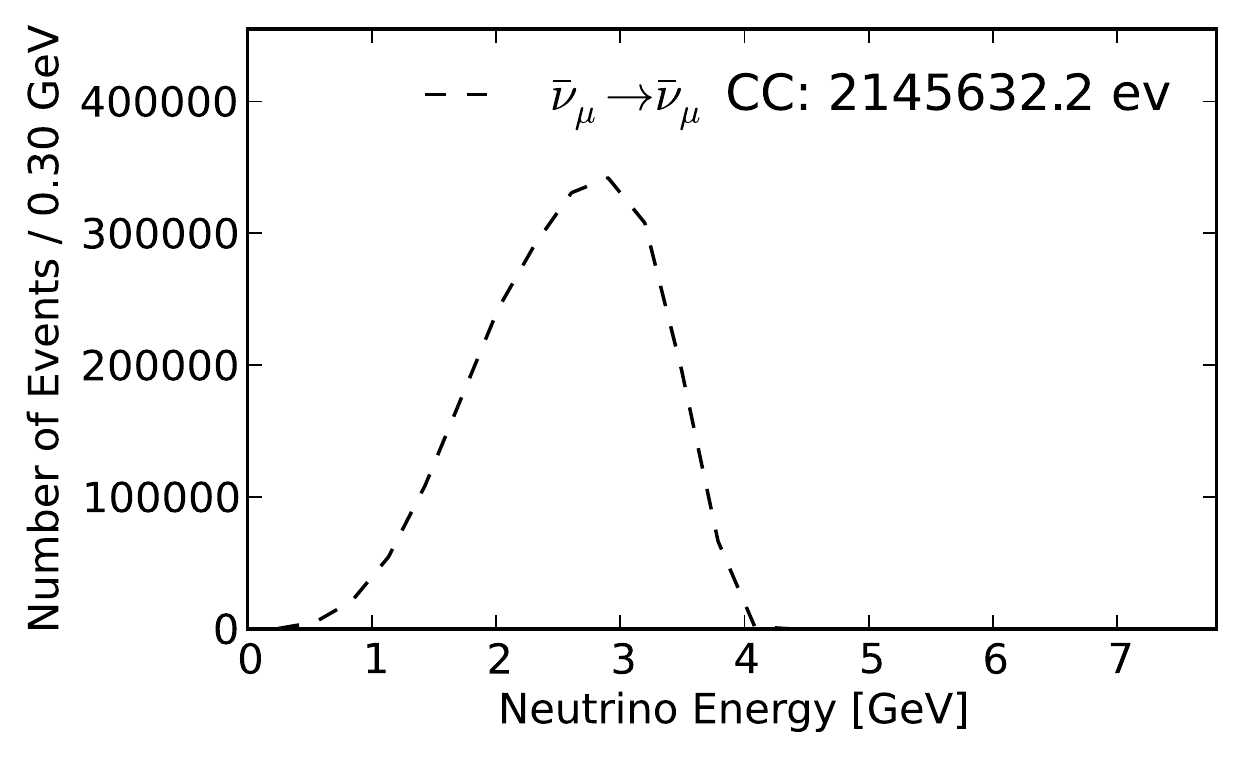} \\
    \end{array}$
  \end{center} 
\caption{$\nu_e$ spectrum at near detector (Left), $\bar{\nu}_\mu$ (Right).}
\label{fig:ND_rates}
\end{figure}

\providecommand{\pip}{\mbox{\boldmath $\pi^+$}} 
\providecommand{\pim}{\mbox{\boldmath $\pi^-$}} 
\providecommand{\piz}{\mbox{\boldmath $\pi^0 $}}
\providecommand{\gam}{\mbox{\boldmath $\gamma$}}
\providecommand{\vz}{\mbox{\boldmath ${\cal V}^0$}}
\providecommand{\mup}{\mbox{\boldmath $\mu^+$}}
\providecommand{\mum}{\mbox{\boldmath $\mu^-$}}
\providecommand{\kap}{\mbox{\boldmath $K^+$}}
\providecommand{\kam}{\mbox{\boldmath $K^-$}}
\providecommand{\kl}{\mbox{\boldmath $K^0_L$}}
\providecommand{\ks}{\mbox{\boldmath $K^0_S$}}
\providecommand{\nova}{\mbox{NO$\nu$A}}
\providecommand{\cohp}{\mbox{\boldmath ${\cal {Coh}} \pi^0$}}
The near detector hall at $\nu$STORM presents opportunities for both oscillation physics and neutrino cross section measurements.  We have assumed that the hall will be located at $\sim$ 50m from the end of the straight.  The neutrino flux at this position has been calculated and the representative number of events (per 100T fiducial mass) for our $10^{21}$ POT exposure is given in Fig.~\ref{fig:ND_rates}, left for $\nu_e$ and right for $\bar{\nu}_\mu$.  
\begin{table}[h]
\begin{minipage}[h]{.48\linewidth}
\begin{tabular}{l|r|}
Channel & $N_\textrm{evts}$ \hspace{5mm}\\
        \hline
$\bar{\nu}_\mu$ NC & 844,793 \\
$\nu_e$ NC & 1,387,698 \\
$\bar{\nu}_\mu$ CC & 2,145,632 \\
$\nu_e$ CC & 3,960,421 \\
\end{tabular}
\caption{Event rates at near detector (for 100T) with $\mu^+$ stored}
\label{tab:Near+}
	 \end{minipage} \hfill
\begin{minipage}[h]{.48\linewidth}
\begin{tabular}{l|r|}
Channel & $N_\textrm{evts}$ \\
        \hline
$\bar{\nu}_e$ NC & 709,576 \\
$\nu_\mu$ NC & 1,584,003 \\
$\bar{\nu}_e$ CC & 1,784,099 \\
$\nu_\mu$ CC & 4,626,480 \\
\end{tabular}
\caption{Event rates at near detector (for 100T) with $\mu^-$ stored}
\label{tab:Near-}
	 \end{minipage} \hfill
\end{table}	 
%
\subsection{For short-baseline oscillation physics}
A near detector is needed for the oscillation disappearance searches and our concept (detailed studies have not yet been done for these channels) is to build a near detector that is identical to SuperBIND, but with approximately 100-200T of fiducial mass.  A muon ``catcher" will most likely be needed in order to maximize the usefulness of the ``as-built" detector mass.  Before a final specification for this near detector can be made, full simulation and analysis for the disappearance channels will have to be done.
\subsection{HIRESMNU: A High-Resolution Detector for $\nu$ interaction studies}
\begin{figure}
\begin{center}
	\includegraphics[width=0.7\textwidth]{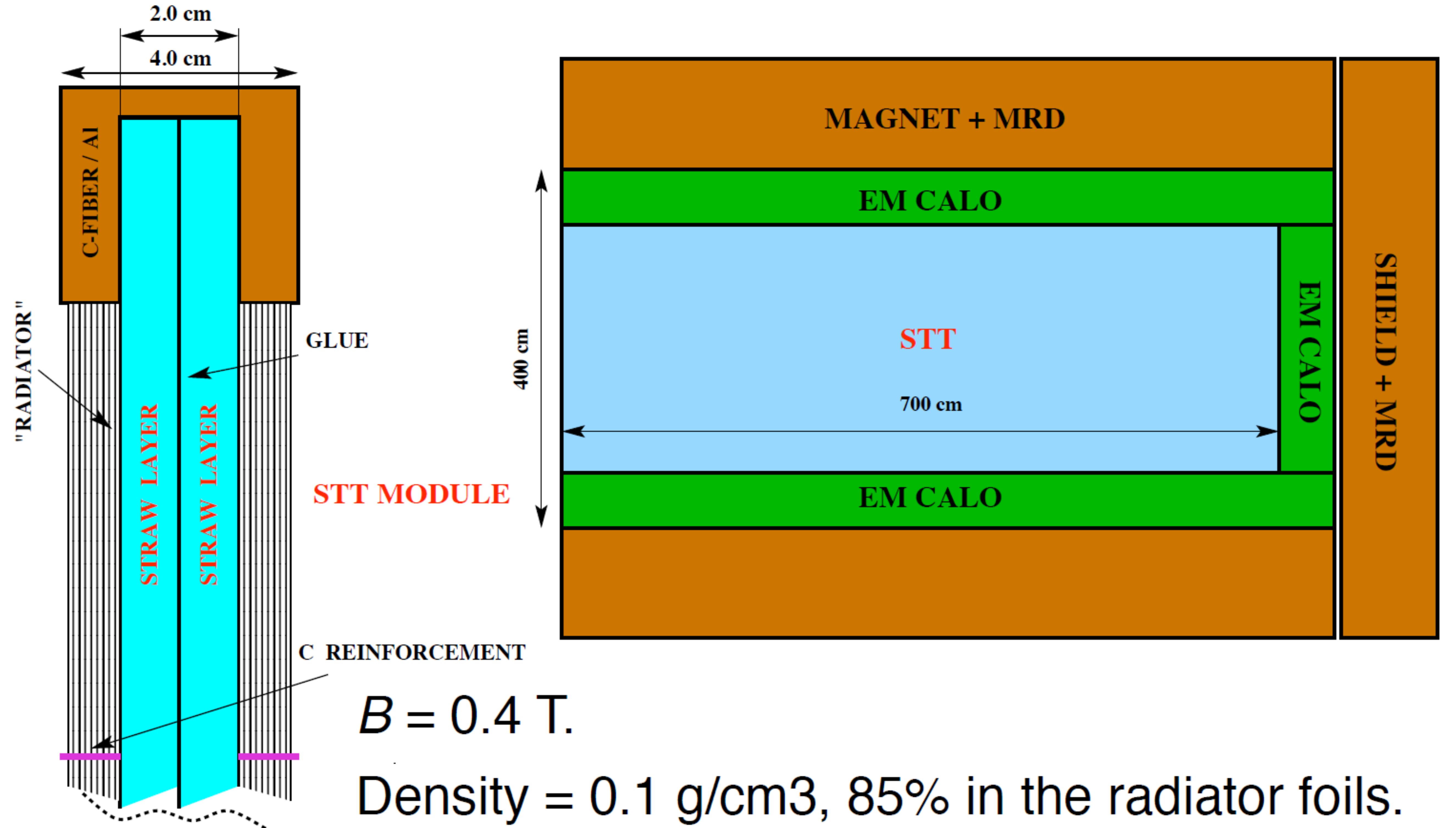}
\caption{Schematic of the ND showing the straw tube tracker (STT), the
electromagnetic calorimeter (ECAL) and the magnet with the muon range
detector (MRD). The STT is based upon ATLAS~\cite{Akesson:2004nj} and 
COMPASS ~\cite{Bychkov:2002xw} trackers. 
Also shown is one module of the proposed straw tube tracker (STT).
Interleaved with the straw tube layers are plastic foil radiators, which
provide 85\% of the mass of the STT.
At the upstream end of the STT are layers of nuclear-target 
for the measurement of cross sections and 
the $\pi^{\circ}$'s on these materials.
}
\label{fig-det-schematic}
\end{center}
\end{figure}

Precision measurements of neutrino-interactions at the 
near-detector (ND) are necessary to ensure 
the highest possible sensitivity for  neutrino oscillation studies (both for $\nu$STORM and
for any future long-baseline neutrino oscillation experiment).
Regardless of the process under study ---  
$\nu_\mu \rightarrow \nu_e$ appearance or $\bar{\nu}_\mu \rightarrow \bar{\nu}_e$ disappearance ---
the systematic error should be less than the corresponding statistical error. 
A near detector concept which will well suit this purpose is 
the high resolution detector, HIRESMNU,  
proposed for the LBNE project ~\cite{Mishra:2010zz}. 
It can fulfill four principal goals:

\begin{enumerate}
	\item Measurement of the absolute and the relative abundance
	  of the  {\bf four} species of neutrinos,
	  $\nu_\mu, \bar{\nu}_\mu, \nu_e, \bar{\nu}_e$
	  as a function of energy (E$_\nu$). 
	  Accurate determination of  the angle and the momentum of the electron 
	  in neutrino-electron neutral current interaction will provide 
	 the  absolute flux.
	\item Determination of the absolute E$_\nu$-scale,  a factor 
	  which determines value of the oscillation-parameter $\Delta m^2$. 
	\item Determination of  $\pi^{\circ}$'s and $\pi^+/\pi^-$'s 
	  produced in the NC and CC interactions. 
	  The pions are the predominant source of background for any 
	  oscillation study. 
	\item Measurement of $\nu$-Nucleus cross-section where 
	  the nuclear target will be that of the far-detector. The cross-section 
	  measurements of exclusive and inclusive CC and NC processes 
	  will furnish a rich panoply of physics relevant for most neutrino research. 
	  Knowing the cross sections at the E$_\nu$ typical of the $\nu$STORM beam is
	  essential for predicting both the signal and the background.
\end{enumerate}
\noindent
Figure~\ref{fig-det-schematic} shows a schematic of this the HIRESMNU design.
The architecture~\cite{Mishra:2010zz} 
derives from the experience of NOMAD ~\cite{Altegoer:1997gv}.
It embeds a $4 \times 4 \times 7$~m$^3$ STT and a surrounding 
4$\pi$ electromagnetic calorimeter (ECAL) in a 
dipole magnet with $B \simeq 0.4$~T.
Downstream of the magnet and additionally within the magnet yoke are
detectors for muon identification.
The STT will have a low average density similar to liquid hydrogen,
about 0.1~gm/cm$^3$, which is essential for the momentum determination 
and ID of electrons, protons, and pions.
The foil layers interleaved with the straw tubes contribute most of the
7~ton fiducial mass.
The foil layers serve both as the mass on which the neutrinos will
interact and as generators of transition radiation (TR), which aids in electron identification.
Its depth in radiation lengths is sufficient for 50\% of the photons from $\pi^{\circ}$ decay to be
observed as $e^+e^-$ pairs, which delivers superior resolution compared
with conversions in the ECAL.
Layers of nuclear-targets will
be deployed at the upstream end of the STT for the determination of 
cross sections on these materials. 
The HIRESMNU delivers the most sensitive systematic constraints 
as studied within the context of future long-baseline $\nu$ experiments. The systematic studies include 
$\nu$-electron scattering, quasi-elastic interactions, $\nu_e/\bar{\nu}_e$-CC, 
neutral-current identification, $\pi^{\circ}$ detection, etc. 
The quoted dimensions, mass, and segmentation of HIRESMNU will be 
further optimized  for $\nu$STORM as the proposal evolves.

\section{Performance}
\label{sec:performance}
\subsection{Event rates}
The number of muon decays ($N_\mu$) for $\nu$STORM can be defined in terms of the following:
\begin{equation}
N_\mu = (\text{POT}) \times (\text{$\pi$ per POT}) \times \epsilon_\text{col} \times \epsilon_\text{trans}  \times 
\epsilon_\text{inj}  \times (\text{$\mu$ per $\pi$}) \times A_\text{dyn} \times \Omega
\end{equation}
\noindent
where $(\text{POT})$ is the number of protons on target,  $\epsilon_\text{col}$ is the collection efficiency,
$\epsilon_\text{trans}$ is the transport efficiency, $\epsilon_\text{inj}$ is the injection efficiency,
$(\text{$\mu$ per $\pi$})$ is the chance that an injected pion results in a muon within the ring acceptance,
$A_\text{dyn}$ is the probability that a muon within the decay ring aperture is within the dynamic aperture, and
$\Omega$ is the fraction of the ring circumference that directs muons at the far detector.
$\nu$STORM assumes $10^{21}$ POT for a 4-5 year run using 60 GeV protons.  From section ~\ref{sec:TnC}, we obtain
(with horn collection) $\simeq$ 0.1$\pi$/pot$\times$collection efficiency.  We have assumed that the transport efficiency,
and the injection efficiency are 0.8 and 0.9, respectively and that the probability that a $\pi$ decay results in a $\mu$ within the acceptance $\times \gamma$c$\tau$ is 0.08.  $\Omega$ is 0.34.  This results in approximately 2 $\times10^{18}$ useful $\mu$ decays.
With a 1kT fiducial mass far detector located at approximately 2 km from the end of the decay ring straight, we have the 
following raw event rates:
\newpage
\begin{center}
Neutrino mode with stored $\mu^+$.
\end{center}
\vspace{-7mm}
\begin{table}[h]
	\centering
\begin{tabular}{c|r|r|r|r}
Channel & $N_\textrm{osc.}$ & $N_\textrm{null}$ & Diff. & $(N_\textrm{osc.} - N_\textrm{null})/\sqrt{N_\textrm{null}}$ \\
        \hline
$\nu_e \to \nu_\mu$ CC & 332 & 0 & $\infty$ & $\infty$ \\ 
$\bar{\nu}_\mu \to \bar{\nu}_\mu$ NC & 47679 & 50073 & -4.8\% & -10.7 \\ 
$\nu_e \to \nu_e$ NC & 73941 & 78805 & -6.2\% & -17.3 \\ 
$\bar{\nu}_\mu \to \bar{\nu}_\mu$ CC & 122322 & 128433 & -4.8\% & -17.1 \\ 
$\nu_e \to \nu_e$ CC & 216657 & 230766 & -6.1\% & -29.4 \\ 
\end{tabular}
\begin{center}
Anti-neutrino mode with stored $\mu^-$.
\end{center}
\begin{tabular}{c|r|r|r|r}
Channel & $N_\textrm{osc.}$ & $N_\textrm{null}$ & Diff. & $(N_\textrm{osc.} - N_\textrm{null})/\sqrt{N_\textrm{null}}$ \\
        \hline
$\bar{\nu}_e \to \bar{\nu}_\mu$ CC & 117 & 0 & $\infty$ & $\infty$ \\ 
$\bar{\nu}_e \to \bar{\nu}_e$ NC & 30511 & 32481 & -6.1\% & -10.9 \\ 
$\nu_\mu \to \nu_\mu$ NC & 66037 & 69420 & -4.9\% & -12.8 \\ 
$\bar{\nu}_e \to \bar{\nu}_e$ CC & 77600 & 82589 & -6.0\% & -17.4 \\ 
$\nu_\mu \to \nu_\mu$ CC & 197284 & 207274 & -4.8\% & -21.9 \\ 
\end{tabular}
	\caption{\label{tab:raw_evt}Raw event rates for $10^{21}$ POT (for stored $\mu^+$ and stored $\mu^-$) for best-fit values for the LSND anomaly figure-of-merit.}
\end{table}
In addition to the  $\mu$ decay beam, we also have a high-intensity
$\pi$ decay neutrino beam, $\parenbar{\nu}_\mu$, from the straight section (at injection into the ring) which can easily be time separated from the $\mu$ decay beam.  This 
$\parenbar{\nu}_\mu$ is roughly the same intensity as the integrated $\parenbar{\nu}_\mu$ beam from the stored $\mu$ decays.
\subsection{Monte Carlo and analysis}
\label{sec:MCana}
\subsubsection{Neutrino event generation and detector simulation}
\label{Sec:event_generation}
\begin{figure}[htbp]
  \begin{center}$
    \begin{array}{c}
      \includegraphics[width=0.7\textwidth]{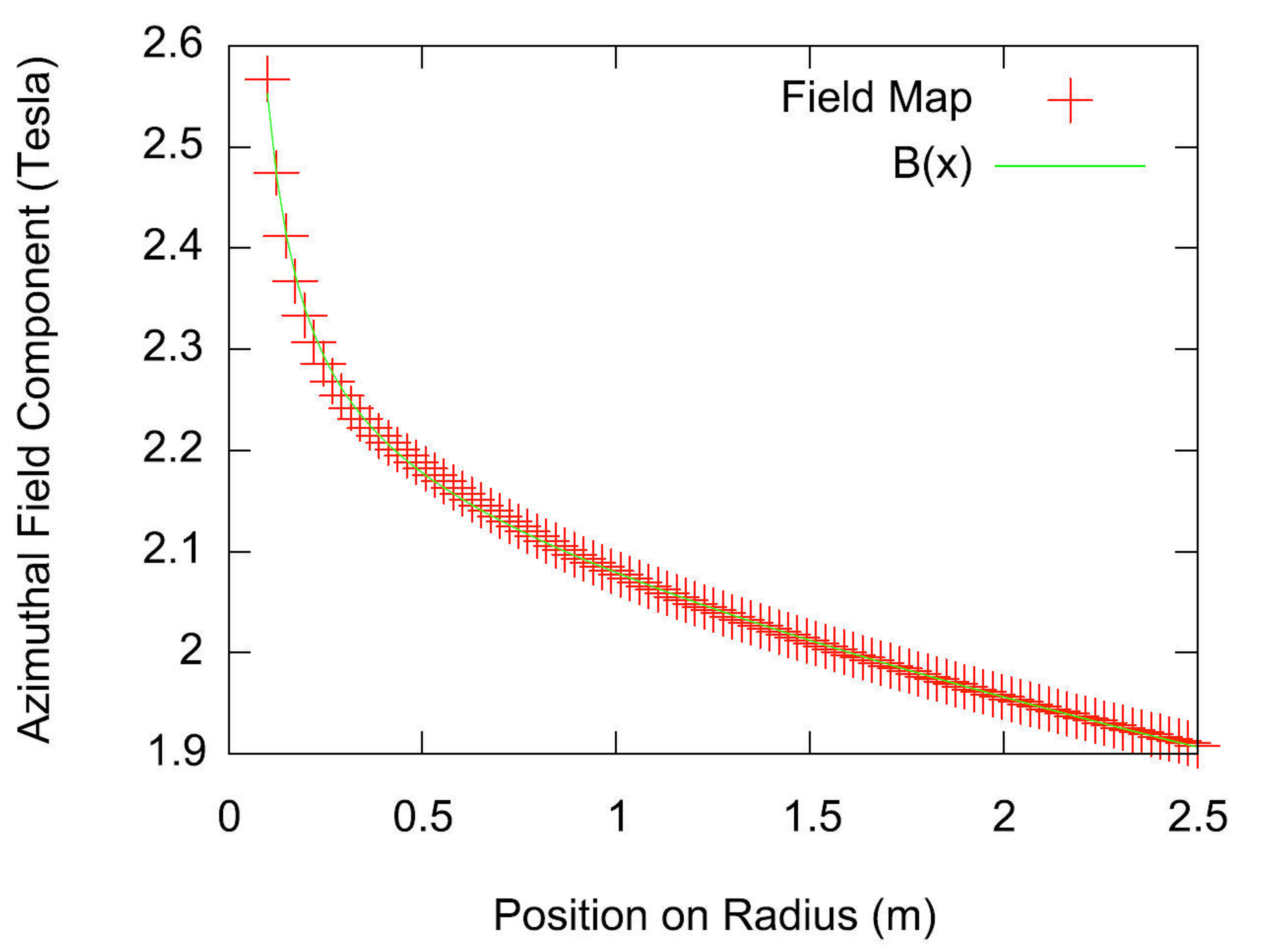} 
    \end{array}$
  \end{center}
  \caption{Radial parameterisation of the toroidal magnetic field in SuperBIND along the 45$^\circ$ azimuth direction.}
  \label{fig:Bparam}
\end{figure}

The Monte Carlo and analysis for the SuperBIND detector is closely based
on the simulations and analysis of the MIND detector for the Interim
Design Report of the International Design Study for a Neutrino factory (IDS-NF) \cite{NF:2011aa}. Generation for all types of interactions was performed using the GENIE
framework~\cite{Andreopoulos:2009rq}. 
The simulation of the generated events was carried out using the GEANT4
toolkit~\cite{Apostolakis:2007zz} (version 4.9.4), with full
hadron shower development and digitization of the events. 
The simulated detector was the SuperBIND detector described in section~\ref{sec:far},
cylindrical in shape with a 5~m diameter and 20~m in length.  Each of the individual modules were composed of alternating 1~cm thick iron plates and 2~cm planes of polystyrene
extruded plastic scintillator in two views (one along the $x$ axis and the other along the $y$ axis).  Simulations with 2~cm iron plates were also carried 
out in order to optimize the geometric configuration. A toroidal magnetic field  is simulated inside the iron. The amplitude of the field is parameterized 
as a function of radius $r$ according to the following:
\begin{equation}
B(r) = B_0+\frac{B_1}{r} + B_2 e^{-Hr},
\end{equation}
with $B_0=1.53$~T, $B_1 = 0.032$~T$\cdot$m, $B_2= 0.64$~T and $H=0.28$~m$^{-1}$. The field and its parametrization along the 
45$^\circ$ azimuth direction are shown in Fig.~\ref{fig:Bparam}.

Events generated for iron and scintillator nuclei are selected according to their relative
weights in the detector and the resultant particles are tracked from a
vertex randomly positioned in three dimensions within a randomly
selected piece of the appropriate material. Physics processes are
modeled using the QGSP\_BERT physics lists provided by GEANT4
\cite{geant4phys}. Secondary particles are required to travel at least 30~mm from their
production point or to cross a material boundary between the detector
sub-volumes to have their trajectory fully tracked. Generally,
particles are only tracked down to a kinetic energy of
100~MeV. However, gammas and muons are excluded from this cut. 

A simplified digitisation model was considered for this
simulation. Two-dimensional boxes with 1~cm edge length -- termed voxels -- represent
view-matched $x$ and $y$ scintillator readout positions. 
The response of the scintillator bars is
derived from the raw energy deposited in each voxel, read out using
wavelength shifting (WLS) fibers with an attenuation length $\lambda =
5$~m, as reported by the Miner$\nu$a
collaboration~\cite{PlaDalmau:2005dp}. Assuming that approximately
half of the energy will come from each view, the deposit is halved and
the remaining energy at each edge in $x$ and $y$ is calculated. This
energy is then smeared according to a Gaussian width $\sigma/E = 6\%$
to represent the response of the electronics and then recombined into
$E_x$, $E_y$ and total energy = $E_x+E_y$ energy deposited per voxel. An output
wavelength of 525~nm, a photo-detector quantum efficiency of
$\sim$30\% and a threshold of 4.7 photo electrons (pe) per view (as in
MINOS~\cite{Michael:2008bc}) were assumed. Any voxel view that is not above the threshold is cut. 

\begin{figure}[htbp]
  \begin{center}
  \includegraphics[width=0.8\textwidth]{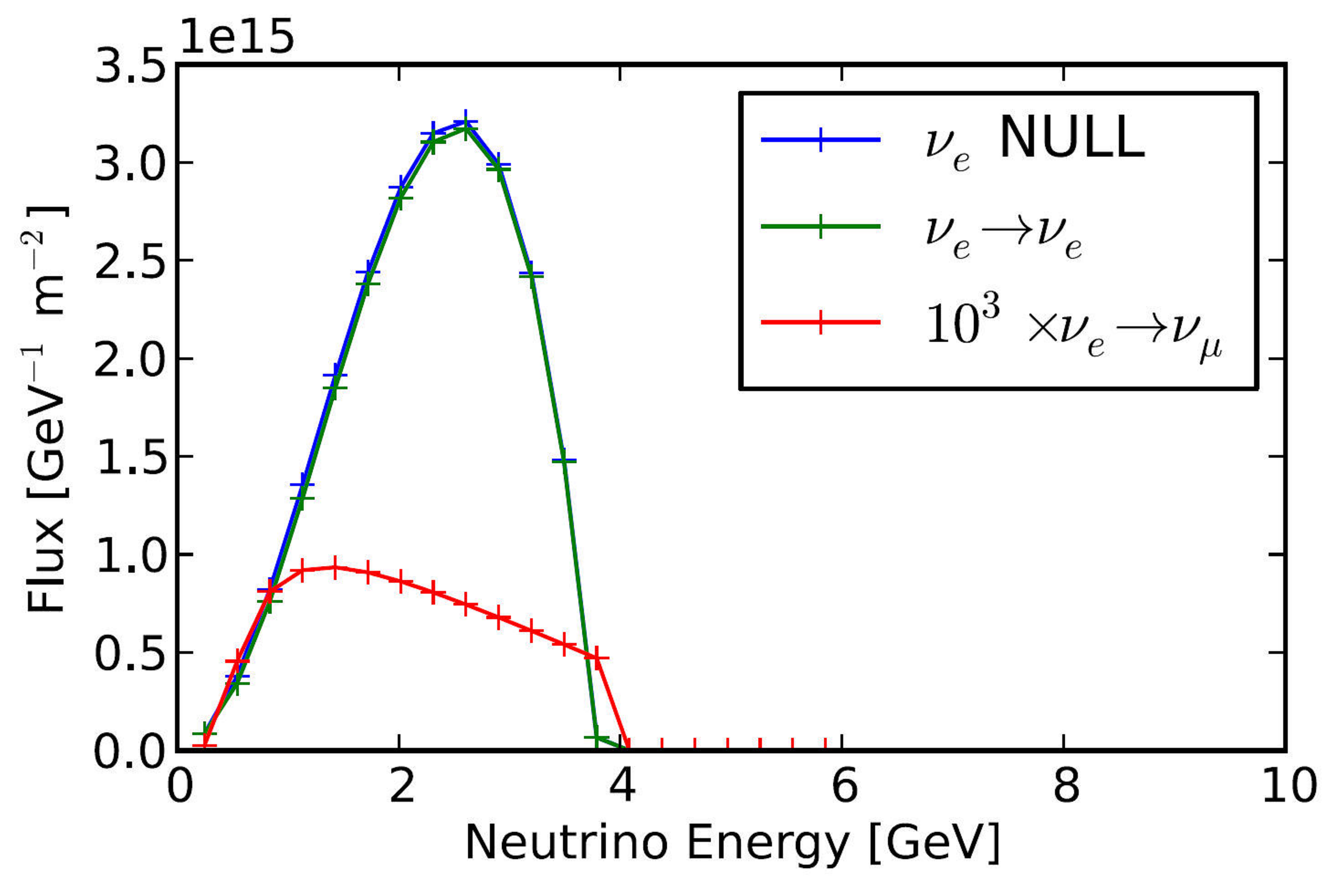}
  \end{center}
  \caption{Neutrino fluxes used to carry out the simulations of the SuperBIND detector.}
  \label{fig:nuflux}
\end{figure}

The simulation was run assuming that the storage ring contains 3.8 GeV/c $\mu^+$, so that the wrong sign muon signal consists of $\mu^-$ tracks from $\nu_{\mu}$ charged current (CC) interactions. The backgrounds consist of mis-identified $\mu^+$ tracks, and tracks constructed from showers generated by $\bar{\nu}_{\mu}$ neutral current (NC) and $\nu_{e}$ CC events. 
The neutrino fluxes were provided as oscillated $\nu_{\mu}$ and un-oscillated $\bar{\nu}_{\mu}$ and $\nu_{e}$ spectra. The exclusive event spectra generated
by GENIE are shown in Fig.~\ref{fig:nuflux}. The appropriate flux spectrum was input into the GENIE simulation to provide the samples of neutrino interaction events passed to the GEANT4 simulation. 

\subsubsection{Event reconstruction}
\label{Sec:RecG4}
The reconstruction package was described in detail in
\cite{Cervera:2010rz} and in the Interim
Design Report of the IDS-NF \cite{NF:2011aa}.
The first stage of the reconstruction includes a clustering algorithm \cite{andrewsthesis}.
The clusters formed from the hit voxels of an event are then passed to the
reconstruction algorithm.
The separation of candidate muons from hadronic activity is achieved
using two methods: a Kalman filter algorithm provided by
RecPack~\cite{CerveraVillanueva:2004kt} and a cellular automaton
method (based on \cite{Emeliyanov_otr/itr-cats:tracking}), both
algorithms are described in detail in~\cite{Cervera:2010rz}. 

Fitting of the muon candidates proceeds using a Kalman filter to fit a
helix to the candidate, using an initial seed estimated from the path length
of the muon track, using the Continuous Slowing Down Approximation (CSDA
\cite{Groom2001183}), and then refitting any successes. 
Neutrino energy is generally reconstructed as the sum of the muon and
hadronic energies, with hadronic reconstruction currently performed
using a smear on the true quantities as described in
reference \cite{Cervera:2010rz}.  The reconstruction of the hadronic energy
$E_{had}$ (in GeV) assumes a resolution $\delta E_{had}$ from the MINOS CalDet
testbeam~\cite{Michael:2008bc,Adamson:2006xv} (although we believe that SuperBIND will do better):
\begin{equation}
  \label{CalDet1}
  \frac{\delta E_{had}}{E_{had}} = \frac{0.55}{\sqrt{E_{had}}} \oplus 0.03.
\end{equation}
The hadronic shower direction vector is also smeared according to the
angular resolution found by the Monolith test-beam~\cite{Bari:2003bt}:
\begin{equation}
  \label{eq:MONang}
  \delta \theta_{had} = \frac{10.4^\circ}{\sqrt{E_{had}}} \oplus \frac{10.1^\circ}{E_{had}}.
\end{equation}

In the case of QE interactions, where there is no hadronic jet, the
neutrino energy reconstruction was carried out using the formula:
\begin{equation}
  \label{eq:quasiEng}
  E_\nu = \displaystyle\frac{m_NE_\mu + \frac{1}{2}\left(m_{N'}^2 - m_\mu^2 - m_N^2\right)}{m_N - E_\mu + |p_\mu|\cos\vartheta} \, ;
\end{equation}
where $\vartheta$ is the angle between the muon momentum vector and
the beam direction, $m_N$ is the mass of the initial state nucleon,
and $m_{N'}$ is the mass of the outgoing nucleon for the interactions
$\nu_\mu + n \rightarrow \mu^- + p$ and $\overline{\nu}_\mu + p
\rightarrow \mu^+ + n$ (see for example \cite{Blondel:2004cx}). 

The iron plate thickness of the detector was studied from the point of view of muon charge identification efficiency. The charge selection efficiency was studied using 1~cm iron plates and  2~cm iron plates. By doubling the thickness of the iron plate, we effectively increase the effective magnetic field between measurements by 50\% so the net charge selection efficiency increases, at the expense of a small increase in the threshold. Both of these effects were studied in detail in the following data analysis section.

\subsection{Data Analysis}
\label{sec:analysis}
The basis for the $\nu_e\rightarrow \nu_\mu$ analysis closely follows the one 
for the MIND detector at a Neutrino Factory \cite{NF:2011aa}, but was adapted for the lower
muon energy of 3.8~GeV. The cuts to reject charged current (CC) and neutral current (NC) backgrounds are organized as follows:
\begin{itemize}
\item Successful reconstruction.\vspace{-3mm}
\item Fiducial volume cut.\vspace{-3mm}
\item Maximum momentum cut.\vspace{-3mm}
\item Fitted proportion of hits allocated to the muon track.\vspace{-3mm}
\item Track quality cuts.\vspace{-3mm}
\item Neutral current rejection cut.\vspace{-3mm}
\end{itemize}

We commence by imposing the reconstruction criteria
from the previous section to guarantee fully reconstructed neutrino
events. We then proceed to impose a fiducial cut, requiring that the
first cluster in a candidate be at least 1~m from the end of the
detector ($z\leq19000$~mm for a 20~m long detector). The isolated clusters that form a muon track candidate are fitted to determine the muon momentum. A maximum value for the reconstructed muon momentum is imposed at 6.1~GeV
 (60\% above the maximum muon momentum) to remove backgrounds caused by 
poorly reconstructed momenta.
Any remaining clusters are assumed to be part of the hadronic component of the event. 
Charged current events have a larger proportion of hits allocated to the muon candidate. 
We only accept those events in which more than
60\% of its clusters are fitted as a muon candidate, to reduce neutral current and electron neutrino background levels.

The track quality cut is based on the
relative error in the inverse momentum of the candidate muon
$\frac{\sigma_{q/p}}{q/p}$, where $q$ is the charge of the muon and
$p$ its momentum. 
A Probability Density Function (PDF) $P\left(\sigma_{q/p}/(q/p)\right)$ is created
for both CC signal and NC background. The log-likelihood ratio $\mathcal{L}_{q/p}$ 
between the two distributions is created. The signal events are
selected as those with a log-likelihood parameter
$\mathcal{L}_{\sigma/p} > -0.5$.

The final cut involves the rejection of neutral current backgrounds, by
exploiting the property that $\nu_\mu$ CC events tend to have greater length than NC
events. Hence, the number of hits, $N_{hit}$, was used to generate Probability Density
Functions (PDF) for charged and neutral current events. The log-likelihood ratio rejection parameter:
\begin{equation}
  \mathcal{L}_1 = \log \left( \frac{P(N_{hit} | CC)}{P(N_{hit} | NC)}
  \right) \, ;
\end{equation}
is used for NC rejection. For the detector geometry with 1 cm thick plates, the chosen cut $\mathcal{L}_{1} >6.5$ allows the background to be rejected to a level below $10^{-3}$. For the case in which we have 2 cm thick plates, the intrinsic NC background is smaller.  This analysis is similar
but simpler than the MIND analysis for a Neutrino Factory \cite{NF:2011aa}. The cuts are summarized in Table~\ref{desc}.

\begin{table}
       \caption{Description of cuts used in the selection of good events from the simulation.}
	\begin{tabular}{|r|l|}
	\hline\hline
	Event Cut & Description \\
	\hline
	Successful Reconstruction & Failed Kalman reconstruction of event removed \\
	Fiducial   				  & First hit of event is more than 1~m from end of detector \\
	Maximum Momentum 	  & Muon momentum less than 1.6$\times E_{\nu}$\\
	Fitted Proportion 		  & 60\% of track nodes used in final fit.\\
	Track Quality 			  & $\log(P(\sigma_{q/p}/(q/p) | CC) / P(\sigma_{q/p}/(q/p)|NC)) > -0.5$\\
	NC Rejection (1 cm plates) & $\log(P(N_{hit} | CC) / P(N_{hit}|NC)) > 6.5$\\
	\hline\hline
	\end{tabular}
	\label{desc}
\end{table}

\begin{figure}[htbp]
  \begin{center}$
    \begin{array}{cc}
      \includegraphics[width=0.49\textwidth]{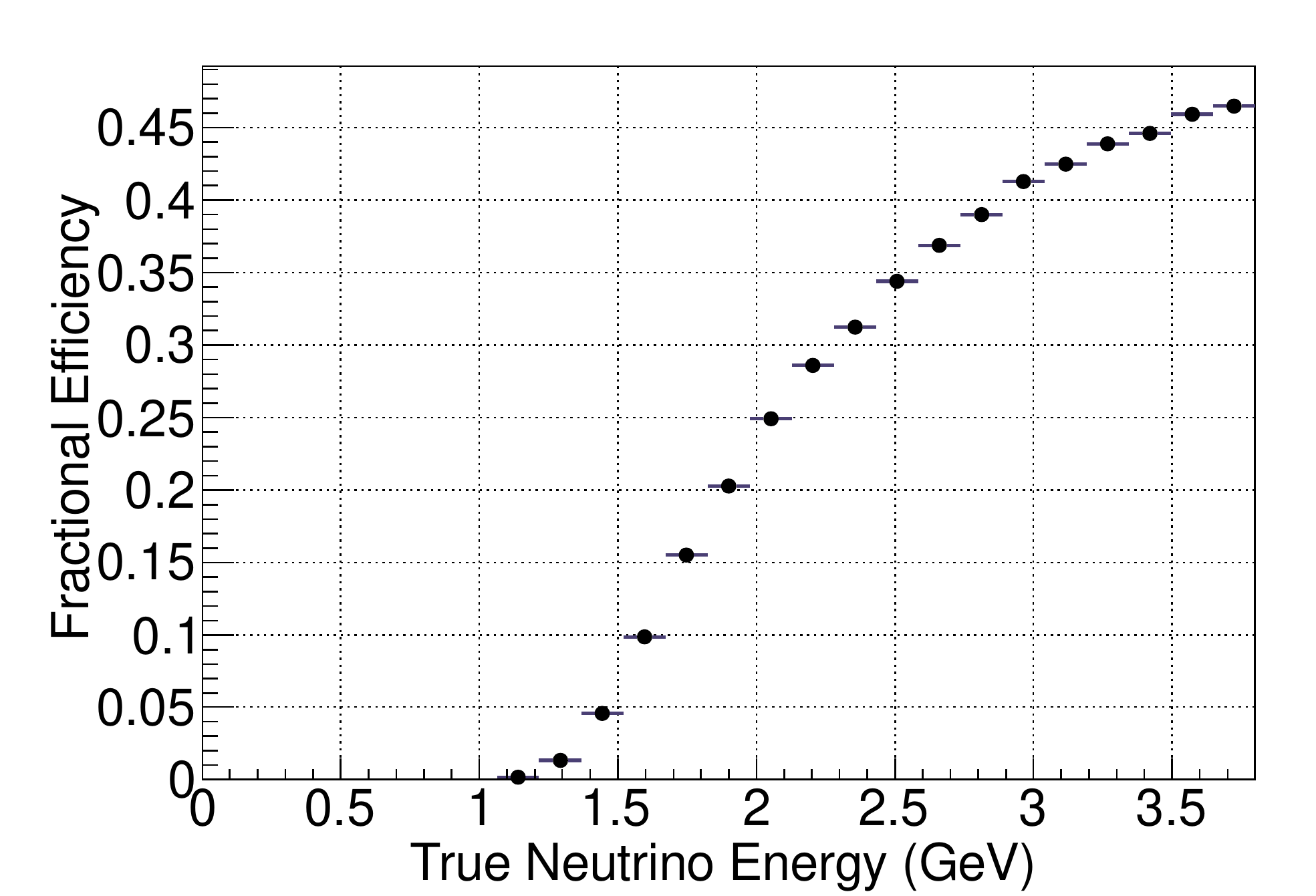}  &
      \includegraphics[width=0.49\textwidth]{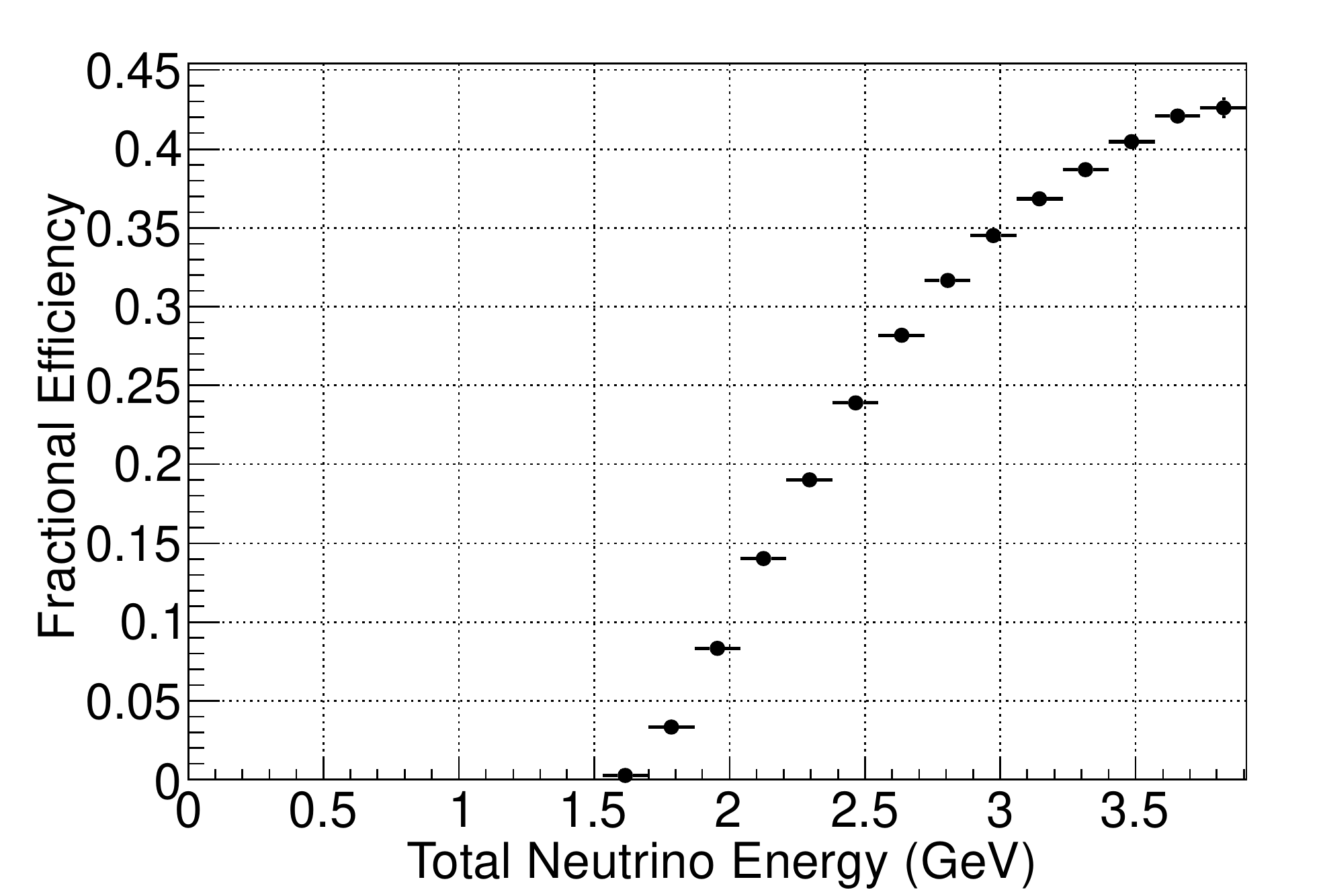} \\
    \end{array}$
  \end{center} 
\caption{Efficiency of detection of a $\mu^-$ signal for a sample of $\nu_{\mu}$ Charge Current interactions 
stopping in a SuperBIND detector with 1~cm iron plates (left) and 2~cm iron plates (right).}
\label{eff}
\end{figure}

\begin{figure}[htbp]
\begin{center}$
    \begin{array}{cc}
      \includegraphics[width=0.49\textwidth]{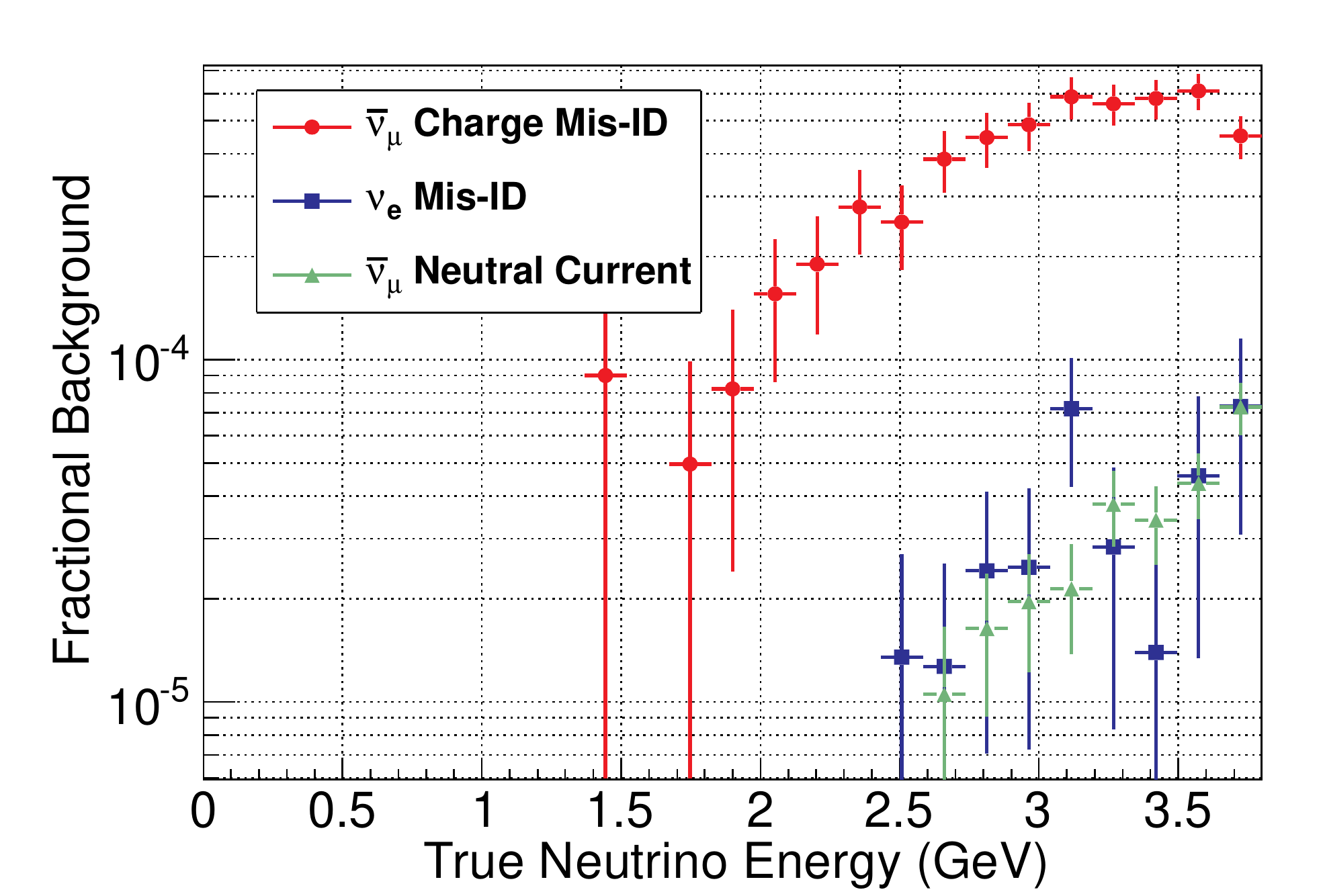}  &
      \includegraphics[width=0.49\textwidth]{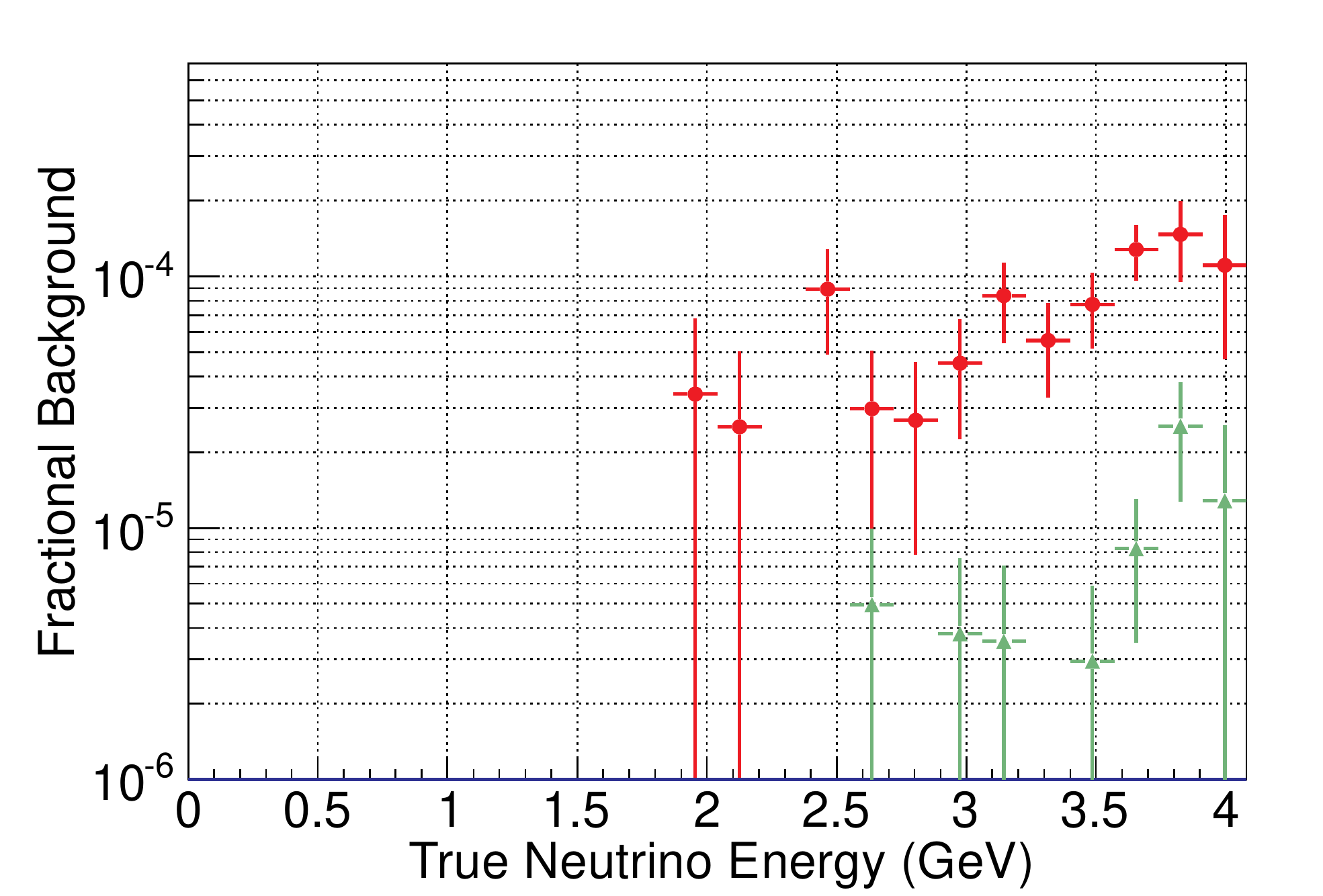} \\
    \end{array}$
  \end{center} 
\caption{Backgrounds for the detection of a $\mu^-$ signal in a SuperBIND detector with 1~cm iron plates
(left) and 2~cm iron plates (right) that will be present when $\mu^+$ are contained in the $\nu$STORM storage ring.}
\label{back}
\end{figure}

The effect of the selection criteria on the signal and background simulations is shown in Table~\ref{cuts}. Figure~\ref{eff} shows the fractional efficiency as a function of neutrino energy after these cuts are applied, for the 1~cm iron plate geometry (left) and 2~cm plate geometry (right). Figure \ref{back} shows the fractional backgrounds for the 1~cm plate (left) and 2~cm plates (right). In  summary, the cuts described in this section lead to an absolute efficiency of 33\% for $\nu_\mu$ CC selection, while reducing the total background to a level of $5\times 10^{-4}$ for the 1~cm plates, while the $\nu_\mu$ CC selection efficiency is 25\%  for 2~cm iron plates, with a total background level less than $7\times 10^{-5}$. This analysis would suggest that 2~cm plates are preferred for the neutrino oscillation $\nu_\mu$ appearance channel. However, this would need to be compared to the $\nu_e$ disappearance channel to determine which of the two geometries would be preferred, so further detector optimizations are needed to be able to make a decision on the optimal geometry.
 \begin{table}
        \caption{ Fraction of events remaining after cuts are applied to simulations of the indicated species in the nominal SuperBIND detector using 1~cm plates when the appearance of a $\mu^-$ in an event is defined as the experimental signal. The final line shows the final event fractions for a detector with 2 cm plates.}
        \begin{tabular}{|r|c|c|c|c|}
        \hline\hline
        & \multicolumn{4}{c|}{Interaction Type and Species} \\
        \cline{2-5}
        Event Cut & $\nu_{\mu}$ CC(\%) & $\bar{\nu}_{\mu}$ CC ($\times 10^3$) & $\nu_{e}$ CC ($\times 10^3$)& $\bar{\nu}_{\mu}$ NC ($\times 10^3$)\\
        \hline
        Successful Reconstruction & 71.9\% & 38.9 &    306 & 99.0 \\
        Fiducial                                   & 69.4\% & 31.0 &    292 & 94.8 \\
        Maximum Momentum           & 68.1\% & 24.2 &    253 & 80.5 \\
        Fitted Proportion                  & 67.3\% & 22.5 &    245 & 75.5 \\
        Track Quality                      & 59.6\% &   7.4 &   42.8 & 18.7 \\
        NC Rejection (1 cm plates)                         & 33.3\% & 0.45 & 0.02 & 0.03 \\
        NC Rejection (2 cm plates)                         & 25.2\% & 0.065 & 0.0 & 0.004 \\
        \hline\hline
        \end{tabular}
        \label{cuts}
\end{table}

%
%
\subsection{Sensitivities}
From Table~\ref{tab:raw_evt} we see that
there are numerous channels in which new physics can be explored.  The statistical significance of NC disappearance is $20\sigma$ and $16 \sigma$ for stored $\mu^+$ and $\mu^-$, respectively, if we combine the 
$\nu_e$ and $\nu_\mu$ NC events together.
Appearance physics via the channel $\nu_e \to \nu_\mu$ gives $\nu$STORM broad sensitivity to sterile physics and directly tests the LSND/MiniBooNE anomaly.  The oscillation probabilities for both appearance and disappearance physics are:
\begin{eqnarray}
\label{eq:prob}\text{P}_{\nu_e \to \nu_\mu} =& 4 | U_{e 4}|^2 |U_{\mu 4}|^2 \sin^2 \left(\frac{\Delta m^2_{41} L}{4 E}\right),\\
\text{P}_{\nu_\alpha \to \nu_\alpha} =& 1 - \left[4 |U_{\alpha 4}|^2 (1 - |U_{\alpha 4}|^2)\right] \sin^2 \left(\frac{\Delta m^2_{41} L}{4 E}\right).
\end{eqnarray}
\subsubsection{Appearance channels}
The appearance signal which the detector is designed for is $\nu_e \to \nu_\mu$;  the CPT conjugate of the LSND anomaly $\bar{\nu}_\mu \to \bar{\nu}_e$.  For nonzero appearance probability in sterile neutrino models, there must simultaneously be both $\nu_e$ and $\nu_\mu$ disappearance since $| U_{e 4}| |U_{\mu 4}| \neq 0$, which allows for testing if the LSND anomaly is due to breaking of Lorentz Invariance.   Both $ U_{e 4}$ and $U_{\mu 4}$ must be small resulting in a double suppressed appearance signal unlike the single suppressed disappearance measurements.  More sensitivity arises from appearance physics than disappearance physics because backgrounds are more suppressed for wrong-sign muon searches.  Assuming oscillations of the type indicated by LSND are present, Table~\ref{tab:raw_evt} shows the event rates between the null hypothesis of no oscillations versus LSND best-fit oscillations.
More details can be found in \cite{Tunnell:2011ya,Tunnell:2012}.

\begin{figure}[htbp]
  \begin{center}
	\includegraphics[width=0.75\textwidth]{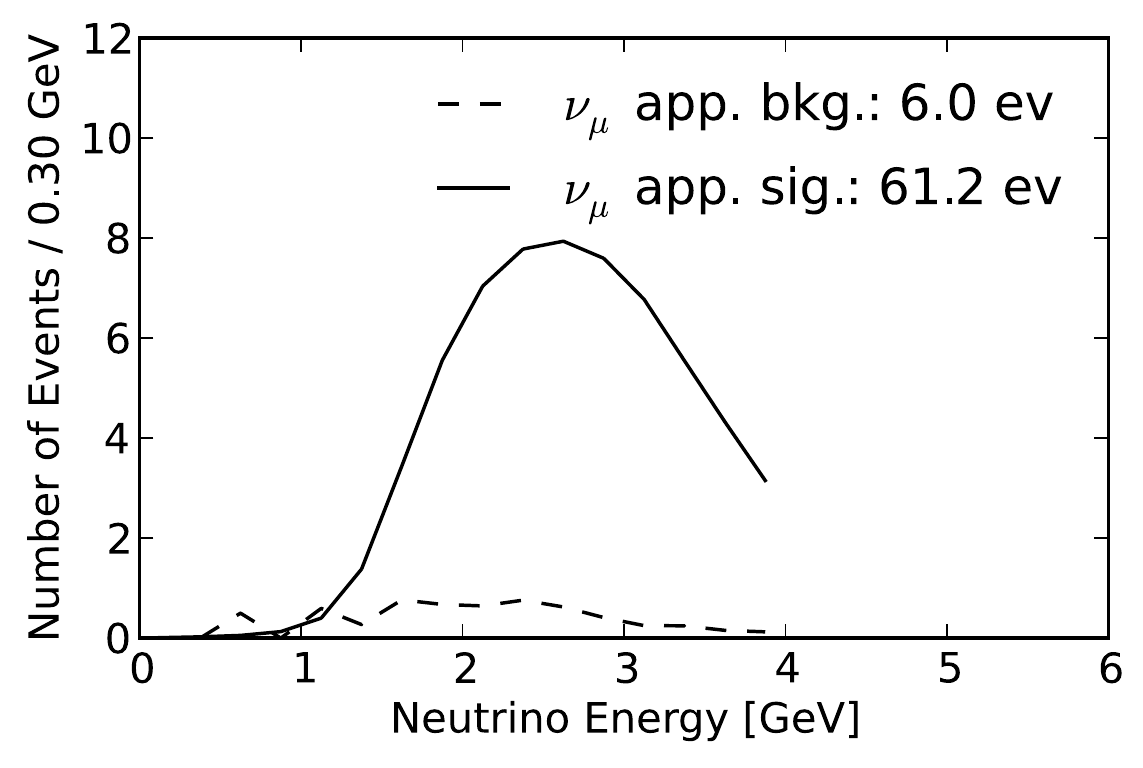}
  \end{center} 
\caption{The expected energy spectrum of signal and background events for stored positive muons.  The energy smearing matrix described in the detector performance section is used.  The fluctuations in the background correspond to fluctuations in the MC-derived matrices.}
\label{fig:appear_ch}
\end{figure}

The raw event rates in Table~\ref{tab:raw_evt} indicate the level of background rejection required to extract the $e \to \mu$ oscillations.  Detector simulation reveal both the energy smearing matrix and the probability that an event is included as signal.  These simulations have been performed for all channels present at this facility (see Section~\ref{sec:analysis}).  The $\chi^2$ shown includes only statistical uncertainties.  Fig.~\ref{fig:appear_ch} gives the spectrum for the expected signal and background levels for stored $\mu^+$ given the analysis from above.
\begin{figure}[htbp]
\begin{center}
		\includegraphics[width=0.8\textwidth]{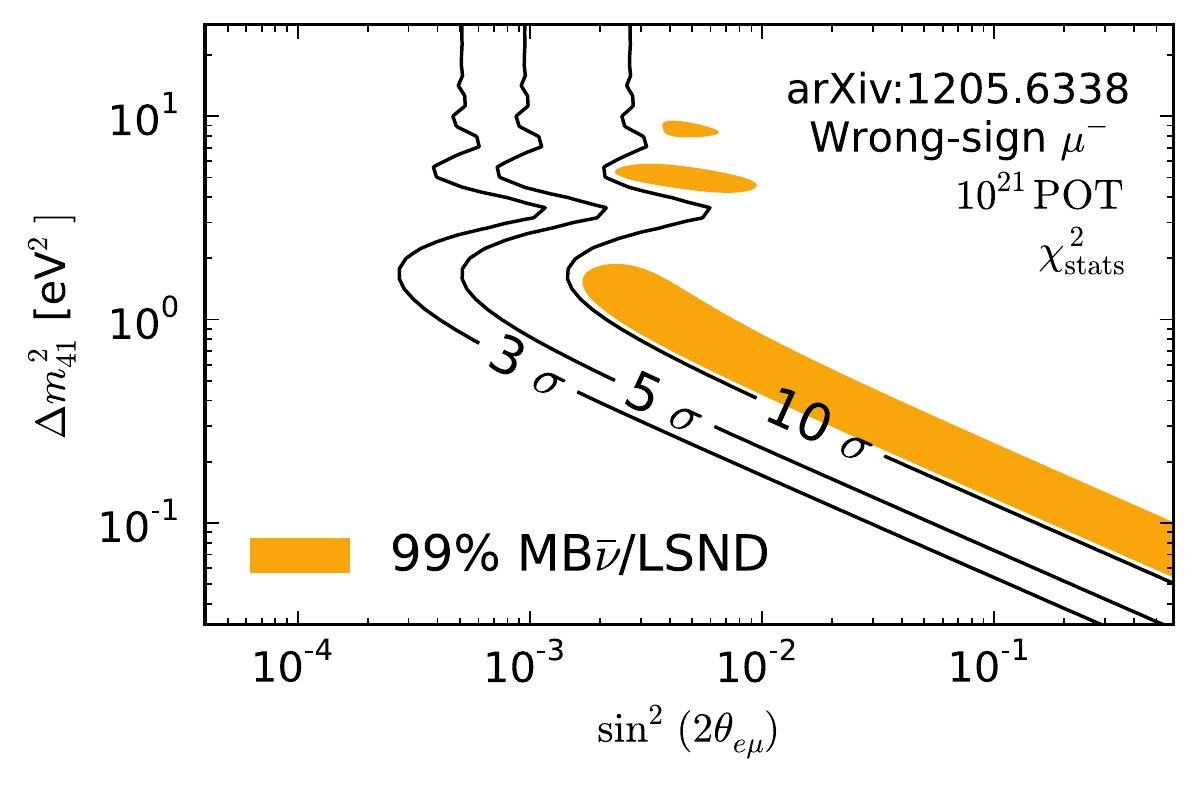}
\end{center}
\caption{ Contour in sterile parameter space associated with $\nu_e \to \nu_\mu$ appearance.  Assumed is $1.8 \times 10^{18}$ stored $\mu^+$ at $p = (3.8 \pm 0.38) \text{ GeV/c}$ and a detector at 2 kilometers with a fiducial mass of 1.3 kilotonne.  A smearing matrix is used corresponding to 2 cm steel plates.  The 150 m integration straight and detector volume are integrated over.  The CPT-conjugate of the LSND best-fit region is shown.}
  \label{fig:contours}
\end{figure}
A sensitivity contour is shown in Fig.~\ref{fig:contours}.  For this figure, $10\sigma$ corresponds to the $\chi^2$ value corresponding to the same p-value as a $10\sigma$ upward Gaussian fluctuation.  This contour shows that in this channel alone $\nu$STORM is able to provide an $10 \sigma$ measurement of the LSND anomaly.  A near detector is not required for the appearance physics analysis, unlike that of the disappearance analysis, given the accelerator instrumentation within the decay ring and that this channel is not systematically limited, thus much higher $\Delta m^2$ can be probed than at previous experiments.
The momentum of 3.8 GeV/c and baseline of 2000 meters were chosen after an optimization 
(Shown in Fig.~\ref{fig:sens_baseline_versus_energy}).  The number of stored muons is independent of ring energy since the 10\% relative acceptance of the ring increases absolutely with energy and counteracts the decrease of pion production at higher energies (Fig.~\ref{fig:piprod}, right).

As the cuts-based detector performance section improves and various cost optimizations are done, there are numerous parameters that can be optimized to compensate and conserve the physics that can be done with this facility.  For example, the optimization of baseline and energy (Fig.~\ref{fig:sens_baseline_versus_energy}) allows one to optimize the baseline depending on site constraints or vary the energy of the ring if the decay ring cost become excessive.  As the cuts-based detector performance improves, the various background rejections (Fig.~\ref{fig:sens_eff_versus_nc} and \ref{fig:sens_eff_versus_cid}) allow for further overall optimizations with respect to physics reach.  The tools have been developed that will allow us to optimize over all components of $\nu$STORM.

\begin{figure}[htbp]
\begin{center}
\includegraphics[width=0.8\columnwidth]{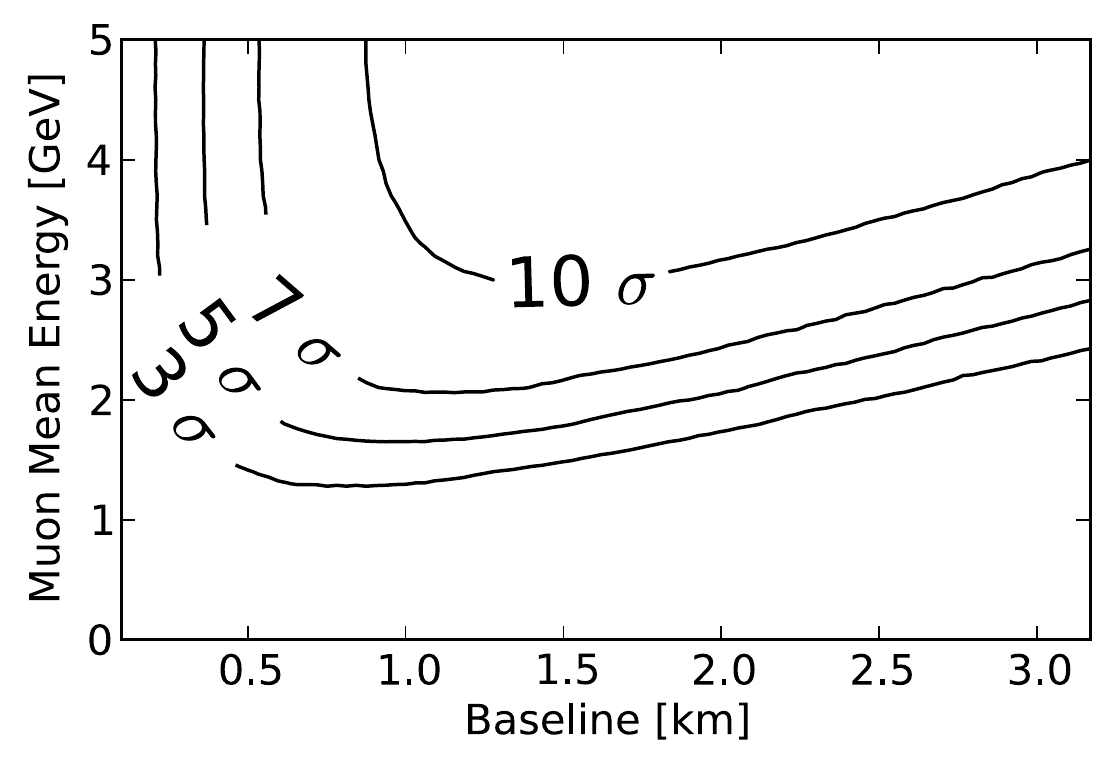}
\end{center}
\vspace*{-0.5cm}
\caption{\label{fig:sens_baseline_versus_energy} A baseline optimization using a total rates statistics-only $\chi^2$, a signal efficiency of 0.5, and background rejection of charge misidentification and NCs at $10^{-3}$ and $10^{-4}$.}
\end{figure}

\begin{figure}[htbp]
\begin{center}
\includegraphics[width=0.8\columnwidth]{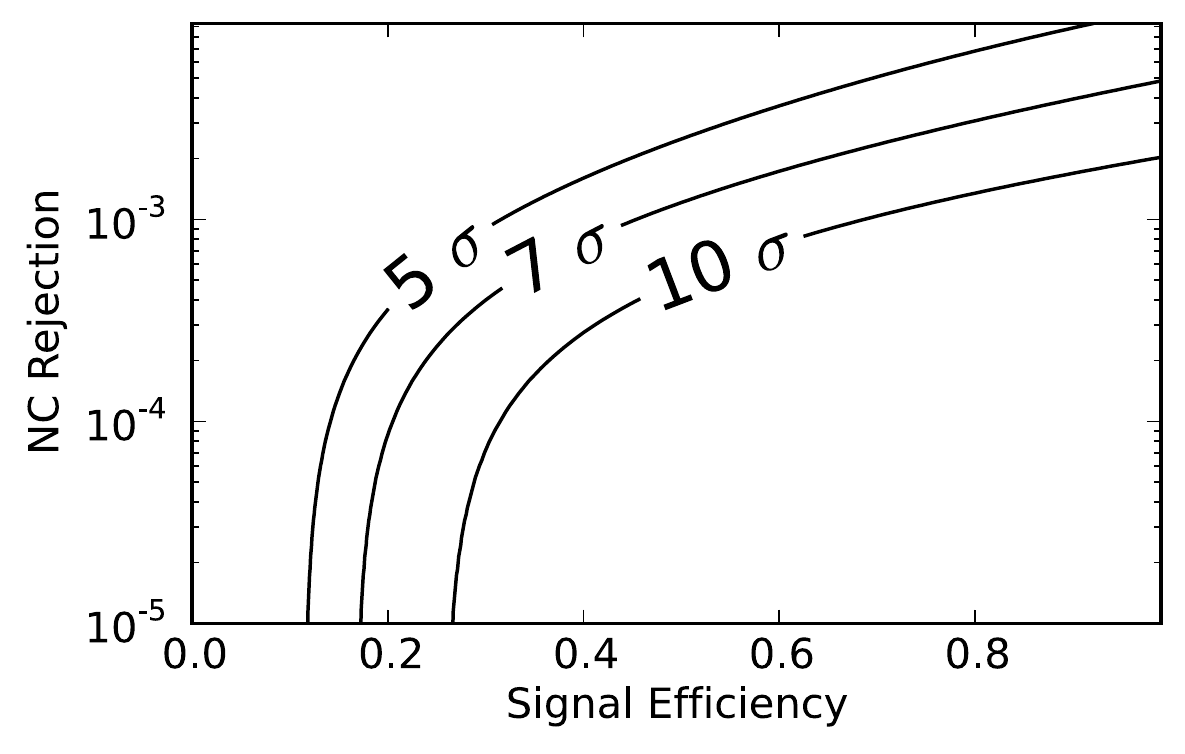}
\end{center}
\vspace*{-0.5cm}
\caption{\label{fig:sens_eff_versus_nc} Tuning the NC rejection cut.  The NC rejection level is shown versus the signal efficiency.  A charge misidentification background of $10^{-4}$ is shown to illustrate when NC backgrounds become statistically significant. A total rates statistics-only $\chi^2$ is used.}
\end{figure}

\begin{figure}[htbp]
\begin{center}
\includegraphics[width=0.8\columnwidth]{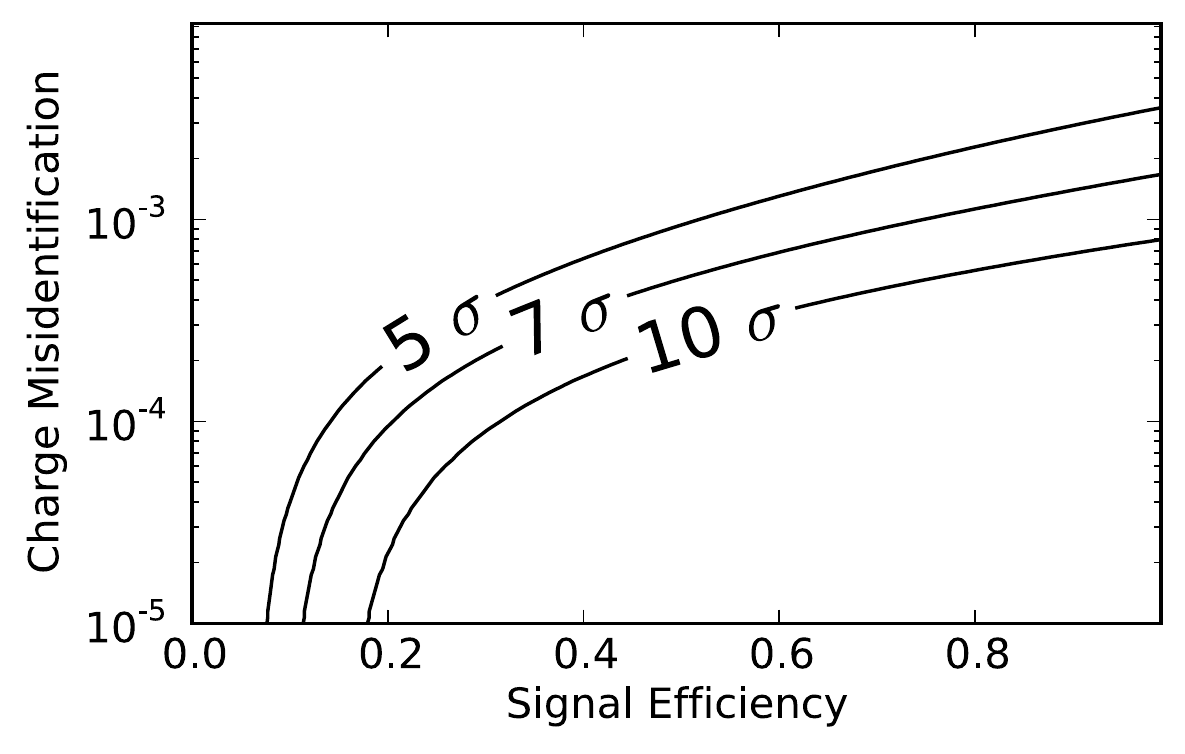}
\end{center}
\vspace*{-0.5cm}
\caption{ \label{fig:sens_eff_versus_cid} Tuning the charge misidentification cut.  The charge misidentification level is shown versus the signal efficiency.  A NC background of $10^{-4}$ is shown to illustrate when charge misidentification backgrounds become statistically significant. A total rates statistics-only $\chi^2$ is used.}
\end{figure}

\begin{figure}[htbp]
\begin{center}
\includegraphics[width=0.8\columnwidth]{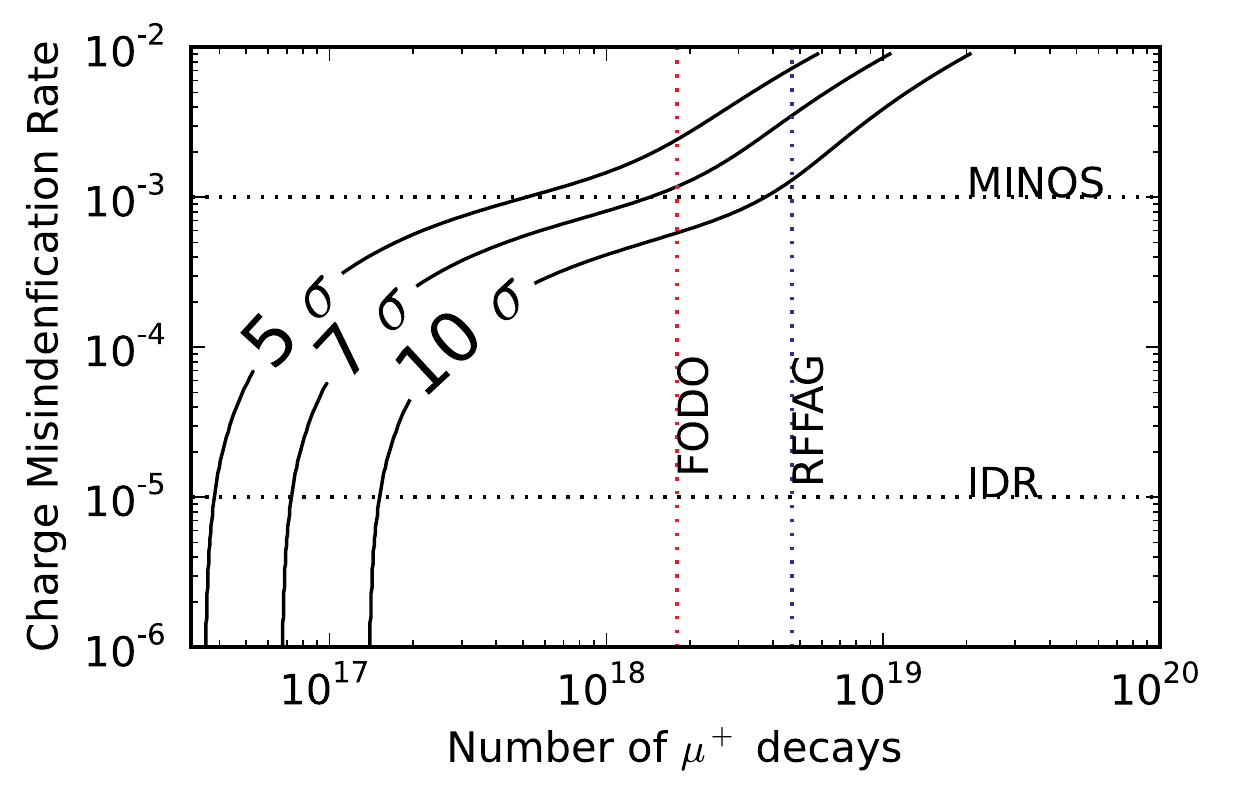}
\end{center}
\vspace*{-0.5cm}
\caption{\label{fig:sens_cid_versus_flux} An optimization between the detector performance and accelerator performance using the charge misidentification rates and number of muon decays as the performance metric.  IDR refers to the Interim Design Report \cite{NF:2011aa} detector performance.  FODO refers to the FODO lattice design that gives $1.8 \times 10^{18}$ useful muon decays whilst FFAG refers to the FFAG design that gives  $4.68 \times 10^{18}$ useful muon decays.  Both accelerators assume a front-end of the main injector at 60 GeV/c.}
\end{figure}

\subsubsection{Disappearance channels}
\begin{figure}[htbp]
\begin{center}
\includegraphics[width=0.48\textwidth]{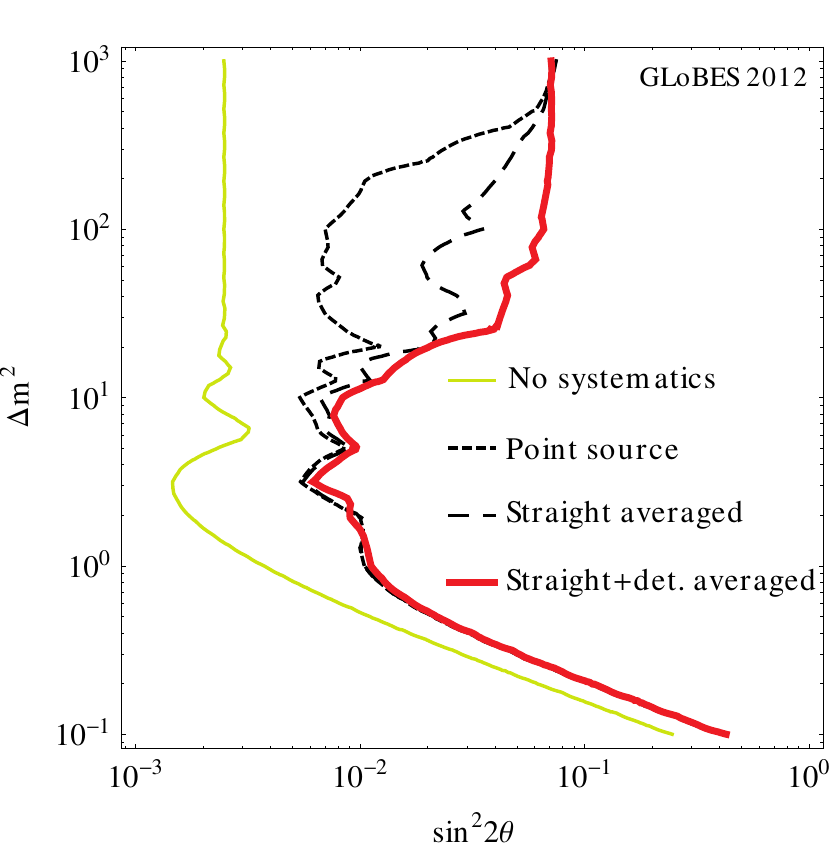} 
\hspace*{0.02\textwidth} \includegraphics[width=0.48\textwidth]{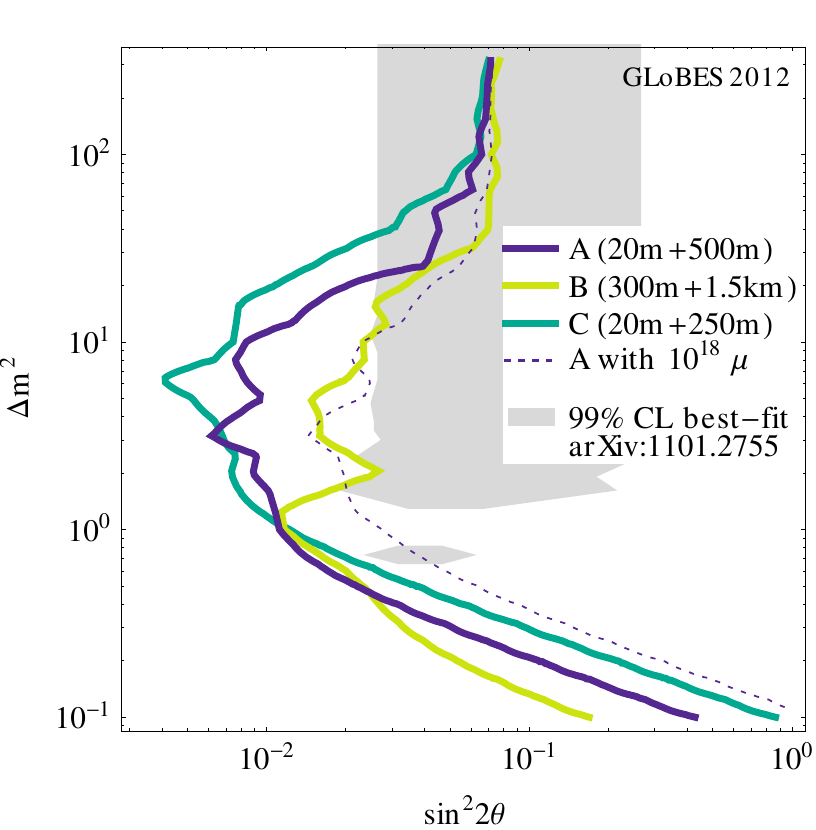}
\end{center}
\caption{\label{fig:disapp} Exclusion region in $\sin^2 2 \theta$-$\Delta m^2$ 
(right hand sides of curves) for $\nu_e$ disappearance for different geometry 
assumptions (left panel) and optimization points (right panel); 90\% CL, 2 d.o.f.
Left panel: The curve ``no systematics'' represents a single detector at $d=500 \,
\mathrm{m}$ using statistics only, whereas the other curves correspond to near-far
detector setups, where the red thick curves include (conservative) full systematics,
including a 10\% shape error, and geometry effects. Right panel: Systematics are fully
included, different two-distance optimization points shown (distances to the end of the
decay straight). Both panels: $E_\mu=2 \, \mathrm{GeV}$, $10^{19}$ useful muon decays per
polarity, $d_1=20 \, \mathrm{m}$ ($200 \, \mathrm{t}$) and $d_2 = 500 \, \mathrm{m}$
($1 \, \mathrm{kt}$), unless noted otherwise.  Note that the curve labeled 
disappearance in  Fig. 2 has to be compared to
the {\em product} of the $\nu_e$ and $\nu_\mu$ disappearance sensitivities.
Figure taken from \Ref~\cite{Winter:2012sk}.}
\end{figure}
Since disappearance measurements are very sensitive to the signal normalization, additional near
 detectors have been proposed in $\bar\nu_e$ disappearance reactor experiments to measure
 $\theta_{13}$~\cite{Minakata:2002jv,Huber:2003pm}. These near detectors are supposed to
 be as similar as possible to the far detectors, where the main purpose is to control
 the uncertainty on the reactor neutrino fluxes. This concept has been well established,
 and can be found in all of the state-of-the-art reactor experiments, such as Double Chooz,
 Daya Bay, and RENO. For $\nu$STORM, the situation is very similar: while the flux is well
 under control, cross sections $\times$ efficiencies must be measured by a near detector. However,
 since oscillations may already take place in the near detector, the oscillation parameters need
 to be extracted in a self-consistent way in a combined near-far fit~\cite{Giunti:2009en}.
 In fact, the near and far detectors may even swap the roles: while for $\Delta m^2 \simeq 1 \,
 \mathrm{eV}^2$, the near detector effectively measures the cross sections and the far detector
 the oscillation, for $\Delta m^2 \gg  10 \, \mathrm{eV}^2$, the near detector measures the oscillations
 and the far detector (where the oscillations average out) the cross sections. 

For the near-far detector combination, there are two crucial issues: the systematics implementation
and the treatment of geometry effects. In order to account for the uncertain cross sections $\times$
efficiencies, one can introduce a large systematic error, which is, however, fully correlated between
the two detectors which measure the same flavors and polarities in the disappearance channels. We
adopt the most conservative case for this systematic: we even assume a completely unknown shape,
i.e., we assume that the cross sections $\times$ efficiencies are unknown to the level of 10\% within
each bin, uncorrelated among the bins, but fully correlated between the near and far detectors
(shape error); for details and further considered systematics see \Ref~\cite{Winter:2012sk}.
Especially for the near detector, geometry effects turn out to be important: the oscillations
will average over the finite decay straight~\cite{Giunti:2009en,Winter:2012sk}, and the beam
divergence, which cannot be avoided at least from the muon decay kinematics, will lead to a
different beam spectrum in the near and far detectors~\cite{Tang:2009na,Winter:2012sk}. These
effects are illustrated in Fig.~\ref{fig:disapp}, left panel, in the two flavor picture: The curve
``Point source'' shows the sensitivity assuming a point neutrino source and a near detector in
the far distance limit, including full systematics. In this curve, a double peak in terms of
$\Delta m^2$ can be clearly seen, coming from the oscillations taking place in the near
($\Delta m^2 \gg 10 \, \mathrm{eV}^2$) or far ($\Delta m^2 \simeq 1 \, \mathrm{eV}^2$) detector.
If, however, the averaging over the decay straight (``Straight averaged'') and the detector geometry
(``Straight+detector averaged'') are taken into account, the large ($\Delta m^2 \gg 10 \,
\mathrm{eV}^2$)  sensitivity vanishes. The $\sin^2 2 \theta$ reach for very large $\Delta m^2$
relies on the external knowledge of systematics, in this case it is limited by the 10\% shape error

As far as the two-baseline optimization is concerned~\cite{Winter:2012sk}, the optimal choice
depends somewhat on the value of $\Delta m^2$. This is illustrated in the right panel of
Fig.~\ref{fig:disapp}, where the sensitivities for several optimization points are shown.
While all of these options perform equally well for $\Delta m^2 \simeq 1 \, \mathrm{eV}^2$,
larger values of $\Delta m^2 \simeq 1 \, \mathrm{eV}^2$ prefer shorter distances (from the
end of the decay straight) for the far detector. The optimization point~A (20~m+500~m) seems
to be a good compromise between the small and large $\Delta m^2$ sensitivities
for $E_\mu=2 \, \mathrm{GeV}$. This is consistent with the optimization for appearance,
but somewhat on the lower end of the optimal baseline range for that. For larger $E_\mu$,
slightly longer far detector distances are preferred, which means that $500 \, \mathrm{m}$ to $800 \,
\mathrm{m}$ seems a reasonable distance range. For the near detector, we find that, in spite
of the geometry effects, as short as possible distances are preferred if the far detector is
in that baseline range.

As for the absolute performance, we show in Fig.~\ref{fig:disapp} (right panel) the 99\% CL
best-fit from one of the global (anomaly) fits in the literature for comparison. It is clear
that $\nu$STORM can exclude  this region for all of the optimization points for $\Delta m^2
\lesssim 10 \, \mathrm{eV}^2$. However, note that either significantly more than $10^{18}$ useful
muon decays per polarity (dashed curve) are needed for that purpose, or muon energies slightly
higher than 2~GeV, as it the case in this document as the central momentum under consideration is 3.8 GeV/c. It can be shown that the proposed setup then has excellent sensitivity to
both $\nu_e$ and $\nu_\mu$ disappearance, both for neutrinos and antineutrinos~\cite{Winter:2012sk},
where the details somewhat depend on the final exposure, detection efficiency, and systematics treatment.

\clearpage
\section{Outlook and conclusions}
\label{sec:appendix}
The physics case for  experiments that search for sterile neutrinos is compelling.  Beyond the hints
from LSND, MiniBooNE and the reactor $\bar{\nu}_e$ flux anomaly, sterile neutrinos arise naturally in
many extensions of the Standard Model.  They appear in GUT models, in the seesaw mechanism and may 
also have an impact in cosmology as they are a possible candidate for DM or hot DM.  Of the 30 or so 
ideas to search for sterile neutrinos that have recently been discussed in the literature, $\nu$STORM is
the only one that can do all of the following:
\begin{itemize}
\item Make a direct test of the LSND and MiniBooNE anomalies.\vspace{-2mm}
\item Provide stringent constraints for both $\nu_e$ and $\nu_\mu$ disappearance to
over constrain $3+N$ oscillation models and to test the Gallium and reactor anomalies
directly.\vspace{-2mm}
\item Test the CP- and T-conjugated channels as well, in order to obtain the relevant clues 
for the underlying physics model, such as CP violation in $3+2$ models.\vspace{-2mm}
\end{itemize}

We have demonstrated the wide range of science that $\nu$STORM can deliver, ranging from probing the existence (or non-existence) of
sterile neutrinos to neutrino interaction physics in support of future programs, to the demonstration of  and test-bed for novel accelerator technology.   The 10 oscillation channels which $\nu$STORM can probe
 allows for the study, in depth and in detail, of the various sterile oscillation scenarios that are theoretically motivated, while simultaneously being the only proposal that can directly test the LSND anomaly at 10 $\sigma$.  A source of both electron and muon neutrinos allow for detailed cross section measurements where the electron neutrino cross section will be particularly important for future long baseline programs.  Experimental R\&D could also be done using this precisely understood neutrino source.  The program that has been proposed is able to do relevant physics on both the short term and long term.
\subsection{Proceeding toward a full Proposal}
In order to present a full proposal to the laboratory (with a defensible cost
estimate), additional scientific and engineering effort will be required.
We have estimated this effort and itemize it in Table~\ref{tab:Effort}.
\begin{table}[h]
\centering
\caption {Estimated effort to produce full proposal}
\label{tab:Effort}
\begin{tabular}{|l|c|}
\hline
Task & $\Sigma$FTE\\
\hline\noalign{\smallskip}
Target Station  					& 0.75     \\
Capture \& transport  				& 1.25     \\
Injection  							& 0.25     \\
Decay ring  						& 2     \\
Far Detector (Engineering) 			& 1    	\\
Far Detector (Sim \& Analysis)		& 2		\\
Near Detector (Engineering) 		& 1    	\\
Near Detector (Sim \& Analysis)\footnote{Note: Much of this effort is in complete synergy with the work on-going for the LBNE near detectors.  And what is given here is likely an overestimate for what will be needed for the $\nu$STORM proposal.}		& 3.5		\\
Costing								& 1		\\
\noalign{\smallskip}\hline
\noalign{\smallskip}\hline
{\bf Total}							&{\bf 12.75}	\\
\noalign{\smallskip}\hline
\end{tabular}
\end{table}

\cleardoublepage
\begin{appendix}
\section{Magnetized Totally Active Detector}
\label{sec:Appen}

We have shown in Sec.~\ref{sec:far} that a magnetized detector is required for $\nu$STORM if
we wish to study the $\parenbar{\nu}_\mu$ oscillation appearance channels and that this naturally lead to the choice of magnetized iron technology.  If one wanted to also look for $\parenbar{\nu}_e$ appearance,
then magnetized totally active detector technology would be an appropriate alternative.
Magnetic solutions for totally active detectors were studied within the International Scoping Study (ISS)~\cite{Abe:2007bi} in the context of investigating how very large magnetic volumes could be produced at
an acceptable cost.  A liquid Argon (LAr) or a totally-active sampling scintillator detector (TASD) could
be placed inside such a volume giving a magnetized totally active detector.
The following technologies were considered:

\begin{itemize}
\item Room Temperature Coils (Al or Cu)
\item Conventional Superconducting Coils
\item High Tc Superconducting Coils
\item Low Temperature Non-Conventional Superconducting Coils
\end{itemize}

Within the ISS much larger detector masses were considered than the 1 kT needed for $\nu$STORM. 
However, we can consider using one of the 10 large solenoids (each 15 m diameter $\times$ 15 m long)
studied in the ISS for use with a 1 kT LAr.  The ISS concept of a ``magnetic cavern" is shown in Fig. 52. 

\begin{figure}[htbp]
  \centering{
    \includegraphics[width=0.6\textwidth]{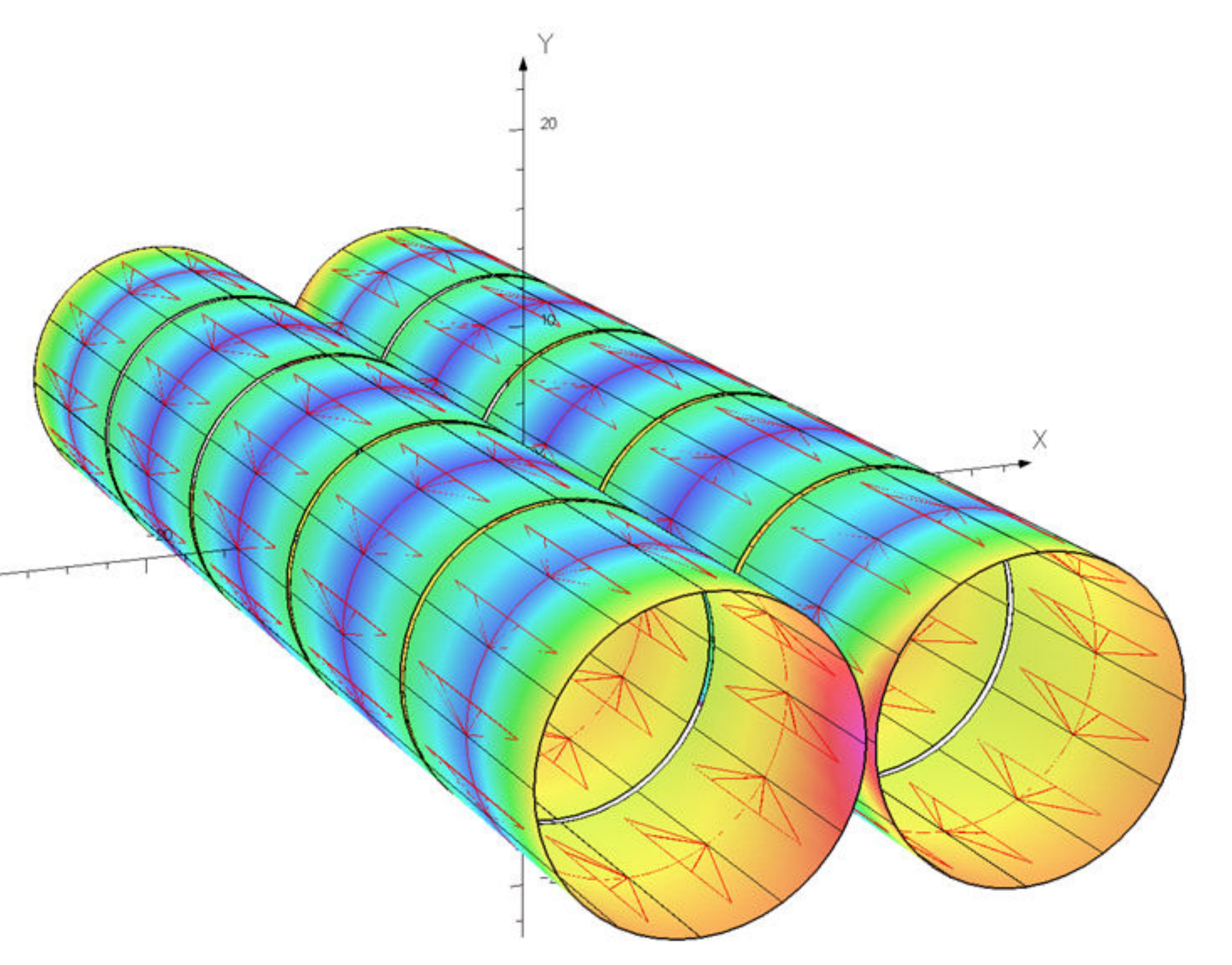}}
  \label{fig:Mag_cavern}
\caption{Magnetic Cavern Configuration}
\end{figure}

\subsection{Conventional Room Temperature Magnets}

In order to get adequate field strength with tolerable power dissipation, conventional
room-temperature coils would have to be relatively thick.  We first considered Al conductor
operating at 150K.  We then determined the amount of conductor necessary to produce a
reference field of only 0.1T.  In order to keep the current density at approximately 100A/cm$^2$,
10 layers of 1 cm$^2$ Al conductor would be required for our 15 m diameter $\times$ 15 m long
reference solenoid.  Using a \$20/kg cost for convention magnets~\cite{Green1}, the estimated cost 
for 1 solenoid is \$5M.  The power dissipation (assuming R=1 $\times$ 10$^{-8}$ Ohm-m) is
approximately 1 MW.  The operating costs for 1 MW of power would
be \$1.5M/year (based on typical US power costs).  The cost of the magnet system including 10
years of operation is then \$20M.  If one includes the cost of cooling the coils to 150K, the costs
increase substantially.  Studies have shown~\cite{Green1} that there is little cost benefit to operating
non-superconducting (Al or Cu) coils at low temperature vs. room temperature.  If we consider that the
power dissipation at room temperature for Al coils triples (vs. 150K operation), then the total magnet
cost increases to \$50M.

\subsection{Conventional Superconducting Coils}
Conventional superconducting solenoids are certainly an option for providing the large magnetic volumes
that are needed.  Indeed coils of the size we are considering were engineered (but never built) for the proposed
GEM experiment at the SSC.  A cylindrical geometry (solenoid) does imply that a fraction of the magnetic 
volume will not be outside the volume of the active detector which will likely be rectangular in cross section.  This is
certainly a disadvantage in terms in the terms of efficient use of the magnetic volume, but would provide 
personnel access paths to detector components inside the magnetic cavern.  It is certainly possible to consider
solenoids of rectangular cross section and thus make more efficient use the magnetic volume, but the engineering 
and manufacturing implications of this type of design have not been evaluated.

Technically, superconducting magnets of this size could be built, but at what cost?
There have been a number of approaches to estimating the cost of a superconducting magnet and we will mention
two of those there.  The first comes from Green and St. Lorant \cite{Green:1993dg}.  They looked at all the magnets that had 
been built at the time of their study (1993) and developed two formulas for extrapolating the cost of a superconducting
magnet: one scaling by stored energy and one scaling by magnetic volume times field.  They are given below:

\begin{equation}
C = 0.5(E_s)^{0.662}
\end{equation}
and
\begin{equation}
C = 0.4(BV)^{0.635}
\end{equation}
where $E_s$ is the stored energy in MJ, B in the field in Tesla, V is the volume in m$^3$ and C is the cost in M\$.  
The formulas given above give a cost for each 15 m diameter $\times$ 15 m long, 0.5T magnet of approximately
\$20M (based on E$_s$) and \$38M (based on magnetic volume). As another
reference point, we used the CMS coil \cite{Herve:2001kv} (B=4T, V=340 m$^3$, Stored energy = 2.7 GJ, Cost  = \$55M).
The Green and St. Lorant formulas give costs for the CMS magnet of  \$93M and \$41M based on stored energy and magnetic volume respectively.
From these data we can make  "Most Optimistic" and "Most Pessimistic" extrapolations for our baseline NF solenoid.  The most optimistic
cost comes from using the formula, based on stored energy and assume that it over-estimates by a factor of 1.7 (93/55) based on the CMS
as built cost.  This gives a cost of  \$14M for each of our NF detector solenoids.  The most pessimistic cost extrapolation comes from using the
formula based on magnetic volume and conclude that it under-estimates the cost by a factor of 1.3 (55/41), based on the CMS as built cost.
This then gives a cost of  \$60M for each of our NF detector solenoids.  There is obviously a large uncertainty represented here.

Another extrapolation model was used by Balbekov {\it et al.}~\cite{Balbekov1} based on a model developed by A. Herve.  The extrapolation
formulae are given below:

\begin{equation}
P_0 = 0.33S^{0.8}
\end{equation}
\begin{equation}
P_E = 0.17E^{0.7}
\end{equation}
and
\begin{equation}
P = P_0 + P_E
\end{equation}

where P$_0$ is the price of the equivalent zero-energy magnet in MCHF, P$_E$ is the price of magnetization, and P is the total price.
S is the surface area (m$^2$) of the cryostat and E (MJ) is the stored energy.  This model includes the cost of power supplies, cryogenics
and vacuum plant.  From the above equations you can see that the model does take into account the difficulties
in dealing with size separately from magnetic field issues.  Balbekov et. al. used three ``as-builts" to derive the coefficients in the above
equations:

\begin{itemize}
\item ALEPH (R=2.65m, L=7m, B=1.5T, E=138MJ, P=\$14M)
\item CMS (R-3.2m, L=14.5m, B=4T, E=3GJ, P=\$55M)
\item GEM (R=9m, L=27m, B=0.8T, E=2GJ, P=\$98M)
\end{itemize}

The GEM magnet cost was an estimate based on a detailed design and engineering analysis.  Using this estimating model we have for
one of the NF detector solenoids: P$_0$ = 0.33(707)$^{0.8}$ = 63MCHF,  P$_E$ = 0.17(265)$^{0.7}$ = 8.5MCHF.  The magnet
cost is thus approximately \$57M (which is close to our most pessimistic extrapolation given above).  One thing that stands out is that 
the magnetization costs are small compared to the total cost.  The mechanical costs involved with dealing with the large vacuum loading
forces on the vacuum cryostat assumed to be used for this magnet are by far the dominant cost.

\subsection{Low Temperature Non-Conventional Superconducting Coils}

In this concept we solve the vacuum loading problem of the cryostat by using the superconducting transmission line (STL)
that was developed for the Very Large Hadron Collider superferric magnets 
\cite{Ambrosio:2001ej}. The solenoid windings now consist
of this superconducting cable which is confined in its own cryostat.  Each solenoid consists of 150 turns 
and requires ~7500 m of cable. There is no large vacuum vessel and access to the detectors can be made through
the winding support cylinder since the STL does not need to be close-packed in order to reach an acceptable
field. We have performed a simulation of the Magnetic Cavern concept using STL solenoids and the results are shown in
Fig. 53.  With the iron end-walls ( 1m thick), the average field in the XZ plane is approximately

\begin{figure}[htbp]
  \centering{
    \includegraphics[width=0.6\textwidth]{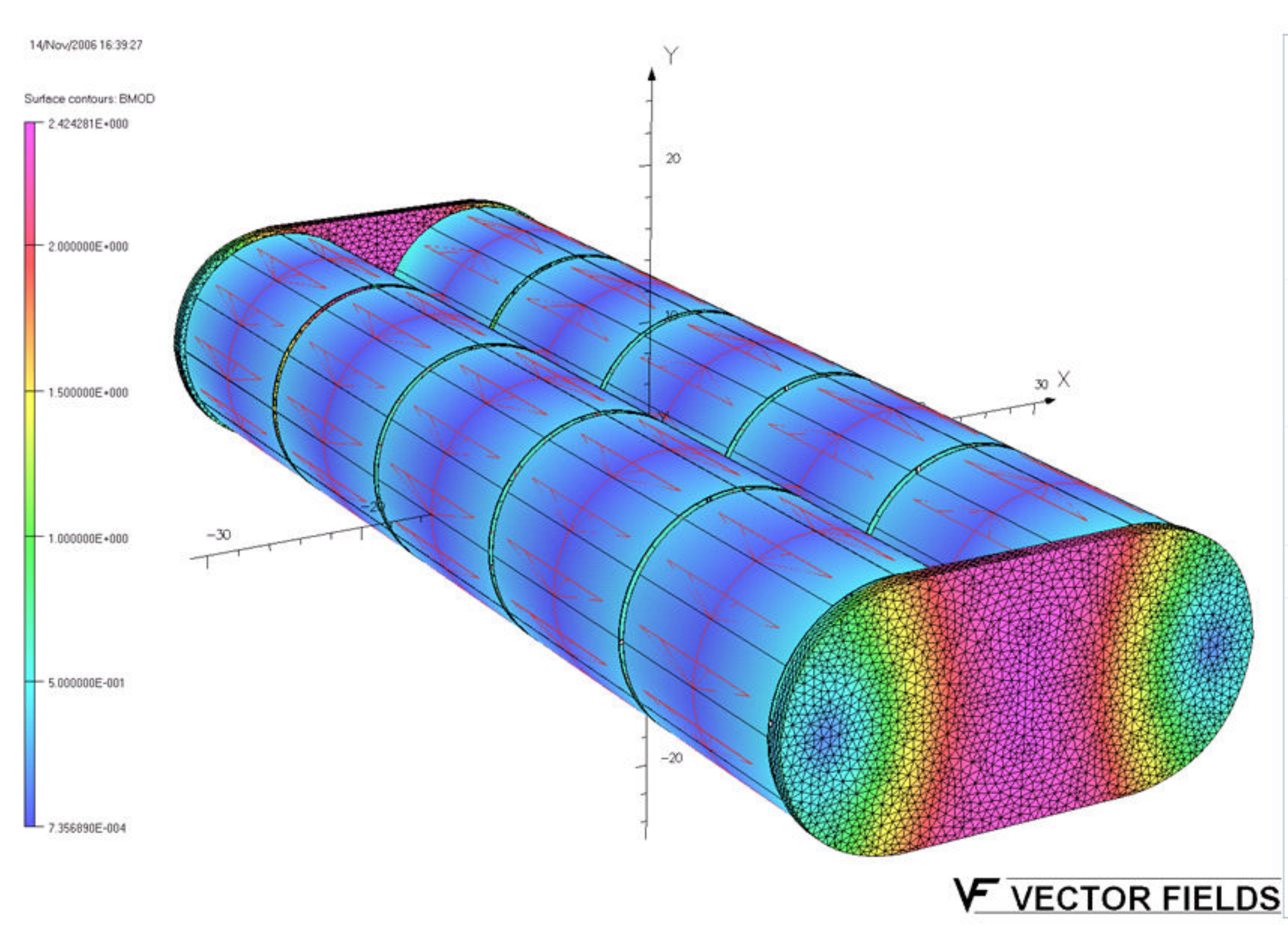}}
  \label{fig:Mag_cav_sim}
\caption{STL Solenoid Magnetic Cavern Simulation}
\end{figure}

0.58 T at an excitation current of 50 kA.   The maximum radial force is approximately 16 kN/m and the maximum axial force
approximately 40 kN/m.  The field uniformity is quite good with the iron end-walls and is shown in Fig. 54.

\begin{figure}[htbp]
  \centering{
    \includegraphics[width=0.6\textwidth]{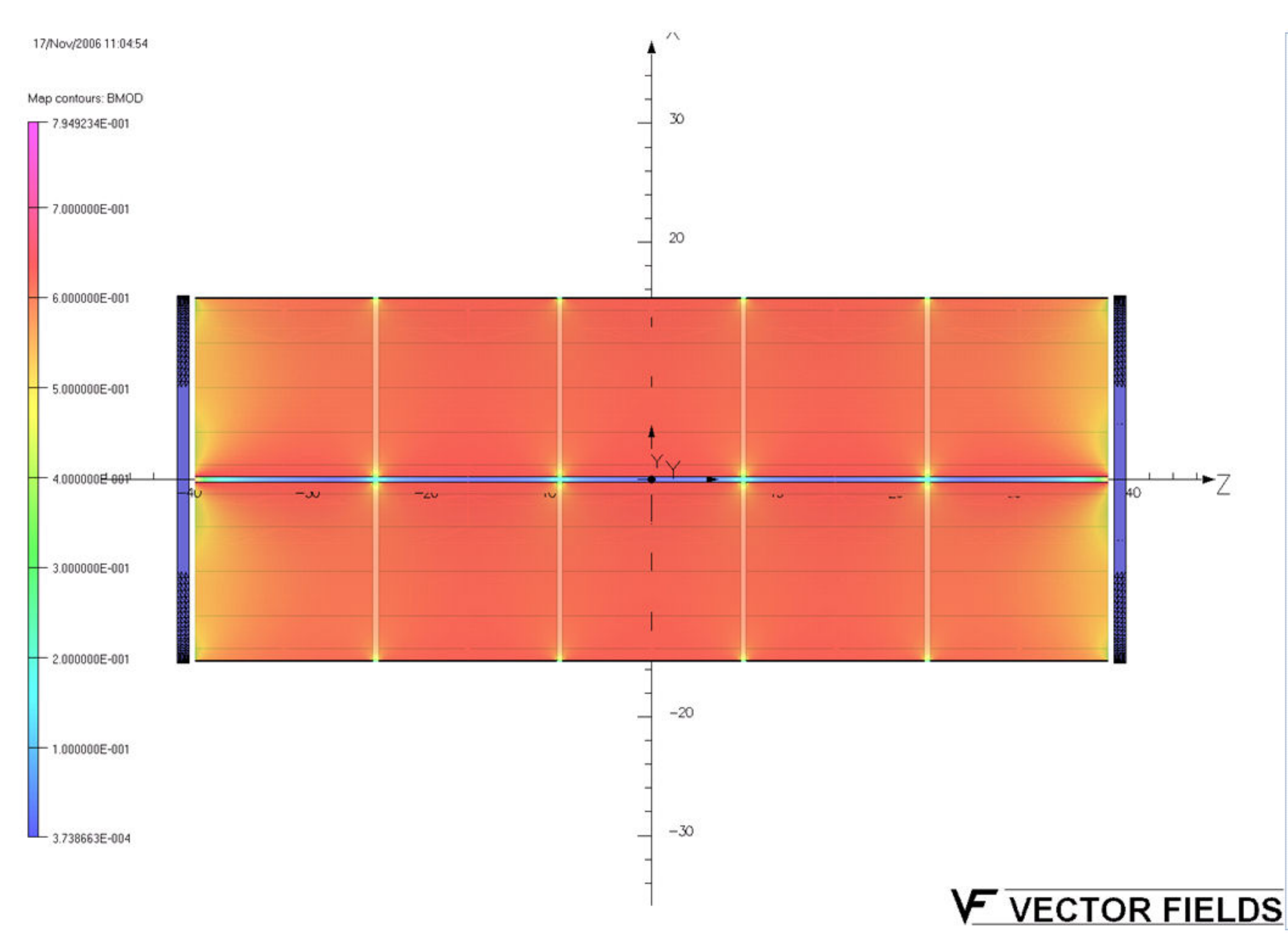}}
  \label{fig:Mag_cav_field_uni}
\caption{STL Solenoid Magnetic Cavern Field Uniformity in XZ plane}
\end{figure}

\subsection{Superconducting Transmission Line}

The superconducting transmission line (STL) consists of a superconducting cable inside a cryopipe cooled 
by supercritical liquid helium at 4.5-6.0 K placed inside a co-axial cryostat. It consists of a perforated Invar 
tube, a copper stabilized superconducting cable, an Invar helium pipe, the cold pipe support system, a thermal 
shield covered by multilayer superinsulation, and the vacuum shell. One of the possible STL designs developed 
for the VLHC is shown in Fig. 55.

\begin{figure}[htbp]
  \centering{
    \includegraphics[width=0.6\textwidth]{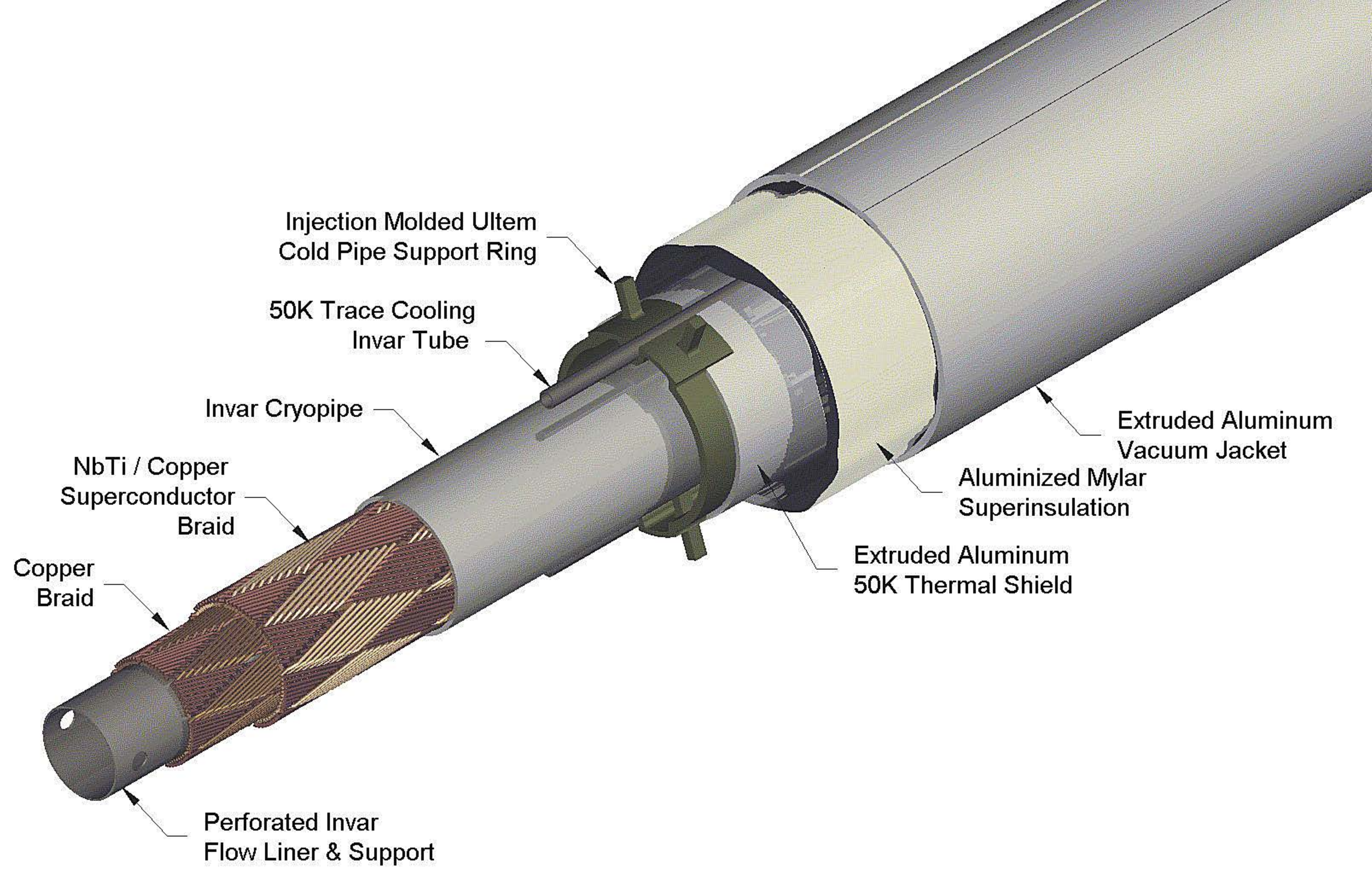}}
  \label{fig:STL}
\caption{Superconducting transmission line}
\end{figure}

The STL is designed to carry a current of 100 kA at 6.5 K in a magnetic field up to 1 T. This provides a 50\% 
current margin with respect to the required current in order to reach a field of 0.5T.  This operating margin
can  compensate for  temperature variations, mechanical or other perturbations in the system. 
The superconductor for the STL could be made in the form of braid or in the form of a two-layer spiral winding using 
Rutherford cable. The braid consists of 288 NbTi SSC-type strands 0.648 mm in diameter and arranged in 
a pattern of two sets of  24 crossing bundles with opposite pitch angle about the tube. A conductor made 
of Rutherford cables consists of 9 NbTi cables that were used in the SSC dipole inner layer. A copper braid is placed inside 
the superconductor to provide additional  current carrying capability during a quench. The conductor is
sandwiched between an inner perforated Invar pipe, which serves as a liquid helium channel, and an outer 
Invar pressure pipe that closes the helium space. Both braided and spiral-wrapped conductors and the 10 
cm long splice between them have been successfully tested with 100 kA transport current within the R\&D program
for the VLHC.  The STL has a 2.5-cm clear bore which is sufficient for the liquid helium flow in a loop up to 10 km in length. 
This configuration allows for cooling each  solenoid with continuous helium flow coming from a helium distribution box. 

The thermal shield is made of extruded aluminum pipe segments, which slide over opposite ends of each 
support spider. The 6.4-mm diameter Invar pipe is used for 50 K pressurized helium. It is placed in 
the cavities at the top and the bottom of both the shield and the supports. The shield is wrapped with 
40 layers of a dimpled super insulation. The vacuum shell is made of extruded aluminum or stainless steel.
Heat load estimates for the described STL are:

\begin{itemize}
\item Support system: 53 mW/m at 4.5 K and 670 mW/m at 40 K
\item Super insulation: 15 mW/m at 4.5 K and 864 mW/m at 40K
\end{itemize}

The estimated cost of the described STL is  approximately \$500/m. Further STL design optimization will be required to adjust the structure to the fabrication and operating conditions of the desired detector solenoids and to optimize its fabrication and operational cost.

\subsection{Conclusions}

Magnetizing volumes large enough to contain upwards of 1kT of LAr or totally active scintillator at fields up to 0.5T with the use of the STL concept would appear to be possible, but would require dedicated R\&D to extend the STL developed for the VLHC to this application.  It eliminates the cost driver of large conventional superconducting coils, the vacuum-insulated cryostat, and has already been prototyped, tested, and costed during the R\&D for the VLHC.  A full engineering design would still need to be done, but this technique has the potential to deliver the large magnetic volume required with a field as high as 1T, with very uniform field quality and at an acceptable cost.
\end{appendix}
\clearpage
\bibliography{%
./01_Overview/references_overview,%
./02_TnE_motivation/references_theory,%
./03_Facility/references_Facility,%
./04_FarDetector/references_far,%
./05_NearDetector/references_near,%
./06_Performance/references_performance,%
./07_OnC/references_OnC,%
./08_Appendix/references_Appendix%
}
%
\end{document}